\documentclass[11pt]{article}

\usepackage{jheppub} 
\usepackage{graphicx} 
\usepackage{caption} 
\usepackage{verbatim} 
\usepackage{mathtools}
\usepackage{url}
\usepackage{setspace}
\usepackage{breqn}
\usepackage{subcaption}
\usepackage{float}
\usepackage{leftidx}
\usepackage{bm}
\usepackage[TS1,T1]{fontenc} 
\usepackage[colorinlistoftodos]{todonotes}
\usepackage{soul}
\usepackage{textcomp}

\usepackage[utf8]{inputenc}
\allowdisplaybreaks

\usepackage{amsmath,amssymb}
\usepackage{array,booktabs,multicol,multirow,tabularx}

\def\D{\Delta}
\def\rhob{\overline{\rho}}
\def\zb{\overline{z}}
\def\wb{\overline{w}}
\def\Wb{\overline{W}}
\def\OO{\mathcal{O}}
\def\OS{\mathcal{O}_s}
\def\OT{\mathcal{O}_t}
\def\GG{\mathcal{G}}
\def\BB{\mathcal{B}}
\def\BB{\mathfrak{B}}
\def\PP{\mathcal{P}}
\def\QQ{\mathcal{Q}}
\def\ms{\mathrm{s}}
\def\mt{\mathrm{t}}
\def\mU{\mathrm{u}}
\def\kv{\mathbf{k}}

\renewcommand{\tfrac}[2]{\textstyle \frac{#1}{#2}}

\title{Mixed Correlator Dispersive CFT Sum Rules}

\author[a]{Anh-Khoi Trinh}

\affiliation[a]{Department of Physics, McGill University, 3600 Rue University, Montr\'eal, QC Canada H3A 2T8}

\emailAdd{anh-khoi.trinh@mail.mcgill.ca}

\abstract{
Conformal field theory (CFT) dispersion relations reconstruct correlators in terms of their double discontinuity.
When applied to the crossing equation, such dispersive transforms lead to sum rules that suppress the double-twist sector of the spectrum and enjoy positivity properties at large twist.
In this paper, we construct dispersive CFT functionals for correlators of unequal scalar operators in position- and Mellin-space.
We then evaluate these functionals in the Regge limit to construct mixed correlator holographic CFT functionals which probe scalar particle scattering in Anti-de Sitter spacetime.
Finally, we test properties of these dispersive sum rules when applied to the 3D Ising model, and we use truncated sum rules to find approximate solutions to the crossing equation.
}

\begin{document} 
\maketitle
\flushbottom

\section{Introduction}
The success of the modern conformal bootstrap stands upon the pillars of high precision
numerical methods \cite{Rattazzi:2008pe,Poland:2018epd}, and 
deeper understanding of the analytic structure of the bootstrap equations (see \cite{Kravchuk:2020scc,Kravchuk:2021kwe} and references therein).
Numerical methods have led to high precision measurements of critical exponents of conformal field theories (CFTs) such as the 3D Ising \cite{ElShowk:2012ht,El-Showk:2014dwa,Kos:2014bka,Caron-Huot:2020ouj} 
and the $O(N)$ models \cite{Kos:2013tga,Kos:2016ysd,Chester:2019ifh},
while analytical methods have proven essential to the study 
of quantum gravity in Anti-de Sitter (AdS) spacetime \cite{Heemskerk:2009pn,Penedones:2010ue,Rastelli:2016nze,Alday:2017vkk,Caron-Huot:2018kta,Goncalves:2019znr} and perturbative CFTs \cite{Kaviraj:2015cxa,Simmons-Duffin:2016wlq}.
In recent years, with the introduction of analytic functionals \cite{Mazac:2016qev,Mazac:2018mdx,Mazac:2018ycv,Mazac:2018qmi,Paulos:2019gtx,Mazac:2019shk,Paulos:2020zxx,Kaviraj:2021cvq} 
and dispersion relation methods \cite{Carmi:2019cub,Caron-Huot:2020adz,Caron-Huot:2021enk,Penedones:2019tng,Carmi:2020ekr,Gopakumar:2021dvg,Meltzer:2021bmb}, 
the gap between numerics and analytics has greatly narrowed revealing new horizons for the bootstrap program.
This paper builds upon this bridge by introducing new analytic dispersive functionals for correlators with unequal external scalar operators.

The numerical bootstrap relies upon a set of linear functionals that act on the crossing equation to determine whether crossing can be satisfied.
The derivative functional has long served as the cornerstone for the numerical bootstrap, but progress in the field has encouraged the pursuit of new methods with more desirable positivity properties and computational efficiency.
A potential candidate with such features is a so-called \textit{dispersive CFT functional} first introduced in \cite{Caron-Huot:2020adz}, which reconstructs functions in terms of their absorptive part \cite{Carmi:2019cub}.
Such a functional can be written as a dispersive transform acting on a stripped correlator $\GG(u,v)$:
\begin{equation} \label{dispersion relation}
\GG(u,v) = \GG^s(u,v) + \GG^t(u,v), \qquad \GG^{s,t}(u,v) = \iint du' dv' K(u,v;u',v') \text{dDisc}_{s,t} [ \GG(u',v') ],
\end{equation}
where we define the double discontinuity $\text{dDisc}[\GG(u',v')]$ in section~\ref{sec:general}, and the cross-ratio variables are
\begin{equation} \label{cross-ratios}
\begin{split}
u=z \zb,  &\qquad v=(1-z)(1-\zb), \\
u'=w \wb, &\qquad v'=(1-w)(1-\wb). 
\end{split}
\end{equation}
In this example, $\GG^{s,t}(u,v)$ can be thought as a functional acting on $\GG(u,v)$\footnote{We will refer to these transforms interchangeably as functionals and dispersive transforms. }.

For equal scalar operators of scaling dimension $\D_\phi$, the authors in \cite{Caron-Huot:2020adz} introduced the collinear $B_{k,v}$ dispersive functional which enjoys the following three key properties:
\begin{itemize}
\item they commute with the OPE,
\item they are sign definite above a twist threshold $\tau\geq 2\D_\phi+ k-2$, where $k\in 2 \mathbb{Z}$,
\item and they return double-zeros when acting on double-twist operators.
\end{itemize}
The first property is referred to as \textit{swappability}, and when combined with the second property, they lead to sum rules which balance the contribution of low twist operators to that of heavy operators:
\begin{equation}
\sum_{\OO_{light}} a_{\OO_{light}}w[ \GG] + \sum_{\OO_{heavy}} a_{\OO_{heavy}} w[\GG] = 0,
\end{equation}
where $a_\OO= f_{12\OO}f_{43\OO}$ is the OPE coefficient square, and $w[\cdot]$ is some dispersive functional. 
If the set of heavy operators consists mostly of double-twist operators, the last property suppresses their contribution to the sum rule effectively allowing one to probe non-perturbative properties of CFTs.
Together, these features represent an enticing addition to the CFT bootstrap toolkit.

When evaluated in the Regge limit, these dispersive functionals are further capable of probing bulk AdS physics.
This insight led to the construction of \textit{holographic} functionals in \cite{Caron-Huot:2021enk} where the collinear $B_{k,v}$ functional served as a seed in their construction.
Such holographic functionals are capable of bounding couplings in AdS, and they become dispersion relations in the flat-space limit \cite{Caron-Huot:2020cmc, Caron-Huot:2021rmr}.

Unfortunately, evaluating these $B_{k,v}$ functionals in position-space can become expensive in certain limits such as for large number of derivatives, or for low twist operators (near the unitarity bound).
An alternative approach is to construct them in Mellin-space \cite{Mack:2009mi,Penedones:2016voo,Gopakumar:2016wkt,Gopakumar:2016cpb,Gopakumar:2018xqi} where they can be computationally more efficient, and where projected functionals such as the $\Phi_\ell$ functional introduced in \cite{Caron-Huot:2020adz} are easier to construct.
This paper builds upon the construction of dispersion relations in both position- and Mellin-space by extending previous results to mixed correlators.

In section~\ref{sec:review}, we briefly review the fundamental features of dispersive CFT sum rules before constructing the 
position-space kernel $B_{\kv;v|mn}^{s,t}$ in section~\ref{sec:mixed pos}, where we further expound on their convergence and positivity properties.
We then use our results to derive holographic dispersive functionals for mixed correlators in section~\ref{sec:pos Regge}.
In section~\ref{sec:mixed mellin}, we perform a corresponding analysis in Mellin-space which benefits from improved computational efficiency and manifest zero structure.
Finally, in section \ref{sec:app}, we apply these functionals to study the 3D Ising model. 
We first evaluate sum rules using our functionals on mixed correlators to verify their convergence and precision properties. 
We then use insight gained from this analysis to build a system of truncated sum rules to estimate approximate solutions to crossing for the 3D Ising model.

\section{Overview of dispersive CFT sum rules} \label{sec:review}
Dispersive transforms repackage information encoded in the usual OPE and crossing equations into dispersive sum rules:
\begin{equation}
\begin{split}
& \sum_{\OO_s} f_{\phi_1\phi_2\OO_s}f_{\phi_4\phi_3 \OO_s} G^s_{\D_{\OO_s},J_{\OO_s}}= \sum_{\OO_t} f_{\phi_3\phi_2\OO_t}f_{\phi_4\phi_1 \OO_t} G^t_{\D_{\OO_t},J_{\OO_t}} \\
& \rightarrow \sum_{\OO_s} f_{\phi_1\phi_2\OO_s}f_{\phi_4\phi_3 \OO_s} \ w[G^s_{\D_{\OO_s},J_{\OO_s}} ] - \sum_{\OO_t} f_{\phi_3\phi_2\OO_t}f_{\phi_4\phi_1 \OO_t} \ w[G^t_{\D_{\OO_t},J_{\OO_t}} ]= 0,
\end{split}
\label{OPE to dispersive sum rules}
\end{equation}
where $f_{ijk}$ are OPE coefficients, $G_{\D,J}^{s,t}$ are $s$- and $t$-channel blocks respectively defined at the beginning of the next section, and $w[\cdot]$ is a dispersive functional with the key property that the action on double-twist operators is suppressed. 
Swappability ensures that the action of the functional commutes with the OPE.
We will briefly sketch how to obtain dispersive sum rules in this section.

A transform is said to be dispersive if it reconstructs the correlator in terms of its absorptive part. For CFTs,
the latter corresponds to the double discontinuity of the correlator.
Therefore, by construction, we seek the following position-space dispersive transform:
\begin{equation}
\GG^{s,t}(u,v) = \iint du' dv' K(u,v;u',v') \text{dDisc}_{s,t} [ \GG(u',v') ],
\end{equation}
where $\GG(u,v)$ is the stripped correlator corresponding to eq.~\eqref{unstripped correlator}.
We aim to derive kernels $K(u,v;u',v')$ valid for any CFT correlator.
To achieve this, one must ensure that such transforms are convergent, in particular in the Regge limit (see section~\ref{sec:convergence} for an extended discussion).
Convergence in the Regge limit is best understood in Mellin-space.

Our dispersive functionals are inspired by boundedness of the Mellin amplitude:
\begin{equation}
\oint \frac{d\ms'}{2\pi i} \frac{M(\ms,\mt)}{\ms - \ms'}  = 0,
\end{equation}
where $\ms,\mt$ are \textit{Mellin-mandelstam} variables.
Boundedness as $\ms,\mt \rightarrow i \infty$ corresponds to boundedness in the ($\mU$-channel) Regge limit.
As shown in \cite{Haldar:2019prg,Penedones:2019tng}, Mellin amplitudes are bounded as follows:
\begin{equation} \label{mellin boundedness spin}
\lim_{|\ms| \rightarrow \infty}|M(\ms,\mt)|_{\ms+\mt \ \text{fixed}}  = |\ms|^J,
\end{equation}
where $J=1$ for physical correlators.
Therefore, Regge bounded Mellin amplitudes can be obtained by an appropriate rescaling:
\begin{equation}
M(\ms,\mt) \rightarrow \frac{M(\ms,\mt)}{\prod \displaylimits_{m,n=0}^{m_{max},n_{max}} (\ms-\ms_m)(\mt-\mt_n)}.
\label{Mellin Regge}
\end{equation}
These rescaled amplitudes lead to spin-$k$ \textit{subtracted} dispersion relations
denoted by subscripts $k$ where $k=n_{max}+m_{max}$.
By performing an inverse Mellin transform, one can obtain a position-space representation of these subtracted dispersion relations valid for any spin-$k$ bounded CFT correlators (by following the nomenclature of eq.~\eqref{mellin boundedness spin}) :
\begin{equation}
\GG_k^{s,t}(u,v) = \iint du' dv' K_k(u,v;u',v') \text{dDisc}_{s,t} [ \GG_k(u',v') ].
\end{equation}

If one applies these dispersive transforms on individual conformal blocks, the result is a so-called \textit{Polyakov-Regge} block denoted as
\begin{equation}
\PP_{k;\D,J}^{s,t}(u,v) = \iint du' dv' \ K_{k}(u,v;u',v') \ \text{dDisc}_{s,t}[G_{\D,J}^{s,t}(u',v') ].
\end{equation}
By expanding the OPE in eq.~\eqref{OPE to dispersive sum rules}, one concludes that such Regge-bounded correlators can be written in terms of a basis of $s$- and $t$-channel Polyakov-Regge blocks:
\begin{equation}
\GG_{k}(u,v) = \sum_{\OS}f_{12\OS}f_{43\OS}\PP_{k;\D_{\OS},J_\OS}^s(u,v) + \sum_{\OT} f_{32\OT}f_{41\OT} \PP_{k;\D_\OT,J_\OT}^t(u,v).
\end{equation}
The above is an example of a dispersive sum rule. 

While the dispersion relation above allows one to fully reconstruct a correlator from its absorptive part, our goal is to constrain CFT data using the least number of assumptions. 
Therefore, one can obtain simpler dispersion relation by taking the $s$-channel collinear limit $u \rightarrow 0$ of these Polyakov-Regge blocks:
\begin{equation}
\PP_{k;\D,J}^{s,t}(u,v) \underset{u\rightarrow 0}{\rightarrow} B_{k;v|mn}^{s,t}.
\end{equation}
For equal external operators, there is a unique collinear expansion.
This is no longer true for mixed correlators where the collinear limit allows for two $u$ expansions:
\begin{equation} \label{correlator s-ch collinear}
\PP(u,v) = \left(u^{\frac{\D_3+\D_4-\D_1-\D_2}{2}}+... \right) f(v) + \left( u^0 + ... \right) g(v),
\end{equation}
where the $\D_i$ dependence is fixed by our convention given in eq.~\eqref{unstripped correlator}.
The ellipses denote subleading powers of $u$. 
We identify the expansions of $u$ by their positive $\D_i$ powers i.e. $(mn)=\{(34),(12)\}$ corresponding to $u^{\frac{\D_m+\D_n-\D_1-\D_2}{2}}$.

Taking the $s$-channel collinear limit leads to a new family of functionals labelled by $(mn)$ and $v$ with their own kernels:
\begin{equation} \label{B transform position}
B_{\kv;v|mn}^{s,t} = \iint du' dv' \BB_{\kv|mn}^{\mathfrak{a},\mathfrak{b}}(v;u',v') \text{dDisc}_{s,t}[G_{\D,J}^{s,t}(u',v') ].
\end{equation}
We will show that these dispersive collinear functionals lead to the following sum rule:
\begin{equation} \label{B sum rule}
\sum_{\OO_s} f_{12\OO_s}f_{43\OO_s} B_{\kv;v|mn}^{s} + \sum_{\OO_t} f_{32\OO_t}f_{41\OO_t} B_{\kv;v|mn}^{t} = \mathbf{1}_{\mU},
\end{equation}
where $\mathbf{1}_{\mU}=1$ is the $\mU$-channel identity\footnote{
Our convention differs from that of \cite{Caron-Huot:2020adz,Caron-Huot:2021enk}: for equal operators, the $\mU$-channel identity is always present. Therefore, they define their sum rules to be normalized as follows:
\begin{equation}
\sum_\OO f_{\phi \phi \OO}^2 B_{k,v}[G_{\D_\OO,J_\OO}^s] = (-1)^{k/2-1}.
\end{equation}
}.
The derivation and analysis of these collinear $B_{\kv;v|34}$ functionals are the main goals of this paper.

As mentioned in the introduction, these dispersive functionals enjoy appealing features.
Firstly, these functionals are constructed in terms of their double discontinuity rather than the full correlator.
Moreover, the presence of the double discontinuity suppresses double-twist operators thereby allowing us to probe non-perturbative features of CFTs.
Finally, they enjoy positivity properties due to the double-zeros that make them desirable for the numerical bootstrap.
That being said,
from eq.~\eqref{B sum rule}, it is apparent that positivity properties can no longer be exploited as straightforwardly for mixed correlators:
individually, the $s$- and $t$-channel functionals enjoy sign definiteness, however the sum does not.
We will verify these statements in this paper.

\section{Position-space functionals} \label{sec:mixed pos}
\subsection{Generalities} \label{sec:general}
We consider the four-point function of unequal scalar operators
with scaling dimensions $\D_i$ written as
\begin{equation} \label{unstripped correlator}
\langle \phi_1(x_1)\phi_2(x_2)\phi_3(x_3)\phi_4(x_4) \rangle = \frac{1}{(x_{13}^2)^{\frac{\D_1+\D_2+\D_3-\D_4}{2}} (x_{24}^2)^{\D_2}} \frac{(x_{34}^2)^{\frac{\D_1+\D_2-\D_3-\D_4}{2}} }{(x_{14}^2)^{\frac{\D_1+\D_4-\D_2-\D_3}{2}}} \GG(u,v).
\end{equation}
where $u,v$ correspond to the usual cross-ratios
\begin{equation}
u=z \zb = \frac{x_{12}^2 x_{34}^2}{x_{13}^2 x_{24}^2}, \qquad v=(1-z)(1-\zb) = \frac{x_{14}^2 x_{23}^2}{x_{13}^2 x_{24}^2}.
\end{equation}

This unconventional prefactor is chosen to fix the $\mU$-channel identity to be $u^{-\frac{\D_2-\D_4}{2}}$; we will highlight certain key distinctions in the following paragraphs.
This stripped prefactor fixes the normalization of the $s$- and $t$-channel conformal blocks as follows:
\begin{subequations} \label{blocks OPE exp}
\begin{align}
G_{\D,J}^{s;\D_1\D_2\D_3 \D_4}(z,\zb) &\sim (z \zb)^{-\frac{\D_1+\D_2}{2}} \times z^{\frac{\D-J}{2}} \zb^{\frac{\D+J}{2}} , \qquad \text{ for } 0<z\ll \zb \ll 1, \\
G_{\D,J}^{t;\D_1 \D_2 \D_3 \D_4}(z,\zb) &= G_{\D,J}^{s; \D_3 \D_2 \D_1 \D_4}(1-z,1-\zb).
\end{align}
\end{subequations}
We will abstain from writing the superscripts $\D_i$ in the future and refer to the definitions above for the $s$- and $t$-channel conformal blocks $G_{\D,J}^{s}$ and $G_{\D,J}^{t}$.
The blocks are further parametrized by the difference in scaling dimensions
\begin{equation} \label{a,b for blocks}
a = \frac{\D_{21}}{2}, \qquad \qquad b=\frac{\D_{34}}{2},
\end{equation}
where $\D_{ij}=\D_i-\D_j$.

Let us record the behaviour of conformal blocks in the $\mU$-channel Regge limit.
The Regge limit is reached by taking $x_1 \rightarrow x_3$ and $x_2 \rightarrow x_4$ along the lightcone; in terms of the cross-ratio variables, we take $z,\zb \rightarrow +i \infty$ while fixing $z/\zb$. In this limit, we find
\begin{equation}
\begin{split} \label{block Regge limit}
&G_{\D,J}^s(z,\zb), G_{\D,J}^t(z,\zb) \sim z^{-2\D_2} + z^{-\left(\D_1+\D_2+\D_3-\D_4\right)}+ z^{-\frac{\D_1+3\D_2+\D_3-\D_4}{2}}, \\
&G_{\D,J}^u(z,\zb)\sim z^{-\D_2+\D_4+j-1}, \\
\end{split}
\end{equation}
Notably, $s$- and $t$-channel conformal blocks are more convergent in the Regge limit than their $\mU$-channel counterpart for arbitrary spin $J$.
We present these limits to highlight the overall $\D_i$ dependence for mixed correlators which will play an important role in studying convergence properties of our dispersive transforms.

Finally, given our convention, the crossing relations now reads
\begin{subequations}
\begin{align}
\GG(z,\zb) &= \GG^{\D_1 \leftrightarrow \D_3}(1-z,1-\zb) &\qquad (\ms \leftrightarrow \mt),  \label{crossing s-t equation}\\
\GG(z,\zb) &=(z \zb)^{\D_2} \GG^{\D_1 \leftrightarrow \D_4}(1/z,1/\zb) &\qquad (\ms \leftrightarrow \mU), \\
\GG(z,\zb) &= ((1-z)(1-\zb))^{\frac{\D_1+\D_2+\D_3-\D_4}{2}} \GG^{\D_1 \leftrightarrow \D_2} \left(\frac{z}{1-z},\frac{\zb}{1-\zb} \right) &\qquad (\mt \leftrightarrow \mU).
\end{align}
\end{subequations}

The stripped prefactor in eq.~\eqref{unstripped correlator} was chosen given the desired properties of our dispersive functionals.
In particular, the latter exhibit manifest crossing symmetry by construction since they are derived in the \textit{crossing region} $\mathcal{R} \times \mathcal{R}$, where $\mathcal{R}=\mathbb{C} \setminus ((-\infty,0] \cup [1,\infty) )$ is the cut plane;
this is further illustrated by the lack of overall $z,\zb$ factors in eq.~\eqref{crossing s-t equation}.
In this domain, the $s$-channel OPE converges in $z \in \mathbb{C} \setminus [1,\infty)$, while the $t$-channel OPE converges in $\mathbb{C} \setminus (-\infty,0]$.
The domain of convergence for both $s$- and $t$-channels are a subset of the Lorentzian lightcones $L_{us}$ and $L_{tu}$ respectively, where the notation highlights the channels where the OPE converges; this notation is further elaborated in section~\ref{sec:convergence} where we discuss the integration region of these dispersive transforms.
We underscore the fact that these domains encode information about the $\mU$-channel OPE.

For a two-variable function $G(w,\wb)$ defined in the crossing region, we define the \textit{double discontinuity} as follows:
\begin{equation}
\begin{split}
\text{dDisc}_s[G(w,\wb)] &=\tfrac{1}{2}\left( G(w_+, \wb_-)+G(w_-,\wb_+)-G(w_-,\wb_-)-G(w_+,\wb_+) \right) \ \text{for } w,\wb<0 , \\
\text{dDisc}_t[G(w,\wb)] &=\tfrac{1}{2}\left( G(w_+, \wb_-)+G(w_-,\wb_+)-G(w_-,\wb_-)-G(w_+,\wb_+) \right) \ \text{for } w,\wb>1.
\end{split}
\end{equation}
When $G(w,\wb)$ is the stripped correlator $\GG(w,\wb)$ on the RHS of eq.~\eqref{unstripped correlator},
one can show that for $w,\wb <0$,
the above reduces to the usual definition of the double discontinuity used in the Lorentzian inversion formula \cite{Caron-Huot:2017vep}:
\begin{equation}
\begin{split}
\text{dDisc}_s G(w,\wb) &\equiv \cos(\pi (a+b) ) G_E(w,\wb) \\
& \qquad - \frac{e^{i\pi(a+b) }}{2} G_E(w,\wb \circlearrowright 0) - \frac{e^{-i\pi (a+b)} }{2} G_E(w,\wb \circlearrowleft 0) .
\end{split}
\end{equation}
A similar relation exists for the double discontinuity of the $t$-channel when $w,\wb>1$.
The relations above are made possible because the stripped correlator is \textit{Euclidean single-valued}  \cite{Caron-Huot:2018kta,Mazac:2019shk}
\footnote{Given a branch point at $(w,\wb)=(p,p)$, a function is Euclidean single-valued if and only if
\begin{equation}
G(w_+,\wb_-) = G(w_-, \wb_+), \qquad w_{\pm} = w \pm i 0.
\end{equation}
For the remainder of this paper, \textit{Euclidean single-valued} objects are simply referred to as being \textit{single-valued}.},
which assigns a Lorentzian interpretation to the discontinuities in the crossing region as originating from the double commutator of four operators in Lorentzian kinematics.
The latter can be interpreted as a scattering matrix where pairs of operators are time-like from each other in different Rindler wedges.
This interpretation makes it clear that it is a positive definite matrix with respect to the two pairs 12 and 34.

Indeed, when acting on a conformal block, the double discontinuity multiplies the blocks by sine functions:
\begin{equation}
\text{dDisc}_s[G_{\D,J}^{s}(w,\wb)] = 2 \sin\left( \tfrac{\D-J-\D_1-\D_2}{2} \pi \right) \sin\left( \tfrac{\D-J-\D_3-\D_4}{2} \pi \right) G_{\D,J}^s(w,\wb).
\end{equation}
An analogous relation exists for the double discontinuity across the $t$-channel cut.
Crucially, if $\D_1+\D_2-\D_3-\D_4 \in 2\mathbb{Z}$, the $s$-channel double discontinuity returns double-zeros when $\D$ takes on $s$-channel double-twist values $\D_{n,J} = \D_1+\D_2 +J+ 2n$ thereby suppressing the double-twist sector of the spectrum, and ensuring that the double discontinuity is positive definite.
This feature plays a key role in probing non-perturbative features of CFTs.
For mixed correlators, simple zeros are allowed which causes the double discontinuity to no longer be positive definite.

\subsection{Deriving the position-space $\BB_{\kv|mn}^{\mathfrak{a},\mathfrak{b}}$ kernel}

After detailing our convention,
we are now ready to derive the dispersion kernel associated with the collinear functional $B_{\kv;v}^{s,t}$.
As discussed in the previous section,
these functionals are defined as the coefficient of the $u \rightarrow 0$ power of the full dispersion kernel in eq.~\eqref{dispersion relation}.
For generic external scalars, the small $u$ limit leads to two expansions as shown in eq.~\eqref{correlator s-ch collinear}. 
We label the functionals by pairs of integers $(mn) \in \{(34),(12) \}$ associated with the corresponding double-twist family.
Subleading $s$-channel double-twist trajectories can be measured by virtue of the soon-to-be presented subtraction schemes.

We seek to compute the kernel associated with eq.~\eqref{B transform position} which we rewrite here for convenience:
\begin{equation}
B_{\kv;v|mn}^{s,t} = \iint du' dv' \BB_{\kv|mn}^{\mathfrak{a},\mathfrak{b}}(v;u',v') \text{dDisc}_{s,t}[G_{\D,J}^{s,t}(u',v') ],
\end{equation}
where the labels will be defined later near eq.~\eqref{B34 sub ab}. To achieve this goal, we will proceed by starting with a \textit{subtracted} Mellin-space dispersion relation.

As discussed in the previous section, boundedness of the correlator in the $\mU$-channel Regge limit is necessary to obtain convergent dispersive sum rules; this in turn translates into the derivation of spin-$k$ bounded kernels.
To derive such a kernel, the authors of \cite{Caron-Huot:2020adz} proposed to apply Cauchy's formula to a rescaled crossing-symmetric Mellin amplitude as shown by eq.~\eqref{Mellin Regge}
where the $\ms$ and $\mt$ poles are located at $s$- and $t$-channel double-twist values.
For equal external operators, there is a single double-twist family $\tau=2\D_\phi-\ell+2m$ for both $s$- and $t$-channels.
For generic correlators, we are free to subtract operators with different double-twist families $\tau_{ij}=\D_i + \D_j - \ell +2m$.
Therefore, we promote the spin-$k$ convergence label to a vector $k \rightarrow \kv = (k_{12}, k_{23}, k_{34}, k_{14})$ where the vector components $k_{ij}$ are associated with the double-twist pairs $\tau_{ij}$ and the sum of the vector components determines the $\mU$-channel Regge decay rate of the kernel.
This can be understood in Mellin space as shown in eq.~\eqref{Mellin Regge}.

We therefore define the following ratios of Pochhammer symbols which will allow us to implement various subtractions schemes:
\begin{equation} \label{Pochhammer6}
P_{\kv}(\ms,\mt;\ms',\mt') = \frac{\left(\tfrac{\D_1+\D_2-\ms}{2} \right)_{k_{12}} \left(\tfrac{\D_2+\D_3-\mt}{2} \right)_{k_{23}} \left(\tfrac{\D_3+\D_4-\ms}{2} \right)_{k_{34}} \left(\tfrac{\D_1+\D_4-\mt}{2} \right)_{k_{14}} }{\left(\tfrac{\D_1+\D_2-\ms'}{2} \right)_{k_{12}} \left(\tfrac{\D_2+\D_3-\mt'}{2} \right)_{k_{23}} \left(\tfrac{\D_3+\D_4-\ms'}{2} \right)_{k_{34}} \left(\tfrac{\D_1+\D_4-\mt'}{2} \right)_{k_{14}}}.
\end{equation}
Zeros in $\ms$ and $\mt$ cancel poles from the Mellin amplitude, while poles in $\ms',\mt'$ strengthens the convergence in the Regge limit.

By construction, $k_{ij} \in \mathbb{Z}$\footnote{In \cite{Caron-Huot:2020adz}, they consider subtraction units of $k \in 2\mathbb{Z}$ which parametrize both $s$- and $t$-channel subtractions. }. 
However, the range of $k_{ij}$ can be further restricted depending on the scaling dimensions of the external operators; this is detailled in section \ref{sec:convergence} when we discuss convergence properties.
Between the two collinear expansion towers in $u$ and the various subtraction combinations labelled by $\kv$, it might appear that the space of dispersive sum rules has greatly expanded. 
However, we will find relations between different subtraction schemes and operator ordering.

\begin{figure}
\centering
\includegraphics[width=0.5\textwidth]{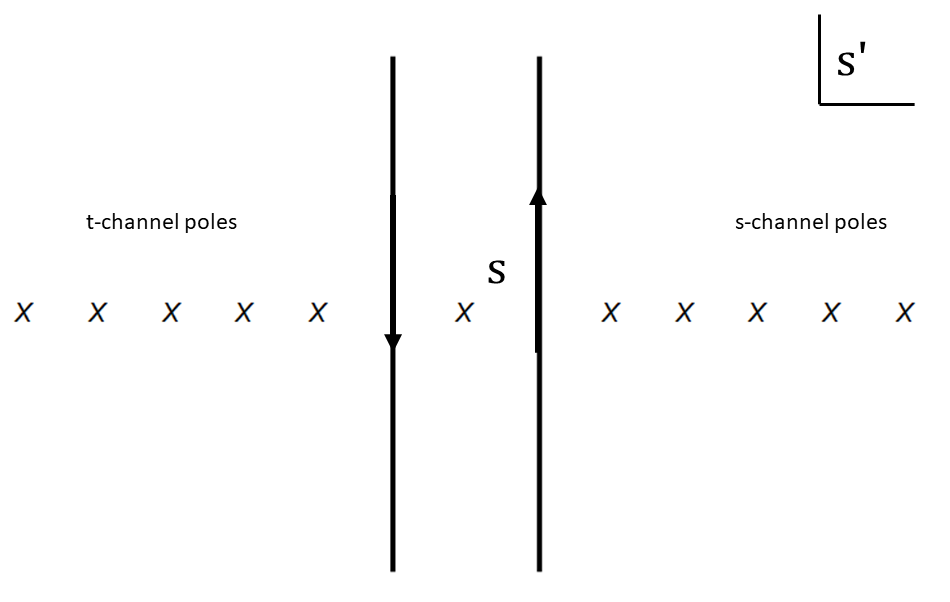}
\caption{Mellin contours for $M_{\kv}^s$ and $M_{\kv}^t$. $s$-channel poles are located at $\ms'=\tau+2m$ while $t$-channel poles are located at $\ms'=s+t-\tau-2m$ for $m\in \mathbb{Z}$. }
\label{Mellin poles}
\end{figure}

We now derive the $\BB_{\kv|mn}^{\mathfrak{a},\mathfrak{b}}$ dispersion kernel for mixed correlators in position-space based on the Mellin-inspired dispersion relation of \cite{Caron-Huot:2020adz}. 
The Mellin-space dispersion relation can be written as a contour integral in the $\ms'$-plane as follows:
\begin{equation} \label{M dispersion original def}
0= \oint \frac{ds'}{2\pi i} \frac{M_{\kv}(\ms,\mt;\ms',\mt')}{\ms-\ms'} =  M_{\kv}^s(\ms,\mt;\ms,\mt') + M_{\kv}^t(\ms,\mt;\ms',\mt'),
\end{equation}
where the $\ms'$-plane is illustrated in fig.~\ref{Mellin poles}, and
\begin{equation} \label{M dispersion def}
\begin{split}
M_{\kv}^s(\ms,\mt;\ms',\mt') &= \int\displaylimits_{\ms+\epsilon-i\infty}^{\ms+\epsilon+i\infty} \frac{d\ms'}{2\pi i} \frac{P_{\kv}(\ms,\mt;\ms',\mt')M(\ms',\mt')}{\ms-\ms'} \\
M_{\kv}^t(\ms,\mt;\ms',\mt') &= \int\displaylimits_{\ms-\epsilon-i\infty}^{\ms-\epsilon+i\infty} \frac{d\ms'}{2\pi i} \frac{P_{\kv}(\ms,\mt;\ms',\mt')M(\ms',\mt')}{\ms-\ms'} =- \int\displaylimits_{\mt+\epsilon-i\infty}^{\mt+\epsilon+i\infty} \frac{dt'}{2\pi i} \frac{P_{\kv}(\ms,\mt;\ms',\mt')M(\ms',\mt')}{\mt-\mt'}.
\end{split}
\end{equation}
In the above, we used the identity $\ms+\mt=\ms'+\mt'$ for a fixed-$\mU$ dispersion relations. 
To sketch the derivation of the kernel, we restrict ourselves to $\kv=0$ for the moment. 
We seek to express the Mellin amplitude $M(\ms',\mt')$ as the inverse Mellin transform of a positions-space double discontinuity $\text{dDisc}_{s}[\GG(u',v')]$. 
The Mellin representation of a correlator $\GG(u,v)$ is defined as follows:
\begin{equation}
\GG(u,v) = \iint_\gamma \frac{d\ms d\mt}{(4 \pi i)^2} \Gamma_{\D_i}^6(\ms,\mt) u^{\frac{\ms-\D_1-\D_2}{2}} v^{\frac{\mt-\D_2-\D_3}{2}} M(\ms,\mt),
\end{equation}
where $\Gamma_{\D_i}^6(\ms,\mt)$ is given by eq.~\eqref{Gamma6}.
We wish to substitute the Mellin amplitude defined by eq.~\eqref{M dispersion def} in the above..
Since the double discontinuity multiplies the block by sine factors with zeros at double-twist values, the Mellin amplitude must include corresponding factors that ``undo" this feature.
We therefore write $M(\ms',\mt')$ evaluated on the $\ms'$-channel poles as
\begin{align}
M(\ms',\mt')\Big|_{\ms'-\text{channel poles}} = \iint \frac{du' dv'}{u' v'} &\frac{u'{}^{\frac{\D_1+\D_2-\ms'}{2}} v'{}^{\frac{\D_2+\D_3-\mt'}{2}} }{2 \sin[\tfrac{\pi}{2}(\ms'-\D_1-\D_2)] \sin[\tfrac{\pi}{2}(\ms'-\D_3-\D_4)]} \qquad \nonumber \\
& \qquad \qquad \qquad \qquad \times \frac{\text{dDisc}_{s}[\GG(u',v')]}{\Gamma^6_{\D_i}(\ms',\mt')}.
\label{M spoles}
\end{align}

After inserting eq.~\eqref{M spoles} into \eqref{M dispersion def}, we can perform an inverse Mellin transform to return to position-space. 
By doing so, we arrive at the position-space dispersion relation of eq.~\eqref{dispersion relation} with the following dispersion kernel:
\begin{align}
K_{\kv}(u,v;u',v') =\frac{1}{u'v'} \iiint \frac{d\ms \ d\mt \ d\ms'}{(4\pi i)^2(2\pi i)} & \frac{\Gamma_{\D_i}^6(\ms,\mt)/\Gamma_{\D_i}^6(\ms',\mt')}{2 \sin[\tfrac{\pi}{2}(\ms'-\D_1-\D_2)] \sin[\tfrac{\pi}{2}(\ms'-\D_3-\D_4)]} \nonumber \\
& \qquad \times \frac{u{}^{\frac{\ms-\D_1-\D_2}{2}} v{}^{\frac{\mt-\D_2-\D_3}{2}}}{u'{}^{\frac{\ms'-\D_1-\D_2}{2}} v'{}^{\frac{\mt'-\D_2-\D_3}{2}}} \frac{P_{\kv}(\ms,\mt;\ms',\mt')}{\ms-\ms'}.
\label{Kernel general definition}
\end{align}
The subscript $\kv$ emphasizes that this \textit{subtracted} kernel reconstructs spin-$\kv$ bounded correlators for any subtraction scheme.
The $t$-channel dispersion kernel is obtained by substituting $\frac{d\ms'}{\ms-\ms'}$ with $\frac{d\mt'}{-\mt+\mt'}$ and by replacing the arguments of the sine functions for those of $t$-channel double-twists.
From eq.~\eqref{M dispersion original def}, it is clear that there is a single kernel for both the $s$- and $t$-channel correlators, and they must be related to each other by an overall minus sign.

In principle, one can compute $K_\kv$ to obtain a dispersion relation analogous to that of \cite{Carmi:2019cub} for spin-$\kv$ bounded mixed correlators.
However,  this can be challenging as discussed in appendix~\ref{app:B kernel}. 
Our goal is more modest: we aim to compute the collinear functional $B_{\kv;v|mn}^{s,t}$ whose kernel is simply eq.~\eqref{Kernel general definition} evaluated on the desired leading $s$-channel pole at $\ms=\D_{m}+\D_n+2k_{mn}$. 
A closed-form expression for the kernel $\BB_{\kv|mn}^{\mathfrak{a},\mathfrak{b}}$ is then obtained by deforming the remaining $\ms'$ and $\mt'$ contours and re-summing the series; see appendix~\ref{app:B kernel} for computational details.
By doing so, we find the following expression for the subtracted kernel valid for spin-$\kv$ bounded mixed correlators:
\begin{equation} \label{B34 sub ab}
\begin{split}
\BB_{\kv|34}^{\mathfrak{a},\mathfrak{b}}(v;u',v') &= \frac{(-1)^{1+k_{12}+k_{34}}}{4 \pi^{3/2} u^{\mathfrak{a}+\mathfrak{b}}  } \left( \frac{u}{u'} \right)^{k_{12}} \frac{v^{\mathfrak{a}+k_{23}} v'{}^{\mathfrak{b}-k_{23}}}{(v v')^{3/2}} \frac{\Gamma(\mathfrak{a}+\mathfrak{b})}{\Gamma(\mathfrak{a}+\mathfrak{b}-\tfrac{1}{2})}(u'-v-v') \chi^{3/2-\mathfrak{a}-\mathfrak{b}}  \\
& \qquad \qquad \times {}_2F_1(\mathfrak{a},\mathfrak{b},-\tfrac{1}{2}+\mathfrak{a}+\mathfrak{b},-\tfrac{1}{\chi}),
\end{split}
\end{equation}
where
\begin{equation} \label{chit var} 
\chi = \frac{4v v'}{(v - (\sqrt{u'}+\sqrt{v'})^2) (v- (\sqrt{u'}-\sqrt{v'})^2)},
\end{equation}
and
\begin{equation} \label{ab frak}
\mathfrak{a}= \frac{1}{2}(\D_1-\D_3+k_{12}-k_{23}-k_{34}+k_{14}), \qquad \mathfrak{b}= \frac{1}{2}(\D_2-\D_4+k_{12}+k_{23}-k_{34}-k_{14}).
\end{equation}
The kernel associated with the expansion powers $(mn)=(12)$ can be similarly derived.
One can show that the following relation holds:
\begin{equation} \label{B12 to B34 relation sub}
\BB_{\kv|12}^{\mathfrak{a},\mathfrak{b}}(v;u',v') = \left( \frac{u'}{u} \right)^{\mathfrak{a}+\mathfrak{b}} \BB_{\kv|34}^{\mathfrak{a},\mathfrak{b}}(v;u',v').
\end{equation}
Therefore, the two $s$-channel collinear expansions encode the same information.

A few comments are in order. Firstly,
we note that the kernels are antisymmetric under the swapping of $w \leftrightarrow \wb$; this is necessary to be consistent with the sum of $B_{\kv;v|mn}^s$ and $B_{\kv;v|mn}^t$ in eq.~\eqref{B sum rule}.
Furthermore, the combination of cross-ratios $\chi$ in eq.\eqref{B34 sub ab} naturally arises in the $u \rightarrow 0$ limit of the full equal operator dispersion kernel.
As shown in \cite{Carmi:2019cub}, 
the equal operator dispersion kernel can be conveniently packaged in terms of the following special combination of cross-ratio variables:
\begin{align}
\frac{16 \sqrt{u u' v v'}}{((\sqrt{u}+\sqrt{v})^2 - (\sqrt{u'}+\sqrt{v'})^2) ((\sqrt{u}-\sqrt{v})^2 - (\sqrt{u'}-\sqrt{v'})^2)} \underset{u\rightarrow 0}{=} 4 \chi \sqrt{\frac{u u'}{v v'}} + O(u).
\end{align}
We expect the special combination of cross-ratios on the LHS to appear in the full mixed correlator Polyakov-Regge block dispersion kernel.

Moreover, we find that the $u$ dependence correctly tracks the corresponding double-twist trajectories:
\begin{equation} \label{kernel leading trajectory}
\BB_{\kv|12}^{\mathfrak{a},\mathfrak{b}}(u,v;u',v') \propto u^{k_{12}}, \qquad \qquad \BB_{\kv|34}^{\mathfrak{a},\mathfrak{b}}(u,v;u',v') \propto u^{\frac{\D_3+\D_4-\D_1-\D_2+2k_{34}}{2}}.
\end{equation}
We see that subtractions along the $\tau_{12}$ and $\tau_{34}$ trajectories probe subleading double-twists trajectories.
We further note that $k_{mn}<0$ is allowed leading to ``anti-subtracted" functionals such as the $B_{-2}$ functional introduced in \cite{Caron-Huot:2021enk} which lead to superconvergent sum rules\cite{Kologlu:2019bco}.

Since $\mathfrak{a},\mathfrak{b} \in \mathbb{R}$, positivity properties are obscured in general.
We will revisit this statement in the next subsection after constraining the space of subtraction schemes and operator ordering by verifying convergence conditions.
However, we can immediately address the range of allowed $s$-channel subtractions: the latter is determined by the gamma function in the numerator of the kernels. This gamma function arises from the contour deformation in eq.~\eqref{M spoles} of the Mellin-Mandelstam variable $\ms$ evaluated on the desired $s$-channel double-twist trajectory. 
It is therefore fixed by the OPE and it is an artefact of taking the $s$-channel collinear limit. 
For generic subtractions, we have
\begin{equation} \label{Gamma function subtractions}
\BB_{\kv|34}^{\mathfrak{a},\mathfrak{b}} \propto \Gamma(\frac{\D_{13}+\D_{24}}{2}+k_{12}-k_{34}), \qquad \BB_{\kv|12}^{\mathfrak{a},\mathfrak{b}} \propto \Gamma(\frac{-\D_{13}-\D_{24}}{2}-k_{12}+k_{34}).
\end{equation}
The gamma functions diverge for non-positive integer arguments, and therefore the kernels are non-divergent under the following conditions:
\begin{equation} \label{s-subtraction gamma condition}
\begin{split}
B_{\kv;v|34}^{s,t}: \ & k_{12} - k_{34} +\frac{\D_1+\D_2-\D_3-\D_4}{2} \notin \mathbb{Z}_{\leq 0} \\
B_{\kv;v|12}^{s,t}: \ & k_{34} - k_{12} +\frac{-\D_1-\D_2+\D_3+\D_4}{2} \notin \mathbb{Z}_{\leq 0}.
\end{split}
\end{equation}
We note that the $\D_i$ dependence drops out when double-twist operators are exchanged in the $s$-channel.
Thus, at least one $s$-channel subtraction is necessary in such cases.
In particular, only the $\langle AABB \rangle$ correlator has unconstrained $s$-channel subtractions.

The conditions above are necessary to obtain non-singular kernels.
The divergence caused by the gamma function is a signature of the $\lim_{z\rightarrow 0} \log(z)$ term in the collinear expansion of the correlator.
Hence, singular kernels can still be useful sum rules since they define the log coefficient in the $s$-channel collinear expansion.
An example of such a functional is the one obtained from the unsubtracted kernel.
Log coefficient sum rules are consistent sum rules by themselves, however they may be less practical since the identity is absent in the sum rule.
In summary, eq.~\eqref{s-subtraction gamma condition} isn't a convergence condition, but rather a condition to derive either log or non-log functionals.
For the reminder of this paper, we focus on non-log functionals in order to leverage the contribution of the identity.

As an additional consistency check, one can verify that the above expression reduces to the equal operator kernel derived in \cite{Caron-Huot:2020adz}
\footnote{In \cite{Caron-Huot:2020adz}, the $B_{k,v}$ functional is defined as 
\begin{equation}
B_{k,v} \equiv B_{k,v}^s + B_{k,v}^t,
\end{equation}
and therefore our results differ by a factor of $1/2$. This factor carries over in future results, notably in the derivation of holographic functionals $C_{\kv;\nu|34}^{s,t}$ in eq.~\eqref{Ckv position} where the equal operator limit differs by a factor of $2$.
} 
in the limit where $\mathfrak{a}=0,\mathfrak{b}\rightarrow k/2$ for $\kv=(k/2,k/2,0,0), k\neq0$ and $(mn)=(34)$
\footnote{To obtain the unsubtracted kernel associated with the collinear functional $B_v^s$ in \cite{Caron-Huot:2020adz}, we must take the limit $p \rightarrow 0$ of the sum of both $\BB_{0|34}^{0,k/2}$ and $\BB_{0|12}^{0,k/2}$ due to the degeneracy of $s$- and $t$-channel double-twist families in the equal operator limit.
}.
Furthermore, we verified the following equivalences:
\begin{equation}
\BB_{(k/2,k/2,0,0)|34}^{0,k/2} = \BB_{(k/2,0,0,k/2)|34}^{k/2,0} = \BB_{(0,k/2,k/2,0)|12}^{-k/2,0} = \BB_{(0,0,k/2,k/2)|12}^{0,-k/2}.
\end{equation}
These equalities foreshadow relations between subtraction schemes.

Lastly, we note that the mixed correlator kernel includes a hypergeometric ${}_2F_1$ while  the equal operator kernel was a simple function of $u',v'$. We explain how the kernel can be simplified for certain pairwise equal operator correlators in appendix~\ref{app:kernel simp}.

\subsection{Convergence and positivity properties: a plethora of subtraction schemes} \label{sec:convergence}
In this subsection, we examine the convergence conditions of these functionals associated with different operator ordering and subtraction schemes.
It will be helpful to consider the three pairwise equal operator cases separately: $\langle AABB \rangle $, $\langle ABBA \rangle $, $\langle ABAB \rangle $.

\begin{figure}[h]
\centering
\includegraphics[width=0.5\textwidth]{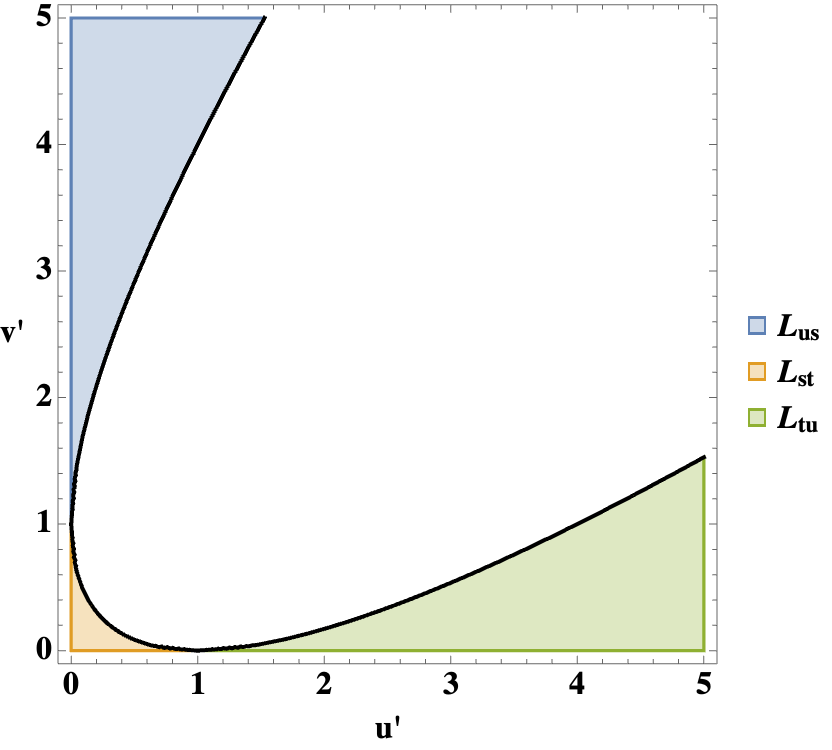}
\caption{Integration region of dispersive transforms. The black line defined as $\sqrt{v'} = \sqrt{u'}+\sqrt{v}|_{v=1}$ separates the Euclidean region (in white) where $w^*=\wb$ and the three lightcones $L_{us}$, $L_{st}$ and $L_{tu}$.}
\label{integration region upvp}
\end{figure}

The integration region for eq.~\eqref{B34 sub ab} is shown in fig.~\ref{integration region upvp}.
As discussed in \cite{Penedones:2019tng,Carmi:2019cub,Caron-Huot:2020adz}, the $s$-channel integration region is determined by $\sqrt{v}'\geq \sqrt{u}'+\sqrt{u}+\sqrt{v}$; this covers a subset of the $L_{us}$ lightcone provided $\sqrt{u}+\sqrt{v}\geq 1$.
Since the $B_{\kv;v|mn}$ functional is defined in the $s$-channel collinear limit, we set $u=0$ in the above.
If this bound is violated, the integration region includes other spacetime regions and the integral is not guaranteed to be sign definite.
The $v' \rightarrow \infty$ limit probes the $\mU$-channel collinear limit.
However, by virtue of the double discontinuity which probes physics on the second sheet, these dispersive transform also captures information from the $\mU$-channel Regge limit.
The integration boundary therefore includes five dangerous limits: the $s$-, $t$- and $\mU$-channel collinear limits, the $\mU$-channel Regge limit, and the boundary determined by $\sqrt{v}'\geq \sqrt{u}'+\sqrt{v}$ which we refer to as the lightcone boundary.
We summarize all convergence conditions for each of these correlators in table~\ref{Tab:convergence} at the end of this subsection.

We first discuss the $\mU$-channel Regge limit.
We remind the reader that a spin-$J$ $\mU$-channel conformal block scales as $u^{-\D_2+\D4+J-1}$ in the Regge limit as shown by eq.~\eqref{block Regge limit}.
In this limit, the subtracted kernel scales as
\begin{equation}
\BB_{\kv|mn}^{\mathfrak{a},\mathfrak{b}}(v;u',v') \underset{\wb \rightarrow +i \infty}{\sim} (\wb)^{-1+\D_2-\D_4-k_{12}-k_{23}-k_{34}-k_{14}}, \qquad w/\wb \text{ fixed},
\end{equation}
and therefore, the Regge boundedness condition is independent of external operator dimensions $\D_i$.

We discussed the relevance of $s$-channel subtractions in the previous subsection where we found that they serve to differentiate between log and non-log functionals, and to track the $s$-channel collinear expansion.
What about $t$-channel subtractions?
We will show that the $\mU$-channel collinear limit fixes the range of allowed $t$-channel subtractions.

\begin{figure}[h]
\centering
\includegraphics[width=\textwidth]{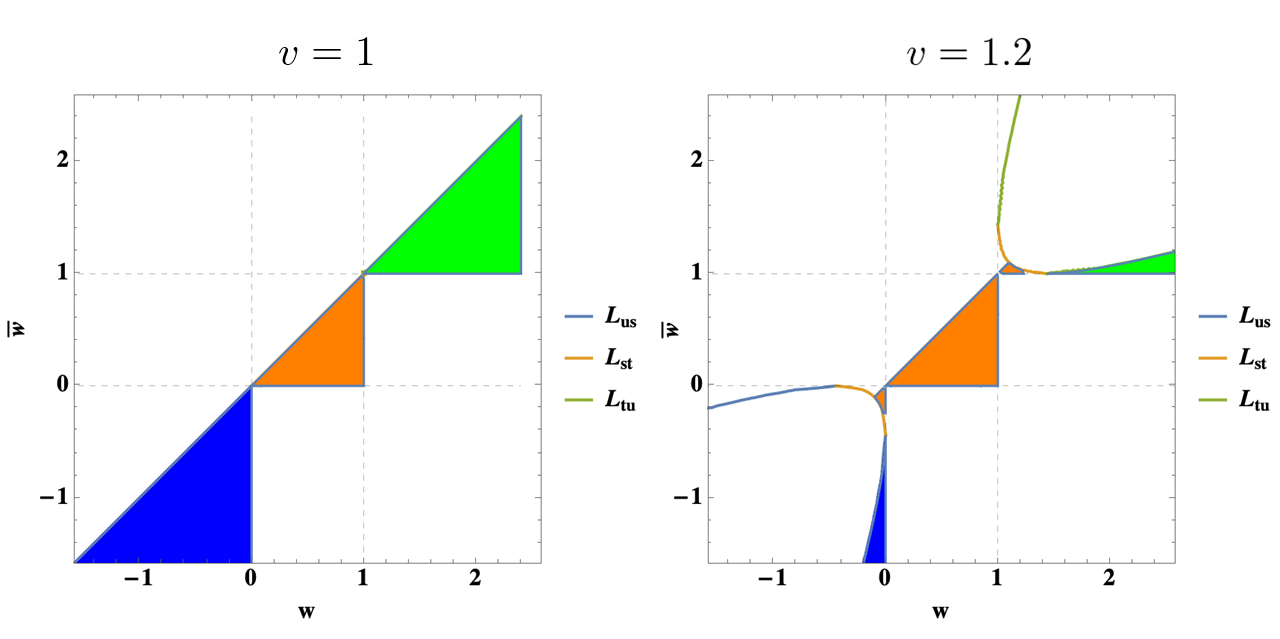}
\caption{The dispersive transform integration region correspond to the area under the curve where we have illustrated the three Lorentzian regions $L_{us} = w,\wb<0$, $L_{st}=w,\wb \in (0,1) $ and $L_{tu} = w,\wb>1$ for $\wb<w$. The left panel corresponds to $v=1$ while the right panel corresponds to $v=1.2$.  The domain of integration lies within the blue ($L_{us}$) and green ($L_{tu}$) regions.}
\label{integration region}
\end{figure}
For our analysis, it is helpful to work in different variables than $u'$ and $v'$.
For convenience, we will map the cross-ratio variables $(w,\wb)$ to new variables defined between the interval $(0,1)$ where the hypergeometric functions are naturally defined.
If $w,\wb<0$, as shown in fig.~\ref{integration region}, we are in the $L_{us}$ lightcone displayed in blue. 
Let $(W,\Wb)=(\tfrac{w}{w-1},\tfrac{\wb}{\wb-1})$ such that the $s$-channel dispersive transform becomes
\begin{equation}
\int_{0}^1 d\Wb \int_0^{1-\frac{1-\Wb}{\left(\sqrt{v}-\sqrt{v-1} \sqrt{\Wb}\right)^2}} dW \BB_{\kv|mn}^{\mathfrak{a},\mathfrak{b}}(W,\Wb) \ \text{dDisc}_{s} [ \GG(W,\Wb)].
\end{equation}
The $t$-channel kinematics are defined for $w,\wb>1$ where the analogous change of variable is $(W,\Wb)=(\tfrac{w-1}{w},\tfrac{\wb-1}{\wb})$. In both cases, the $\mU$-channel collinear limit is mapped to $\Wb=1$ while the $s$- and $t$-channel collinear limits are mapped to $W=0$.

Let us evaluate the kernel at the boundaries of integration. 
In these kinematics, as described in the appendix of \cite{Caron-Huot:2020adz}, the lightcone boundary can be evaluated by analytic continuation from a regime where the integrand is convergent.
Therefore, convergence is unaffected by the lightcone boundary limit.

We record for convenience the behaviour
of the $s$-channel collinear limit $W \rightarrow 0$ of $B_{\kv;v|mn}^s$ where the domain is in the lightcone $L_{u}$:
\begin{equation} \label{s-ch collinear kernel}
L_{us}: \quad \BB_{\kv|34}^{\mathfrak{a},\mathfrak{b}}(W,\Wb) \underset{W \rightarrow 0}{\sim} (W\Wb)^{-k_{12}},  \quad \BB_{\kv|12}^{\mathfrak{a},\mathfrak{b}}(W,\Wb) \underset{W \rightarrow 0}{\sim} (W\Wb)^{\frac{\D_1+\D_2-\D_3-\D_4}{2}-k_{34}}.
\end{equation}
Due to the antisymmetry in $W\leftrightarrow \Wb$, the kernel behaves similarly in the $t$-channel collinear limit $\Wb \rightarrow 0 $ limit.
To probe the $t$-channel collinear limit of $B_{\kv;v|34}^t$, we note that the change of variable for $W,\Wb$ is different in the lightcone $L_{tu}$ such that both kernels behave as follows:
\begin{equation} \label{t-ch collinear kernel}
L_{tu}: \qquad \BB_{\kv|34}^{\mathfrak{a},\mathfrak{b}}(W,\Wb), \BB_{\kv|12}^{\mathfrak{a},\mathfrak{b}}(W,\Wb) \underset{W \rightarrow 0}{\sim} (W\Wb)^{-k_{23}} + (W\Wb)^{\frac{\D_{21}+\D_{34}}{2}-k_{14}}.
\end{equation}
We note that for equal operators, the subtraction units $k_{ij}$ become degenerate.
After accounting for the contribution of the conformal blocks from eq.~\eqref{blocks OPE exp},
we conclude that the dispersive transform is guaranteed to be convergent and sign definite according to table~\ref{tab:s and t collinear}.
\begin{table}[h]
\centering
\begin{tabular}{c| c c}
$(mn)$ & $B_{\kv;v|mn}^{s}$ & $B_{\kv;v|mn}^{t}$ \\ \hline 
& & \\
$(34)$ & $\tau>\D_1+\D_2+2k_{12}-2$ &  $ \tau >\min(\Delta _1{+}\Delta _4{+}2 k_{14},\D_2{+}\D_3{+}2k_{23}) -2$  \\
 & &  \\
$(12)$ & $\tau >\D_3+\D_4+2k_{34}-2$ & $ \tau >\min(\Delta _1{+}\Delta _4{+}2 k_{14},\D_2{+}\D_3{+}2k_{23}) -2$  \\
\end{tabular}
\caption{Collinear limit of the $B_{\kv;v|mn}^{s,t}$ functional's integrand. In the equal operator limit, the $s$- and $t$-channel conditions become degenerate.}
 \label{tab:s and t collinear}
\end{table}
Below these thresholds, the functionals are defined by analytic continuation and therefore, they are no longer guaranteed to be sign definite.
Implementing the analytic continuation is subtle. The reader is encouraged to consult appendix~D of \cite{Caron-Huot:2020adz} to learn more.

Since the kernel is antisymmetric in $W \leftrightarrow \Wb$, we describe the final limit of interest: the $\mU$-channel collinear limit $\Wb \rightarrow 1$. 
To determine the convergence conditions,  we probe the $\mU$-channel collinear limit by taking $x_2 \rightarrow x_4$ and $x_1 \rightarrow x_3$.
The behaviour of the correlator can be easily assessed by considering eq.~\eqref{unstripped correlator}:
by working in radial coordinates $\rho$ where $\rho \rightarrow 1$ is the $\mU$-channel collinear limit, the stripped factor is polynomially singular\footnote{Explicitly, we work in a configuration where all the points $x_i$ lie in a 2-plane with complex coordinates $y=x^1+i x^2$, $\overline{y}=x^1-i x^2$, and the operators are positioned at 
\begin{equation}
\langle \phi_1(-\rho,-\rhob) \phi_2(\rho,\rhob) \phi_3(-1,-1) \phi_4(1,1) \rangle,
\end{equation}
such that $\rho \rightarrow 1$ corresponds to the $\mU$-channel collinear limit.
}
\begin{equation}
\langle \phi_1 \phi_2 \phi_3 \phi_4 \rangle \sim (1-\rho)^{-\frac{\D_1+3\D_2+\D_3-\D_4}{2}} \GG(\rho,\rhob).
\end{equation}
The stripped correlator can then be bounded using Cauchy-Schwarz \cite{Kravchuk:2020scc,Kravchuk:2021kwe}. 
We therefore conclude that
\begin{equation}
| \GG(\rho,\rhob) | \leq C (1-\rho)^{\D_2-\D_4},
\end{equation}
for some $C>0$.

In this limit, the kernels behave as
\begin{equation} \label{B Kernel Collinear}
\begin{split}
\BB_{\kv|34}^{\mathfrak{a},\mathfrak{b}}(\rho,\rhob)/(u^{-\frac{\D_2-\D_4}{2}} ) &\underset{\rho \rightarrow 1}{\sim} (1-\rho)^{-\D_1-3\D_2+2\D_4+2k_{34}-1} \left( (1-\rho)^{\D_2+\D_3+2k_{23}} + (1-\rho)^{\D_1+\D_4+2k_{14}} \right),  \\
\BB_{\kv|12}^{\mathfrak{a},\mathfrak{b}}(\rho,\rhob)/(u^{-\frac{\D_2-\D_4}{2}} ) &\underset{\rho \rightarrow 1}{\sim} (1-\rho)^{-2\D_2-\D_3+2k_{34}-1} \left( (1-\rho)^{\D_2+\D_3+2k_{23}} + (1-\rho)^{\D_1+\D_4+2k_{14}} \right),
\end{split}
\end{equation}
where we have divided by the $\mU$-channel identity to make manifest the convergence conditions.
We note that the symmetry between the $k_{23}$ and $k_{14}$ subtractions are due to hypergeometric function in eq.~\eqref{B34 sub ab}.
Together, the last two equations determine convergence conditions in the $\mU$-channel collinear limit.

By combining all of the previous discussions, we can derive the convergence conditions summarized in the following table for non-log functionals.
\begin{table}[h]
\centering
\begin{tabular}{c||c|c}
Functional & $s$-channel subtractions & $t$-channel subtractions \\ \hline
& & \\
$B_{\kv;v|34}^{s,t}$ & $k_{12} {-} k_{34} +\D_1+\D_2-\D_3-\D_4 \notin \mathbb{Z}_{\leq 0}$ & \shortstack{$2k_{14}>\D_2-\D_4-2k_{34}$ \\ $2 k_{23}>\D_1-\D_3-2k_{34}$} \\
& & \\
$B_{\kv;v|12}^{s,t}$ & $k_{34} {-} k_{12} -\D_1-\D_2+\D_3+\D_4 \notin \mathbb{Z}_{\leq 0}$ & \shortstack{$2k_{14}>-\D_2+\D_4-2k_{12}$ \\ $2 k_{23}>-\D_1+\D_3-2k_{12}$} 
\end{tabular}
\caption{Convergence conditions for non-log functionals determined by subtraction units $k_{ij} \in \mathbb{Z}$. $s$-channel subtractions determine whether the functionals are log or non-log, while $t$-channel subtractions determine convergence conditions. The $t$-channel condition becomes eq.~\eqref{convergence ABAB t-ch} for the $\langle ABAB \rangle$ correlator if we consider the $\GG-1$ correlator.  Lastly, we note that ``anti-subtractions" are possible for $k_{ij}<0$.}
\label{Tab:convergence}
\end{table}
We note that the $t$-channel condition can be relaxed for the $\langle ABAB \rangle$ correlator if we consider the $\GG-1$ correlator.
In this case where the $t$-channel convergence conditions are marginally satisfied for $k_{ij}=0$,  we assume that the subleading $\mU$-channel collinear singularity comes from a $\mU$-channel block such that the functionals $B_{\kv;v|mn}^{s,t}$ converge provided that
\begin{equation} \label{convergence ABAB t-ch}
2k_{14}>-\min(\D_A,\D_B)-2k_{mn}, \qquad 2 k_{23}>-\min(\D_A,\D_B)-2k_{mn}.
\end{equation}
In other words, $t$-channel subtractions are unnecessary for the $\langle ABAB \rangle$ correlator unless $k_{mn}<0$.
Furthermore, the convergence conditions in table~\ref{Tab:convergence} illustrates how different operator ordering and subtraction schemes may be related.
For example, applying a dispersive transform with the kernel $\BB_{(r_1,k_{23},r_2,k_{14})|34}$ to the $\langle AABB \rangle$ correlator is equivalent to using $\BB_{(r_2,k_{23},r_1,k_{14})|12}$ kernel for the $\langle BBAA \rangle$ correlator; this follows from eq.~\eqref{B12 to B34 relation sub}.
We will test the convergence conditions in table~\ref{Tab:convergence} by exploring different subtraction schemes in section~\ref{sec:Ising sum rule} for mixed correlators of the 3D Ising model.

Let us consider an example to illustrate these convergence conditions.
We consider the $B_{\kv;v|34}^s$ functional acting on the $\langle ABBA \rangle$ correlator with external operators $\sigma$ and $\epsilon$ from the 3D Ising model.
The collinear limit of the integrand behaves as follows:
\begin{equation}
\begin{split}
\BB_{\kv|34}\langle \sigma \epsilon \epsilon \sigma \rangle & \sim (1-\rho)^{-1.9+2k_{14}+2k_{34}} + (1-\rho)^{-0.1+k_{23}+2k_{34}} \\
\BB_{\kv|34}\langle \epsilon \sigma \sigma \epsilon \rangle & \sim (1-\rho)^{-0.1+2k_{14}+2k_{34}} + (1-\rho)^{-1.9+k_{23}+2k_{34}}
\end{split}
\end{equation}
A single $t$-channel subtraction along the $\tau_{14}$ double-twist trajectory is necessary for the $\langle\sigma \epsilon \epsilon \sigma \rangle$ while a subtraction is necessary along the $\tau_{23}$ trajectory for the $\langle \epsilon \sigma \sigma \epsilon \rangle$.
This example highlights relations between operator ordering and subtraction schemes.

We further underscore the possibility for ``anti-subtractions" according to table~\ref{Tab:convergence}, i.e. $k_{ij}<0$.
In particular, for equal operators, the $\kv=(0,-1,-1,0)$ subtraction scheme, or equivalently $(0,0,-1,-1)$, applied to the $B_{\kv;v|34}$ functional corresponds to the anti-subtracted functional $B_{-2}$ introduced in \cite{Caron-Huot:2021enk}.
We will show that anti-subtractions change the Mellin-space domain of convergence.

Finally, we revisit the positivity properties of our dispersive transforms.
First, we note that the hypergeometric function in eq.~\eqref{B34 sub ab} is sign definite in the integration range, and therefore
the integral is guaranteed to be positive according to the inequalities shown in table~\ref{tab:s and t collinear}.
As previously discussed, the double discontinuity defines a positive definite matrix in terms of the pairs of external operators 12 and 34.
We therefore consider each pairwise equal operator correlator separately.

For the $\langle ABAB \rangle$ correlator, the double discontinuity returns a $\sin^2$ and hence it is positive definite in both channels.
Given the convergence conditions, we further conclude that the dispersive transform is also positive. Thus, any negativity stemming from this such sum rules must be due to an operator below the first double-twist value $\D_{A}+\D_B+\ell$, or they are due to the OPE coefficient which is proportional to $f_{BA\OO}=f_{AB\OO}(-1)^{\ell_\OO}$.
Thus, even and spin trajectories are individually sign definite, but their sum is not.
Moreover, as opposed to the other two cases, the sum rule adds up to the $\mU$-channel identity rather than vanishing.

For the $\langle ABBA \rangle$ correlator, the double discontinuity also returns a $\sin^2$ in the $s$-channel, but they are the product of two independent $\sin$ functions in the $t$-channel.
Thus, the $s$-channel sector of the sum rule is positive for $\tau>\D_A+\D_B+2k_{ij}-2$ for a number of $s$-channel subtractions $k_{ij}$; it is no longer sign definite in the $t$-channel.
The $\langle AABB \rangle$ correlator behaves similarly but with $s$- and $t$-channel positivity properties exchanged.
A key different between the $\langle AABB \rangle$ and $\langle ABBA \rangle$ correlators is the allowed range of $s$-channel subtractions to produce non-log functionals.
These properties will be verified in section~\ref{sec:Ising sum rule} where we evaluate sum rules of various correlators of the 3D Ising model.

To summarize this section, 
$s$-channel subtractions track the $s$-channel collinear expansion and therefore distinguish between log and non-log functionals.
On the other hand, $t$-channel subtractions are necessary to control the behaviour of the correlator in the $\mU$-channel collinear limit.
These conditions are summarized in table~\ref{Tab:convergence}.
Positivity on the other hand is controlled by $s$- and $t-$channel double-twist operators and their corresponding subtractions.
Given our newfound understanding of these subtraction schemes and convergence properties, we will restrict ourselves to the $B_{\kv;v|34}$ functional for the rest of this paper.

\subsection{Mixed correlator holographic functionals $C_{\kv;\nu|mn}$} \label{sec:pos Regge}
In \cite{Caron-Huot:2021enk}, the authors introduced a dispersive functional $C_{k,\nu}$ which localizes physics in AdS spacetime motivated by previous work conducted in flat-space \cite{Caron-Huot:2021rmr}.
The goal was to relate AdS and CFT quantum numbers to probe local physics in bulk AdS.
This is achieved by localizing wavepackets at fixed impact parameter in the bulk-point limit.
The idea is to parametrize $2 \rightarrow 2$ scattering by labelling states with conserved quantities such as the impact parameter $b=2J/m$ conjugate to the transverse momentum $u=-p^2$.
In AdS, pairs of particles with large center-of-mass energy $\D/R_{AdS}$ separated by a geodesic distance of $\beta R_{AdS}$ carry total angular momentum
\begin{equation}
J = \D \tanh(\frac{\beta}{2}).
\end{equation}
We therefore see that local physics, described by the bulk point limit where the RHS vanishes, is obtained by taking $J/\D \ll 1$ for $\D \gg 1$. 
The Regge limit is reached by taking $\D$ large for fixed $J/\D$.
We review other parameters of this dictionary in table~\ref{tab:dictionary AdS/CFT}.
\begin{table}[h]
\centering
\begin{tabular}{c c c}
\hline
 & AdS$_{d+1}$ & CFT$_d$ \\ \hline
angular momentum & $J$ & $J$ \\ \\
energy & $m$ & $(\D-J-d+1)(\D+J-1)$ \\ \\
impact paramter & $\beta = \cosh^{-1}(\eta_{\text{AdS}})$ & $ \log \tfrac{\D+J-1}{\D-J+d-1}$ \\ \\
Regge limit & $m\gg 1$ at fixed $\beta$ & $\D \gg 1$ at fixed $J/\D$ \\ \\
bulk-point limit & $m \gg 1$ at fixed $J$ & $\D \gg 1$ at fixed $J$ \\ 
\hline
\end{tabular}
\caption{Dictionary between AdS and CFT quantum numbers.}
\label{tab:dictionary AdS/CFT}
\end{table}

These so-called holographic functionals $C_{k,\nu}$ exhibit the following properties:
\begin{itemize}
\item In the Regge limit where $\D$ is the largest parameter, they are proportional to AdS harmonic functions $\PP_{\tfrac{2{-}d}{2}{+}i\nu}(\eta_{AdS})$:
\begin{equation}
C_{k,\nu}[\D,J] \propto \frac{\PP_{\tfrac{2{-}d}{2}{+}i\nu}(\tfrac{\D^2+J^2}{\D^2-J^2})}{m^{2k}} \times \left(1+ O\left( \frac{\nu^2}{m^2}, \frac{1}{m^2} \right) \right),
\end{equation}
where the Gegenbauer spin $J=\tfrac{2-d}{2}+i \nu$ is fixed.

\item In the bulk-point limit where $\nu \in [0,m)$, they encode the contribution of massive states in a partial wave expansion:
\begin{equation}
C_{k,\nu}[\D,J] \propto \PP_J(1-\tfrac{2\nu^2}{m^2}) \times \left( 1 + O\left( \frac{J^2}{m^2} \right) \right),
\end{equation}
where $uR^2_{AdS} = -\nu^2$.

\item The action on light operators matches that of the flat-space dispersion relation in the large $\nu$ (small impact parameter) limit
\begin{equation}
-C_{k,\nu}|_{light} =-\mathcal{C}_{k,u}  \times \left( 1 + O\left( \frac{1}{\nu^2} \right) \right).
\end{equation}
\end{itemize}
Such fonctionals lead to holographic sum rules providing an additional pathway to bootstrap quantum gravity in AdS from CFTs.
While such a bootstrap exercise extends beyond the goal of this paper,
given our goal to derive mixed correlator dispersive CFT functionals, we will generalize the previous construction to mixed correlators.

We first sketch the derivation of these holographic functionals $C_{k,\nu}$ based on the work of \cite{Caron-Huot:2021enk}:
\begin{enumerate}
\item Given the physical functional $B_{k,v}$, we write position-space dispersive transform in terms of the radial and angular coordinates $r$ and $\eta$
\begin{equation}
B_{k,v} = \iint dr \  d\eta \ \BB_{k}(v;r,\eta) \text{dDisc}[G_{\D,J}(r,\eta)] \propto \int \frac{\eta d\eta}{(\eta^2-v)^{\frac{3-k}{2}} } \Pi_{k,\eta} + O(\Pi_{k+2,\eta}),
\end{equation}
where we have expanded the integrand in terms of Regge moments
\begin{equation}
\Pi_{k,\eta}[\GG] = \int_{0}^{r_{max}(\eta)} dr r^{k-2} \text{dDisc}_s \GG(r,\eta).
\end{equation}

\item We can extract a Regge moment functional by performing an inverse transform on $B_{k,v}$:
\begin{equation}
\Pi_{k,\eta} [G_{\D,J}] \propto \int_{\eta^2}^\infty \frac{dv}{[(v-\eta^2)^{\frac{k+1}{2}}]_+} B_{k,v},
\end{equation}
where the $+$ subscript denotes a distribution where we have subtracted singular terms around $v=\eta^2$.
Unfortunately, this functional is delocalized in impact parameter and therefore fails to describe bulk AdS physics.

\item To solve this, one can localize in transverse impact parameter $\eta$ by performing harmonic analysis on the transverse space $H_{d-1}$ in AdS:
\begin{equation}
C_{k,\nu} \propto \int_1^{\infty} [d\eta] \PP_{\tfrac{2-d}{2}+i\nu}(\eta) \int_{\eta^2}^{\infty} \frac{dv}{[(v-\eta^2)^{\frac{k+1}{2}}]_+} B_{k,v},
\end{equation}
where the proportionality factor is fixed by matching the functional with the flat-space dispersion relation $\mathcal{C}_{k,u}$.
\end{enumerate}
We will now extend this position-space derivation to mixed correlators; a Mellin-space derivation will be detailed in section \ref{sec:mellin regge}.

Let us define the radial coordinates from \cite{Hogervorst:2013sma} adapted in the $\mU$-channel as follows:
\begin{equation}
\rho = \frac{1}{(\sqrt{1-w}+\sqrt{-w})^2}, \qquad \bar{\rho} = \frac{1}{(\sqrt{1-\bar{w}} + \sqrt{-\bar{w}})^2},
\end{equation}
We define the radial and angular coordinates as
\begin{equation}
r \equiv \sqrt{\rho \bar{\rho}}, \qquad \eta \equiv \frac{\rho+\bar{\rho}}{2\sqrt{\rho \bar{\rho}}}.
\end{equation}
In these new coordinates, the dispersive transform reads
\begin{equation} \label{Bkv Regge Moment expansion}
\begin{split}
B_{\kv;v|34}^{s,t} &= \int \displaylimits_{\sqrt{v}}^{\infty} d\eta \int \displaylimits_{0}^{r_{max}(\eta)} dr \ \BB_{\kv|34}^{\mathfrak{a},\mathfrak{b}}(v;\eta,r) \ \text{dDisc}_{s,t}[\GG], \\
&= \int \displaylimits_{\sqrt{v}}^{\infty} d \eta \ \ \Big(H_0(v,\eta)\Pi_{\kv,0,\eta}^{s,t}[\GG] + H_1(v,\eta)\Pi_{\kv,1,\eta}^{s,t}[\GG] + O(\Pi_{\kv,2,\eta}^{s,t}[\GG]) \Big),
\end{split}
\end{equation}
where $r_{max}=\eta - \sqrt{\eta^2-1}$.
Similar to the equal operator case, we define 
Regge moments of mixed correlators $\Pi_{\kv,n,\eta}^{s,t}$ as the following expansion of the kernel in the $\mU$-channel Regge limit $r \rightarrow 0$:
\begin{equation}
\Pi_{\kv,n,\eta}^{s,t}[\GG] \equiv \int \displaylimits_0^{r_{max}(\eta)} dr \ r^{-\D_2+\D_4+k_{12}+k_{23}+k_{34}+k_{14}+n-2} \ \text{dDisc}_{s,t}[\GG].
\label{Regge moments}
\end{equation}
We use a slightly different notation than in \cite{Caron-Huot:2021enk} since different subtraction schemes will play a role in our analysis; $n$ labels the moment \textit{relative} to the total spin convergence.

The next step is to invert eq.~\eqref{Bkv Regge Moment expansion} to extract the Regge moment functionals.
To achieve this, we explicitly write the expansion of $B_{\kv;v|34}$ up to leading order in the Regge moment:
\begin{equation}
\label{Bkernel position Regge moment}
\begin{split}
B_{\kv;v|34}^{s,t} &=\int \displaylimits_{\sqrt{v}}^{\infty} d \eta \ \frac{(-1)^{1+k_{12}+k_{34}} }{2 \pi^{3/2}v^{k_{12}}} \frac{\eta \ (4 \sqrt{v})^{-\D_2+\D_4+k_{12} + k_{14}+k_{23}+k_{34}}}{(\eta^2-v)^{\frac{3}{2}-\mathfrak{a}-\mathfrak{b}}} \frac{\Gamma(\mathfrak{a}+\mathfrak{b})}{\Gamma(-\tfrac{1}{2}+\mathfrak{a}+\mathfrak{b})}  \\
&\qquad \times {}_2F_1(\mathfrak{a},\mathfrak{b},-\tfrac{1}{2}+\mathfrak{a}+\mathfrak{b},1-\tfrac{\eta^2}{v}) \ \Pi_{\kv,0,\eta}^{s,t} + O(\Pi_{\kv,1,\eta}^{s,t}),
\end{split}
\end{equation}
where $\mathfrak{a},\mathfrak{b}$ is given by eq.~\eqref{ab frak}.
We note that the $\eta$ dependence in the first line is analogous to the equal operator case where the authors in \cite{Caron-Huot:2021enk} found $\tfrac{\eta}{(\eta^2-v)^{(3-k)/2}}$. 
Furthermore, the equal operator limit is given by $(\mathfrak{a},\mathfrak{b}) \rightarrow (0,k/2)$ or $(k/2,0)$.
The major distinction between the two cases is the presence of the hypergeometric function in the second line above.
Fortunately, the $\eta$ integral can be done analytically\footnote{
The integral can be reduced to the following form under an appropriate change of variables:
\begin{equation}
\int \displaylimits_{0}^{\infty} dx \ (1+x)^d x^{c} \ {}_2F_1(a,b,c+1,-x).
\end{equation}
}
and the result is simply a polynomial in $v$ with additional gamma functions.
Therefore, like the equal operator case, an inverse transform exists where for some power $p$ of $\eta^p$, we have
\begin{equation}
\begin{split}
&\int\displaylimits_{\sqrt{v}}^{\infty} d \eta \ \frac{\eta^{1+p}}{(\eta^2-v)^{\frac{3}{2}-\mathfrak{a}-\mathfrak{b}}} {}_2F_1(\mathfrak{a},\mathfrak{b},-\tfrac{1}{2}+\mathfrak{a}+\mathfrak{b},1-\tfrac{\eta^2}{v}) \\
& \qquad = \frac{v^{\frac{1}{2}(-1+p)+\mathfrak{a}+\mathfrak{b}}}{2} \frac{\Gamma({-}\tfrac{1}{2}+\mathfrak{a}+\mathfrak{b})\Gamma(\tfrac{1}{2}-\mathfrak{a}-\tfrac{p}{2})\Gamma(\tfrac{1}{2}-\mathfrak{b}-\tfrac{p}{2})}{\Gamma(\tfrac{1-p}{2}) \Gamma(-\tfrac{p}{2})}
\end{split}
\end{equation}
and an inverse exists of the form
\begin{equation}
\begin{split}
&\int\displaylimits_{\eta^2}^{\infty} d v \ \frac{v^{\frac{1}{2}(-1+p)+\mathfrak{a}+\mathfrak{b}} }{(v-\eta^2)^{\frac{1}{2}+\mathfrak{a}+\mathfrak{b}}} \left( \frac{v}{\eta^2} \right)^{-\mathfrak{b}} {}_2F_1(\tfrac{1}{2}-\mathfrak{b},-\mathfrak{b},\tfrac{1}{2}-\mathfrak{a}-\mathfrak{b},1-\tfrac{v}{\eta^2}) \\
& \qquad = \eta^p \frac{\Gamma(\tfrac{1}{2}-\mathfrak{a}-\mathfrak{b})\Gamma(\tfrac{1}{2}-\tfrac{p}{2})\Gamma(-\tfrac{p}{2})}{\Gamma(\tfrac{1-p}{2}-\mathfrak{a}) \Gamma(\tfrac{1-p}{2}-\mathfrak{b})}.
\end{split}
\label{v inverse integral}
\end{equation}
By combining the two relations above, we obtain an inverse transform for the leading Regge moment:
\begin{equation} \label{inverse Regge moment}
\int \displaylimits_{\eta^2}^{\infty} d v \ \frac{2v^{\frac{1}{2} \left(\Delta _2-\Delta _4+k_{12}-k_{14}-k_{23}-k_{34}\right)} \left(v/\eta^2 \right)^{-\mathfrak{b}}}{(v-\eta^2)^{\frac{1}{2}+\mathfrak{a}+\mathfrak{b}}} \frac{ {}_2F_1(\tfrac{1}{2}-\mathfrak{b},-\mathfrak{b},\tfrac{1}{2}-\mathfrak{a}-\mathfrak{b},1-\tfrac{v}{\eta^2})}{\Gamma(\tfrac{1}{2}-\mathfrak{a}-\mathfrak{b})\Gamma(-\tfrac{1}{2}+\mathfrak{a}+\mathfrak{b})}
B_{\kv;v|34}^{s,t} \sim \Pi_{\kv,0,\eta}^{s,t}
\end{equation}

The final step in our derivation consists of localizing in impact parameter which is achieved by performing harmonic analysis in the transverse space $H_{d-1}$.
We proceed as in the equal operator case by first projecting the Regge moment functional against $SO(d-1,1)$ partial waves
\begin{equation}
\widehat{\Pi}_{\kv;0,\nu|34}^{s,t} = \int_1^\infty [d\eta] \PP_{\tfrac{2-d}{2}+i\nu}(\eta) \Pi_{\kv,0;\eta|34}^{s,t}, \qquad \text{where} \ [d\eta] \equiv 2^{d-2}(\eta^2-1)^{\frac{d-3}{2}} d\eta,
\label{inverse eta integral}
\end{equation}
where
\begin{equation}
\PP_J = {}_2F_1(-J,J+d-2,\tfrac{d-1}{2},\tfrac{1-x}{2}).
\end{equation}
Thus, our holographic functional reads
\begin{equation}
\begin{split}
C_{\kv;\nu|34}^{s,t} &= (-1)^{k_{12}+k_{34}+1} 4^{1+\D_2-\D_4-k_{12}-k_{23}-k_{34}-k_{14}} \pi^{3/2} a_{\D_i,k_{ij}}(\nu) \int \displaylimits_1^{\infty} [d\eta] \PP_{\tfrac{2-d}{2}+i \nu} (\eta)  \\
& \times  \int \displaylimits_{\eta^2}^{\infty} d v \ \frac{\left(v/\eta^2 \right)^{-\mathfrak{b}}v^{\frac{1}{2} \left(\Delta _2-\Delta _4+k_{12}-k_{14}-k_{23}-k_{34}\right)} }{(v-\eta^2)^{\frac{1}{2}+\mathfrak{a}+\mathfrak{b}}} \frac{ {}_2F_1(\tfrac{1}{2}-\mathfrak{b},-\mathfrak{b},\tfrac{1}{2}-\mathfrak{a}-\mathfrak{b},1-\tfrac{v}{\eta^2})}{\Gamma(\tfrac{1}{2}-\mathfrak{a}-\mathfrak{b})\Gamma(\mathfrak{a}+\mathfrak{b})}  \ B_{\kv;v|34}^{s,t},
\end{split}
\label{Ckv position}
\end{equation}
where
the normalization factor $a_{\D_i,k_{ij}}(\nu)$ is defined as
\begin{equation} \label{a_normalization}
a_{\D_i,k_{ij}}(\nu) =  \frac{\pi^{\frac{d-3}{2}} \left( \prod_{i=1}^4 \sqrt{\Gamma(\D_i)\Gamma(\D_i+1-d/2)} \right)/\left(2^{d-4}\Gamma(\tfrac{d-1}{2}) \right) }{\gamma_{\Delta _1+\D_3+k_{12}+k_{14}+k_{23}-k_{34}-1} (\nu) \gamma_{\D_2+\D_4+k_{12}+k_{14}+k_{23}-k_{34}-1} (\nu) },
\end{equation}
and
\begin{equation} \label{gamma nu}
\gamma_a(\nu) \equiv  \Gamma(\tfrac{1+a-d/2-i\nu}{2})\Gamma(\tfrac{1+a-d/2+i\nu}{2}).
\end{equation}
This normalization factor is fixed by relating the Mellin amplitude to the flat-space scattering amplitude based on the convention in \cite{Penedones:2010ue}:
\begin{equation}
M(\ms,\mt) = \frac{1}{8\pi^{d/2}\prod_{i=1}^4 \sqrt{\Gamma(\D_i)\Gamma(\D_i+1-d/2)} } \int \displaylimits_0^\infty d\beta \beta^{\frac{\sum_i \D_i}{2}-d/2-1} e^{-\beta} \mathcal{M}_{\text{flat}}(2\beta \ms, 2 \beta \mt),
\end{equation}
and the $\gamma_a(\nu)$ dependence will be justified by our derivation of the Mellin-space functional in section~\ref{sec:mellin regge} where we will further elaborate on the $\D_i$ dependence. 
One can verify that eq.~\eqref{Ckv position} matches the result of \cite{Caron-Huot:2021enk} in the equal operator limit.

A similar $\gamma_{\D_i}(\nu)$ factor was also present in the equal operator case. 
We expect this factor to be the only $\nu$-dependent term in the Regge moment $\widehat{\Pi}_{\kv;0,v|34}^{s,t}$:
\begin{equation}
\widehat{\Pi}_{\kv;0,v|34}^{s,t} \propto \gamma_{\Delta _1+\D_3+k_{12}+k_{14}+k_{23}-k_{34}-1} (\nu) \gamma_{\D_2+\D_4+k_{12}+k_{14}+k_{23}-k_{34}-1} (\nu)   \PP_{\tfrac{2-d}{2}+i \nu}(\eta).
\end{equation}
Physically, these gamma functions are necessary to ``undo" the smearing of these dispersive transforms in order to localize the scattering processes in the bulk.
This insight is made manifest in Mellin-space.
Moreover, because these dispersion relations are derived in a fixed-$\mU$ convention, the $\D_i$ should reflect the exchange of $\mU$-channel double-twists.

What about higher order Regge moments? By expanding the $\BB_{\kv|34}^{\mathfrak{a},\mathfrak{b}}$ kernel to subleading order in $r$, we find
\begin{equation}
\begin{split}
B_{\kv;v|34}^{s,t} \bigg|_{r^1} &= \int \displaylimits_{\sqrt{v}}^{\infty} d \eta \ \frac{(-1)^{1+k_{12}+k_{34}} }{2\pi^{3/2}v^{k_{12}}} \frac{\eta \ (4 \sqrt{v})^{-\D_2+\D_4+k_{12} + k_{14}+k_{23}+k_{34}}}{(\eta^2-v)^{\frac{1}{2}(-3+\mathfrak{a}+\mathfrak{b})}} \frac{\Gamma(\mathfrak{a}+\mathfrak{b})}{\Gamma(-\tfrac{1}{2}+\mathfrak{a}+\mathfrak{b})}  \\
&\quad \times \bigg(-\frac{8 \mathfrak{a} \mathfrak{b}(\eta^2-v) \eta^2}{v(-1+2\mathfrak{a}+2\mathfrak{b})} {}_2F_1(1+\mathfrak{a},1+\mathfrak{b},\tfrac{1}{2}+\mathfrak{a}+\mathfrak{b},1-\tfrac{\eta^2}{v}) \\
& \quad  + (-2v+ 2\eta^2 (2\mathfrak{a}-2k_{12}+k_{23})) \ {}_2F_1(\mathfrak{a},\mathfrak{b},-\tfrac{1}{2}+\mathfrak{a}+\mathfrak{b},1-\tfrac{\eta^2}{v}) \bigg) \ \Pi_{\kv,1,\eta}^{s,t}.
\end{split}
\end{equation}
In contrast to the equal operator case, there is no unique inverse transform that can extract all Regge moments.
The hypergeometric functions compels us to use variations of eq.~\eqref{inverse Regge moment} on higher order Regge moments; we leave the study of higher Regge moments for mixed correlators to future work.

\section{Mellin-space functionals} \label{sec:mixed mellin}

The position-space dispersion relation in the previous section was inspired by the Mellin-space dispersion relation of eq.~\eqref{M dispersion original def}.
It is therefore tempting to work directly in Mellin-space by decomposing the Mellin amplitude $M(\ms,\mt)$ in terms of the absorptive part of the $s$- and $t$-channel sectors as follows:
\begin{align}
& \text{Position-space} \qquad & \qquad & \qquad \text{Mellin-space} \nonumber \\
\GG_{\kv}(u,v) &= \GG_{\kv}^{s}(u,v) + \GG_{\kv}^{t}(u,v) \qquad & \leftrightarrow \qquad & M_{\kv}(\ms,\mt)= M_{\kv}^s(\ms,\mt) + M_{\kv}^t(\ms,\mt), 
\label{position vs Mellin}
\end{align}
This is indeed possible and we present the Mellin-representation of the Polyakov-Regge blocks and 
the $\widehat{B}_{\kv;\mt|mn}^{s,t}$ functionals in this section.

\subsection{Deriving the $\widehat{B}_{\kv;\mt|mn}^{s,t}$ functional} \label{sec:mellin derivation}

The Mellin representation of a correlator $\GG(u,v)$ can be written as
\begin{equation} \label{G Mellin representation}
\GG(u,v) = \iint_\gamma \frac{d\ms d\mt}{(4 \pi i)^2} \Gamma_{\D_i}^6(\ms,\mt) u^{\frac{\ms-\D_1-\D_2}{2}} v^{\frac{\mt-\D_2-\D_3}{2}} M(\ms,\mt),
\end{equation}
where $\ms$ and $\mt$ are \textit{Mellin-Mandelstam} variables, $\gamma=\gamma_\ms \otimes \gamma_\mt$ is a contour determined by $\Gamma_{\D_i}^6(\ms,\mt)$ which is defined as
\begin{equation} \label{Gamma6}
\Gamma_{\D_i}^6(\ms,\mt) = \Gamma(\tfrac{\D_1+\D_2-\ms}{2})\Gamma(\tfrac{\D_3+\D_4-\ms}{2})\Gamma(\tfrac{\D_2+\D_3-\mt}{2})\Gamma(\tfrac{\D_1+\D_4-\mt}{2})\Gamma(\tfrac{\ms+\mt-\D_1-\D_3}{2})\Gamma(\tfrac{\ms+\mt-\D_2-\D_4}{2}).
\end{equation}
These Mellin-Mandelstam variables are related to the conventional Mellin variables introduced by Mack in \cite{Mack:2009mi} as follows:
\begin{align}
\gamma_{jj}=-\D_j, \qquad \gamma_{12} = \tfrac{\D_1+\D_2}{2} -\tfrac{\ms}{2}, &\qquad \gamma_{13}=\tfrac{\D_1+\D_3}{2} - \tfrac{\mU}{2}, \qquad \gamma_{14}=\tfrac{\D_1+\D_4}{2} - \tfrac{\mt}{2}, \\
\ms + \mt + \mU = \sum_{i=1}^4 \D_i \qquad & \leftrightarrow \qquad \sum_i \gamma_{ij}= 0.
\end{align}

Using the above, one can write the Mellin-representation of Polyakov-Regge blocks as
\begin{equation} \label{PP Mellin representation}
\PP_{\kv;\D,J}^{s,t}(u,v) = \iint_\gamma \frac{d\ms d\mt}{(4 \pi i)^2} \Gamma_{\D_i}^6(\ms,\mt) u^{\frac{\ms-\D_1-\D_2}{2}} v^{\frac{\mt-\D_2-\D_3}{2}} \PP_{\kv;\D,J}^{s,t}(\ms,\mt).
\end{equation}
The Mellin-space representation of a Polyakov-Regge block is explicitly written in eq.~\eqref{PR mixed s}-\eqref{PR mixed t}.
The $s$-channel collinear limit corresponds to evaluating the Mellin amplitude at a $s$-channel double-twist value.
We thus define the $\widehat{B}_{\kv;\mt|mn}^{s,t}$ Mellin-space functionals as the residue of a Polyakov-Regge block on the $s$-channel pole located at $s=\D_{m}+\D_n+2k_{mn}$.
Singularities from the gamma functions in $\Gamma_{\D_i}^6(\ms,\mt)$ will cancel zeros from the Mellin-space Polyakov-Regge block to match the OPE expansion.
The position-space functionals are then obtained by performing a one-variable inverse Mellin transform\footnote{We absorb a factor of $\Gamma(\tfrac{\D_{m'}+\D_{n'}-\D_m-\D_n}{2} +k_{m'n'}-k_{mn})$ in our definition of $\widehat{B}_{\kv;v|mn}^{s,t}$ in eq.~\eqref{Bkt to Bkv} when compared to the equal operator definition given by eq.~(3.41) in \cite{Caron-Huot:2021enk}. Here, $(m'n')$ correspond to the complementary $s$-channel double-twist family.
}:
\begin{align}
\widehat{B}_{\kv;\mt|mn}^{s,t} &= \oint \displaylimits_{\ms=\D_m+\D_n+2k_{mn}} \frac{d\ms}{4\pi i} u^{\frac{\ms-\D_1-\D_2}{2}} \frac{\PP_{\kv; \D,J}^{s,t}(\ms,\mt)}{\Gamma_{\D_i|mn}^4(\mt)} \ \bigg|_{u=1} , \label{Bkt definition}\\
B_{\kv;v|mn}^{s,t} &=  \int_{\gamma_\mt^{mn}}  \frac{d\mt}{4\pi i} \ v^{\frac{\mt-\D_2-\D_3}{2}} \Gamma_{\D_i|mn}^4(\mt) \widehat{B}_{\kv;\mt|mn}^{s,t}, \label{Bkt to Bkv}
\end{align}
where $\gamma_\mt^{mn}$ denotes the contour integral in $\mt$ only, and $\Gamma_{\D_i|mn}^4(\mt)$ is a product of four gamma functions reduced from $\Gamma_{\D_i}^6(\ms,\mt)$ and $P_{\kv}$ after evaluating the residue at $\ms=\D_m+\D_n+2k_{mn}$:
\begin{equation} \label{Gamma4}
\begin{split}
\Gamma_{\D_i|mn}^{4}(\mt) &= \Gamma(k_{23} + \tfrac{\D_2+\D_3-\mt}{2}) \Gamma(k_{14} + \tfrac{\D_1+\D_4-\mt}{2}) \\
& \qquad \times \Gamma(\tfrac{\D_m+\D_n +2k_{mn} + \mt - \D_1-\D_3}{2}) \Gamma(\tfrac{\D_m+\D_n +2k_{mn} + \mt - \D_2-\D_4}{2} ).
\end{split}
\end{equation}
We recall that $(mn)=\{(12),(34)\}$ in our convention.
The contour $\gamma_\mt^{mn}$ is determined by $\Gamma_{\D_i|mn}^4(\mt)$:
\begin{equation} \label{mellin t contour}
\begin{split}
\Re(\gamma_\mt^{mn}) &=  \max( -\D_m-\D_n -2k_{mn} +\D_1+\D_3,-\D_m-\D_n -2k_{mn} + \D_2+\D_4)\\
& \qquad <\Re(\mt)< \min(\D_1{+}\D_4{+}k_{14},\D_2{+}\D_3{+}k_{23}), \\
\Im(\gamma_{\mt}^{mn}) &= \Im(\mt) \in \mathbb{R}.
\end{split}
\end{equation}
We find it helpful to explicitely extract the $\Gamma_{\D_i|mn}^{4}(\mt)$ factor in eq.~\eqref{Bkt definition} to be consistent with the notation in \cite{Caron-Huot:2020adz}.
The $s$- and $t$-channel collinear limit thresholds of table~\ref{tab:s and t collinear} fall within the range above.
Operators below those threshold are defined by analytic continuation where in Mellin-space, they are poles crossing the contour.

The position-space functionals are computed by deforming the $\mt$-contour to the right where it vanishes at the boundary due to boundedness condition in the Regge limit. 
Therefore, only the lower bound above must be considered carefully.
In practice, to obtain the position-space functional $B_{\kv;v|34}^{s,t}$, we integrate eq.~\eqref{Bkt to Bkv} numerically along the imaginary axis within the range above.
The $\gamma_\ms$ contour can also be determined by $\Gamma_{\D_i}^{6}$, however we will not need it since we are only interested in computing $B_{\kv;v|mn}^{s,t}$ rather than performing the full inverse Mellin transform to obtain the position-space Polyakov-Regge blocks.
We will now write explicit formulae for the Mellin-space functional $\widehat{B}_{\kv;\mt|mn}^{s,t}$ appearing in eq.~\eqref{Bkt to Bkv}.

Our starting point will be eq.~\eqref{M dispersion def} which we re-write here for convenience:
\begin{equation} \label{Gsub as dispersive Mellin}
\GG_{\kv}^{s,t} = \iint_\gamma \oint \frac{d\ms d\mt}{(4 \pi i)^2} \frac{d\ms'}{2\pi i} P_\kv(\ms,\mt;\ms',\mt') \Gamma_{\D_i}^6(\ms,\mt) u^{\frac{\ms-\D_1-\D_2}{2}} v^{\frac{\mt-\D_2-\D_3}{2}} \frac{M^{s,t}(\ms',\mt')}{\ms-\ms'},
\end{equation}
where we remind the reader that for fixed $\mU$-channel dispersion relation, we have $\ms+\mt=\ms'+\mt'$.
Our primary object of interest will be the product of Pochhamer symbols with the Mellin amplitude $P_{\kv} \times M^{s,t}(\ms',\mt')$.

The defining feature of Mellin amplitudes is their pole structure: 
each exchanged operator give rise to an infinite sequence of poles at $\tau_\OO + 2m$ for $m \in \mathbb{Z}^{0+}$.
This statement holds true for both $s$- and $t$-channel exchanged operators $\OS$ and $\OT$ respectively corresponding to poles at $\ms=\tau_{\OS}+2m$ and $\mt=\tau_\OT+2m$.
Moreover, the residue evaluated on these poles is fixed by the OPE:
\begin{equation}
\lim_{\ms \rightarrow \tau_\OO+2m} M_0(\ms,\mt) \sim f_{12\OO}f_{43\OO} \frac{\QQ_{\D_\OO,J_\OO}^m(\mt)}{\ms-\tau_\OO-2m},
\end{equation}
where $\QQ_{\D,J}^m(\mt)$ are kinematical objects closely related to so-called \textit{Mack polynomials} 
\cite{Mack:2009mi}.  
Explicit formulae for
unequal external operator Mack polynomials are recorded in appendix~\ref{app:Mack} for convenience.

Mack polynomials are degree $J$ polynomials with simple zeros at $s$-channel double-twist values; these zeros coalesce into double-zeros for equal external operators, which are necessary to cancel  the double poles from $\Gamma_{\D_i}^6$ in eq.~\eqref{G Mellin representation} in order to produce the correct contribution to the OPE.
For mixed correlators, $\Gamma_{\D_i}^6$ has a single pole in each channel which produces a finite result when combined with the zero from the Mack polynomial.

Let us write explicit formulae for the Mellin-space $\widehat{B}_{\kv;t|mn}^{s,t}$ functional.
We start with the Mellin-space dispersion relation of eq.~\eqref{M dispersion original def} such that we expand the Mellin amplitude in the integrand of eq.~\eqref{M dispersion def} in terms of both $s$- and $t$-channel pole contributions:
\begin{equation}
\begin{split}
M^s(\ms',\mt')&=\sum \limits_{m=0}^{\infty} \frac{\QQ_{\D,J}^m(\mt'+\ms'-\tau-2m-\D_2-\D_3)}{\ms'-\tau-2m}. \\
M^t(\ms',\mt') &= \sum \limits_{m=0}^{\infty}  \frac{\QQ_{\D,J}^m(\mt'+\ms'-\tau-2m-\D_1-\D_2)}{\mt'-\tau-2m}.
\end{split}
\end{equation}

Polyakov-Regge blocks are then obtained by closing the $\ms'$ contour towards the boundary to sum over all the contributions of exchanged operators.
Since we are working in a fixed$-\mU$ convention, poles corresponding to operators in the $s$- and $t$-channels are located at $\ms'=\tau+2m$ and $\ms'=\ms+\mt-\tau-2m$ respectively.
Formally, we have
\begin{equation}
\begin{split}
\PP_{\kv;\D,J}^s (\ms,\mt) &= \oint \displaylimits_{\ms'=\tau+2m} \frac{d\ms'}{2\pi i} P_\kv(\ms,\mt;\ms',\mt') \frac{M^{s}(\ms',\mt')}{\ms-\ms'}.
\end{split}
\end{equation}
We can further write subtracted Polyakov-Regge blocks in Mellin-space as follows:
\begin{align} 
\PP_{\kv;\D,J}^{s}(\ms,\mt)&= \sum \limits_{m=0}^{\infty} P_{\kv} (\ms,\mt; \tau+2m, \ms+\mt-\tau-2m) \frac{\QQ_{\D,J}^m(\ms+\mt-\tau-2m-\D_2-\D_3)}{s-\tau-2m}, \label{PR mixed s} \\
\PP_{\kv;\D,J}^{t}(\ms,\mt) &=- \sum \limits_{m=0}^{\infty} P_{\kv} (\ms,\mt; \ms+\mt-\tau-2m, \tau+2m) \frac{\QQ_{\D,J}^m(\ms+\mt-\tau-2m-\D_1-\D_2)}{\tau+2m-\mt}. \label{PR mixed t}
\end{align}
The Mellin-space functionals $\widehat{B}_{\kv;t|mn}^{s,t}$ are obtained by evaluating the residue of the Polyakov-Regge blocks as prescribed by eq.~\eqref{Bkt definition}.

Let us sketch how the Pochhammer sybols $P_\kv$ and the Mack polynomials conspire to produce vanishing or finite results.
The integrand of the $B_{\kv;v|mn}^{s}$ functional is given by
\begin{equation} \label{zeros or else for B}
B_{\kv;v|mn}^s = \int_{\gamma_t^{mn}} \frac{d\mt}{4\pi i} \oint \displaylimits_{\ms=\D_{m}{+}\D_n{+}2k_{mn}} \frac{d\ms}{4\pi i} \oint \displaylimits_{\ms'=\tau{+}2m} \frac{d\ms}{2\pi i} \frac{\Gamma_{\D_i}^{6}(\ms,\mt) \QQ_{\D_{\OO},J_{\OO}}^m(\ms+\mt-\D_2-\D_3)}{(\ms-\ms')(\ms'-\tau_{\OO}-2m)} P_\kv \times (...),
\end{equation}
where the ellipses are non-singular terms.
Let us consider $P_0=1$ first.
Recall that the Mack polynomials associated with $B_{\kv;v|mn}^s$ have simple zeros on both $s$-channel trajectories $\D_1+\D_2+2n$ and $\D_3+\D_4+2n$.
Evaluating the residue above converts the pole at $\ms=\tau+2m$ into one at $\D_m+\D_n+2k_{mn}=\tau+2m$.
For double-twist operators along the $\tau_{mn}$ $s$-channel trajectory, the simple zero in the Mack polynomial cancels this pole. 
However, the integrand is absent of poles in the other $s$-channel trajectory $(m'n') \neq (mn)$ and therefore it vanishes in general.
Thus, to obtain a finite result for exchanged operators along the other $s$-channel trajectory, we can add poles using the ratio of Pochhammer symbols $P_{\kv}$. 
The zero structure described here allows us to decompose  any functional $\widehat{B}_{\kv;\mt|mn}^{s,t}$ in terms of double-twist analytic functionals \cite{Mazac:2018mdx,Mazac:2018ycv,Mazac:2018qmi,Mazac:2019shk,Caron-Huot:2020adz}:
\begin{equation} \label{B analytic funct exp}
\widehat{B}_{\kv;\mt|mn}^{s} \bigg|_{\D=\D_o+\ell} = \widehat{a}_{\ell|mn,\D_o}^{\D_i,s}(\mt) + \widehat{b}_{\ell|mn,\D_o}^{\D_i,s}(\mt) \; (\D-\D_o-\ell) + O((\D-\D_o-\ell)^2),
\end{equation}
where $\D_o$ takes on double-twist values.

There is a key difference between the equal operator and the mixed correlator cases:
equal operator subtracted collinear functionals $B_{k;v}$ can be expanded into a basis of analytic functionals with double-zeros on double-twists $n>0$.
This is no longer the case for mixed correlators unless we restrict ourselves to pairwise equal operators. 
For the $\langle AABB \rangle$ and $\langle ABBA \rangle$ correlators, one channel may be expanded into double-zero functional basis elements, while the other channel may be expanded into a basis of functions with simple zeros.
We elaborate on this distinguishing feature in appendix~\ref{app:projection}.

\subsection{Mellin-space representation of holographic functionals} \label{sec:mellin regge}

Before ending this section, we derive the Mellin representation of the mixed correlator holographic functional $C_{\kv;\nu|34}^{s,t}$ introduced in section~\ref{sec:pos Regge}.
To do so,
one must substitute the Mellin-space integral representation of $B_{\kv;v|34}^{s,t}$ from eq.~\eqref{Bkt to Bkv} into eq.~\eqref{Ckv position}.

Firstly, the $v$-integral which extracts Regge moments can be done straightforwardly using eq.~\eqref{v inverse integral} by acting on the power of $v$ from the Mellin integral representation of $B_{\kv;v|34}^{s,t}$:
\begin{equation}
\begin{split}
&\int \displaylimits_{\eta^2}^{\infty} d v \ \frac{\pi ^{3/2} 2^{1+2 \left(\D_2-\D_4-k_{12}-k_{14}-k_{23}-k_{34}\right)}\left(v/\eta^2 \right)^{-\mathfrak{b}} v^{(\mt-\D_2-\D_3)/2}}{(v-\eta^2)^{\frac{1}{2}+\mathfrak{a}+\mathfrak{b}} v^{-\frac{1}{2} \left(\D_2-\D_4+k_{12}-k_{14}-k_{23}-k_{34}\right)}} \frac{ {}_2F_1(\tfrac{1}{2}-\mathfrak{b},-\mathfrak{b},\tfrac{1}{2}-\mathfrak{a}-\mathfrak{b},1-\tfrac{v}{\eta^2})}{\Gamma(\tfrac{1}{2}-\mathfrak{a}-\mathfrak{b}) \Gamma(\mathfrak{a}+ \mathfrak{b}) }   \\
& \qquad = \frac{2^{-\Delta _1+\Delta _2-2 \Delta _4-3 k_{12}-3 k_{14}-3 k_{23}-k_{34}+\mt+3}  \pi ^2 \eta ^{-\Delta _1-\Delta _2-k_{12}-k_{14}-k_{23}+k_{34}+\mt+1}  }{\Gamma \left(\frac{1}{2} \left(2 k_{12}-2 k_{34}+\Delta _1+\Delta _2-\Delta _3-\Delta _4\right)\right)} \\
& \qquad \times \frac{\Gamma \left(\Delta _1+\Delta _2+k_{12}+k_{14}+k_{23}-k_{34}-\mt-1\right)}{ \Gamma \left(\frac{1}{2} \left(-\mt+2 k_{23}+\Delta _2+\Delta _3\right)\right) \Gamma \left(\frac{1}{2} \left(-\mt+2 k_{14}+\Delta _1+\Delta _4\right)\right)}
\end{split}
\end{equation}
The $\eta$-integral, which is just a Laplace transform of a Gegenbaueur function, can also be computed analytically; 
we recast the result from \cite{Caron-Huot:2021enk} for convenience:
\begin{equation}
\int_1^{\infty} [d\eta] \PP_J(\eta)\eta^{-X} = \frac{2^{X+d-4}\Gamma(\tfrac{d-1}{2})}{\sqrt{\pi}\Gamma(X)} \Gamma(\tfrac{X+J}{2}) \Gamma(\tfrac{X+2-d-J}{2}).
\end{equation}
Importantly, $J=\tfrac{2-d}{2}+i \nu$ such that we obtain a $\gamma_a(\nu)$ factor defined by eq.~\eqref{gamma nu}.

Substituting eq.~\eqref{Bkt to Bkv} into \eqref{Ckv position}, and using the above, we obtain the following result for the Mellin representation of the $C_{\kv;\nu|34}^{s,t}$ functional:
\begin{equation} \label{Ckv mellin}
\begin{split}
C_{\kv;\nu|34}^{s,t} &=(-1)^{k_{12}+k_{34}+1} 2^{d+2 \Delta _2-2 \Delta _4-2 k_{12}-2 k_{14}-2 k_{23}-2 k_{34}-1} \frac{a_{\D_i,k_{ij}}(\nu)  \pi^{3/2} \Gamma(\tfrac{d-1}{2})}{\Gamma(\tfrac{\Delta _1+\Delta _2-\Delta _3-\Delta _4+2 k_{12}-2 k_{34}}{2})}  \\
& \times \int_{\tilde{\gamma}_{\mt}^{34}} \frac{d\mt}{4\pi i} \Gamma(\tfrac{\mt-\Delta _2+\Delta _3}{2} ) \Gamma(\tfrac{\mt-\Delta _1+\Delta _4}{2}) \gamma_{\Delta _1+\Delta _2+k_{12}+k_{14}+k_{23}-k_{34}-\mt-1}(\nu) \widehat{B}_{\kv;\mt|34}^{s,t}
\end{split}
\end{equation}
This calculation justifies the $\D_i$ dependence of $\gamma_a(\nu)$ in $a_{\D_i,k_{ij}}(\nu)$ as defined by eq.~\eqref{a_normalization}: 
in Mellin-space, the $\gamma_a(\nu)$ factor naturally arises to smear the integrand.
Furthermore, the $\D_i$ dependence of $\gamma_a(\nu)$ in the integrand is determined by the $s$-channel double-twist families where the functional is evaluated;
its arguments would be $\D_3+\D_4+k_{34}-k_{12}+k_{23}+k_{14}-1-\mt$ for $(mn)=(12)$.
To localize the action of the functional in the AdS bulk, we must ``unsmear" the integral appropriate $\gamma_a(\nu)$ factors. 
The latter are fixed by the lower bound of the integration region set by $\Gamma^4_{\D_i|34}(\mt)$ which encodes information from the two $\mU$-channel double-twist families.
Lastly, the domain of convergence of the integral has changed: there is only a lower bound on the real part of $\mt$.
We leave the study of these mixed correlator holographic functionals for future work.

\section{Applications to the 3D Ising model}
\label{sec:app}

Dispersive functionals have proven to be a valuable tool for the CFT bootstrap program \cite{Carmi:2020ekr,Caron-Huot:2021enk}:
the suppression of double-twist operators allows one to probe non-perturbative properties of CFTs, while
their positivity properties are an enticing feature for the numerical bootstrap.
In this section, we extend such explorations by applying our mixed correlator functionals to the 3D Ising model.

In section~\ref{sec:Ising sum rule}, our first goal is to determine the optimal subtraction schemes for each correlator i.e. to determine a sum rule that is balanced by the least number of operators.
We will then use this information to derive approximate solutions to crossing for the 3D Ising model in section~\ref{sec:approximate sol}.

Below, all computations are evaluated with the collinear functional $B_{\kv;v|34}$, the derivative functional $\partial_v B_{\kv;v|34} \Big|_{v=1} \equiv B_{\kv;v|34}'$, and the $\Phi_2$ functional evaluated at $v=1$ where the
latter was introduced in \cite{Caron-Huot:2020adz}. 
Itt is constructed from a crossing-symmetric subtraction scheme $\kv=(1,1,0,0)$ and its features are reviewed in appendix~\ref{app:projection}. 
Unfortunately, the $\Phi_2$ functional becomes computationally expensive at high spin which represents a barrier for high precision measurements (see the discussion for the $\langle \epsilon \epsilon \epsilon \epsilon \rangle$ correlator below).
As for the derivative functional, it is best understood in Mellin-space where it is clear that
\begin{equation} \label{derivative functional}
B_{\kv;v|34}^{'s,t} = \int_{\gamma_{\mt}^{34}} \ \frac{d\mt}{4\pi i} \ \frac{\mt-\D_2-\D_3}{2} v^{\frac{t-\D_2-\D_3}{2}} \Gamma_{\D_i|34}^{4}(\mt) \widehat{B}_{\kv;\mt|34}^{s,t},
\end{equation}
thus it can be viewed as a higher spin Mack polynomial.

One could further use the collinear functional for different values of $v$.
We explored the effects of varying $v$ for the $\langle \sigma \sigma \sigma \sigma \rangle$ correlator but found derivative functionals to be more effective;
a short discussion about $v\neq1$ is recorded in appendix~\ref{app:varying v}.

We set $k_{34}=0$ to probe the leading double-twist trajectory, and we further restrict ourselves to spin-2 convergent non-log functionals (without using anti-subtractions) in order to include the identity in the OPE.
Figures in section~\ref{sec:Ising sum rule} displaying the action of the functional for varying twist $\tau$ and fixed spin $J$ are normalized with respect to the mean field theory OPE coefficients.
Moreover, values shown in the tables in section~\ref{sec:approximate sol} include the OPE coefficients as well.

\subsection{Evaluating 3D Ising model sum rules} \label{sec:Ising sum rule}
To test the accuracy of our functionals, we will evaluate mixed correlator sum rules using the 3D Ising model data provided in \cite{Simmons-Duffin:2016wlq}.
Since dispersive functionals suppress the contribution of double-twist operators, one can hope to truncate the OPE sum to a small subset of operators.
We therefore present our results by comparing truncated sums to the total sum of all the following operators where their values are retrieved from \cite{Simmons-Duffin:2016wlq}: the stable $\mathbb{Z}_2$ even and odd operators, and the $[\sigma \sigma]_{0,J}$, $[\epsilon\epsilon]_{0,J}$, $[\sigma\sigma]_{1,J}$ and $[\sigma\epsilon]_{0,J}$ double-twist families.
Heavier operators and subleading double-twist families are not considered.
We summarize the most effective truncated sum rules in table~\ref{tab:sum rule summary} where we have separate the table into two parts: only sum rules in the first part of the table will be used to derive approximate solutions to crossing in section~\ref{sec:approximate sol}, while sum rules in the second part of the table either include new operators $\epsilon', [\sigma\epsilon]_{0,2}$, or they are linearly dependent to those in the first part.
\begin{table}[h]
\centering
\begin{tabular}{|c|c|c|c|c|}
\hline
$\langle \phi_A \phi_B \phi_C \phi_D \rangle$ &Functional & Subtraction & Sum rule & Error \\ \hline
$\langle \sigma \sigma \sigma \sigma \rangle$ & $B_{\kv;1|34}$ & $(1,1,0,0)$ & $\mathbf{1}+ \underline{ \epsilon} + T_{\mu \nu}- \mathbf{1}_\mU$ =0 &$0.008 $ \\
& $B_{\kv;1|34}'$ & $(1,1,0,0)$ & $\mathbf{1} +\underline{ \epsilon} + T_{\mu \nu}=0$ &$0.002 $ \\
 & $\Phi_2$ & $(1,1,0,0)$ & $\epsilon + \underline{T_{\mu \nu} }=0$ &$0.06$ \\
$\langle \epsilon \sigma \sigma \epsilon \rangle$ & $B_{\kv;1|34}$ & $(1,1,0,0)$ & $\underline{ \sigma}+ \mathbf{1}+\epsilon = 0$ & $-0.03 $  \\
& $B_{\kv;1|34}'$ & $(1,1,0,0)$ & $\underline{ \sigma}+ \mathbf{1}+\epsilon = 0$ & $0.09 $  \\
$\langle \sigma \sigma \epsilon \epsilon \rangle$ & $B_{\kv;1|34}$ & $(1,1,0,0)$ & $\underline{ \sigma}+ \epsilon+T_{\mu \nu} = 0$ & $-0.004 $ \\ 
& $B_{\kv;1|34}'$ & $(1,1,0,0)$ & $\underline{\sigma}+  \epsilon+ T_{\mu \nu} = 0$ & $-0.02 $ \\ \hline \hline
$\langle \epsilon \epsilon \epsilon \epsilon \rangle$ & $B_{\kv;1|34}'$ & $(1,1,0,0)$ & $\mathbf{1}+\underline{T_{\mu \nu} }+\epsilon + \epsilon'=0$ & $0.06$ \\
& $B_{\kv;1|34}'$ & $(2,0,0,0)$ & $\mathbf{1}+T_{\mu \nu}+\epsilon + \underline{\epsilon'}=0$ & $-0.01$ \\
$\langle \epsilon \epsilon \sigma \sigma \rangle$ & $B_{\kv;1|34}$ & $(0,1,0,1)$ & $\underline{ \sigma}+ \mathbf{1}+\epsilon = 0$ & $-0.03 $  \\ 
$\langle \epsilon \sigma \epsilon \sigma \rangle$ & $B_{\kv;1|34}$ & $(2,0,0,0)$ & $-\underline{ \mathbf{1}_{\mU}}+\sigma+ [\sigma \epsilon]_{0,2} = 0$ & $0.03 $  \\ \hline
\end{tabular}
\caption{Most effective truncated sum rules with errors rounded to one significant number based on tabulated results presented later in this section. Errors are normalized with respect to the largest contribution to the sum rule denoted by the underlined operator in the 4th column. }
\label{tab:sum rule summary}
\end{table}

In hindsight, it is not surprising that the most effective functionals have crossing-symmetric subtraction schemes:
the $s$-channel restriction for non-log functionals is highly constraining for most correlators, and additional $t$-channel subtractions enhances the contribution of low twist operators with twist $\tau<\D_i+\D_j+2k_{ij}-1=\tau^*$.
Since heavy operators in $s$- and $t$-channel cancel amongst each other due to the suppression endowed by the dispersive nature of our functionals, only light operators below $\tau^*$ dominate the sum rule.
Nonetheless, different subtraction schemes lead to functionals with widely different behaviours as shown below and therefore, they might still prove useful as a basis to carve out the space of allowed CFTs.
One prominent barrier to such an approach is to construct a positive definite combination of these functionals;
it may be possible to construct a mixed correlator matrix with desirable positivity properties as was done in \cite{Kos:2014bka}.

\subsubsection{$\langle \sigma \sigma \sigma \sigma \rangle$}
We first present sum rules associated with the correlator $\langle \sigma \sigma \sigma \sigma \rangle$.
This correlator includes the identity, the $\mU$-channel identity and even $\mathbb{Z}_2$ operators. 
Therefore, our sum rule reads
\begin{equation} \label{sum rule ssss}
\mathbf{1} - \mathbf{1}_{\mU} + \sum_{\OO \in \mathbb{Z}_2^+} f_{\sigma\sigma \OO}^2 \omega [ \GG ] = 0.
\end{equation}
For equal operators, we need at least one $s$-channel subtraction to obtain a non-log functional.
Although table~\ref{Tab:convergence} might suggest that we require at least one subtraction in the $t$-channel, by evaluating the functional on $\GG -\mathbf{1}_{\mU}$, we can relax the $t$-channel convergence conditions to allow for $k_{23}=k_{14}=0$.
Therefore, we are left with the $(k_{12},k_{23},k_{34},k_{14})=(1,1,0,0)$ and $(2,0,0,0)$ subtraction schemes.
Numerical results are presented in table \ref{tab:ssss operators}.

\begin{table}[h]
\centering
\begin{tabular}{|c|c|c|c||c|c|c|c|c|}
\hline
$\OO$ & $\mathbb{Z}_2$ & $J_{\OO}$ & $\tau_{\OO}$ & $\Phi_2$ & $B_{(1,1,0,0);1|34}$ & $B_{(2,0,0,0);1|34}$ & $B_{(1,1,0,0);1|34}'$ & $B_{(2,0,0,0);1|34}'$ \\ \hline
$-\mathbf{1}_\mU$ & $+$ & 0 & 0 & 0 &$ -0.478$ & $ -0.470$ & 0 & 0 \\
$\mathbf{1}$ & $+$ & 0 & 0 & 0 &$ -0.478$ & $ -0.470$ & $1$ & $-0.979$ \\
$T_{\mu \nu}$ & $+$ & 2 & 1 & $1$ & $-0.036$ & $-0.018$ & $-0.014$ & $0.011$ \\
$\epsilon$ & $+$ & 0 & $1.412$ & $-0.944$ & $1$ & $1$ & $-0.984$ & $1$  \\
$\epsilon'$ & $+$ & 0 & $3.82968$ &  &   & $-0.015$  &  & $-0.016$ \\ \hline
\multicolumn{4}{|c||}{Truncated sum} & $0.056$ &$0.008$ &$0.027$ & $0.002$ & $0.016$ \\
\multicolumn{4}{|c||}{Sum of all operators} & $0.04$ &$-0.0008$ &$0.01$ & $0.0004$ & $-0.005$  \\ \hline
\end{tabular}
\caption{Contribution of each operator to the $\langle \sigma \sigma \sigma \sigma \rangle$ sum rule with $v=1$; these results include the contribution of the OPE coefficient square $f_{\sigma \sigma \OO}^2$. The values are normalized with respect to the largest contribution to the sum rule in order to be comparable to the $\Phi_2$ sum rule.  All other operators contribute to order $O(10^{-3})$ and less. Finally, the sum of all operators must add up to 0 in accordance with eq.~\eqref{sum rule ssss}.}
\label{tab:ssss operators}
\end{table}

The $B_{\kv;1}$ sum rule balances the $\epsilon$ operator against both the $\mU$-channel identity and the $s$- and $t$-channel identity. 
The $(2,0,0,0)$ subtraction scheme marginally spreads the contribution of all operators across the OPE sum.
In contrast, the $\Phi_2$ functional mostly balances the stress-tensor against the $\epsilon$ operator; we were able to reproduce Table 1 of \cite{Caron-Huot:2020adz} which we write here again for convenience.
We have further included the derivative sum rule which mostly balances the identity against the $\epsilon$ operator.
We therefore conclude that different subtraction schemes yield negligeable changes the precision of the $\langle \sigma \sigma \sigma \sigma \rangle$ sum rule.

\begin{figure}[h]
\centering
\includegraphics[width=\textwidth]{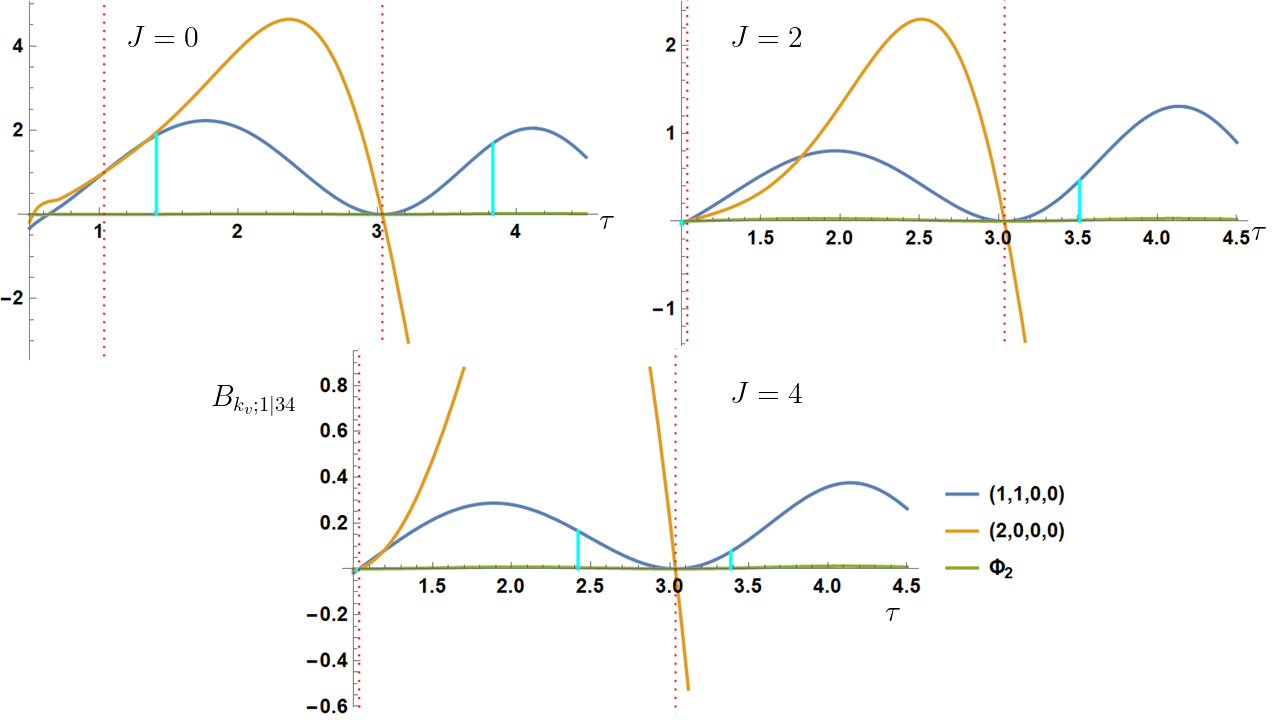}
\caption{Action of $B_{\kv;1|34}$ for the $\langle \sigma \sigma \sigma \sigma \rangle$ correlator acting on conformal blocks $G_{\D,J}$ of different spin $J$, twist $\tau=\D-J$ and subtraction schemes $\kv=(1,1,0,0)$ and $(2,0,0,0)$ represented as blue and orange curves respectively. Red dotted vertical lines denote double-twist values while vertical cyan lines denote the Ising model operators  appearing in the OPE of $\langle \sigma \sigma \sigma \sigma \rangle$.}
\label{spinGraph ssss}
\end{figure}

Let us understand how the values in table~\ref{tab:ssss operators} are obtained by describing the behaviour of the functionals as a function of twist at fixed spin $J$.
Since $s$-channel subtractions convert double zeros into simple zeros, the $(2,0,0,0)$ subtraction scheme is negative definite for $\tau> 2\D_\sigma +2$ and it is positive between the leading and subleading double-twist i.e. for $2\D_\sigma< \tau < 2\D_\sigma+2$ as shown in fig.~\ref{spinGraph ssss}.
We further observe that the absolute value of the $B_{\kv|1}$ functional is larger for the $(2,0,0,0)$ subtraction scheme despite the fact that the relative contribution of each operator is similar to the $(1,1,0,0)$ subtraction scheme when normalized with respect to the contribution of the $\epsilon$ operator.
This larger action is suggestive of a slower OPE convergence rate which is confirmed by the green and red curves in the left panel of fig.~\ref{fig:eeee convergence}.
To understand the difference between the $(2,0,0,0)$ and $(1,1,0,0)$ subtraction schemes at large twist, let us analyze the mellin-space functionals:
\begin{equation}
\begin{split}
\widehat{B}_{(1,1,0,0);\mt} &\propto (2\D_\phi -\mt),  \\
\widehat{B}_{(2,0,0,0);\mt} &\propto \frac{(\Delta -J)^2+4 \D_\phi^2-2 \D_\phi (\Delta +\mt-2-J)+\mt^2-(\Delta +2-J) \mt}{(-\Delta +2 \D_\phi+2\ J-2) (-\Delta +\mt+J-2)}.
\end{split}
\end{equation}
We note the $\D$ and $\mt$ dependence in the above which enhances the action of the position-space functional for larger twist.
From this example, we learn that an ideal subtraction scheme would suppress the contribution of larger twist operators for fixed spin $J$ relative to the low twist sector.

\subsubsection{$\langle \epsilon \epsilon \epsilon \epsilon \rangle$} \label{sec:Ising eeee}
\begin{figure}[h]
\centering
\begin{subfigure}{0.49 \textwidth}
\includegraphics[width=\linewidth]{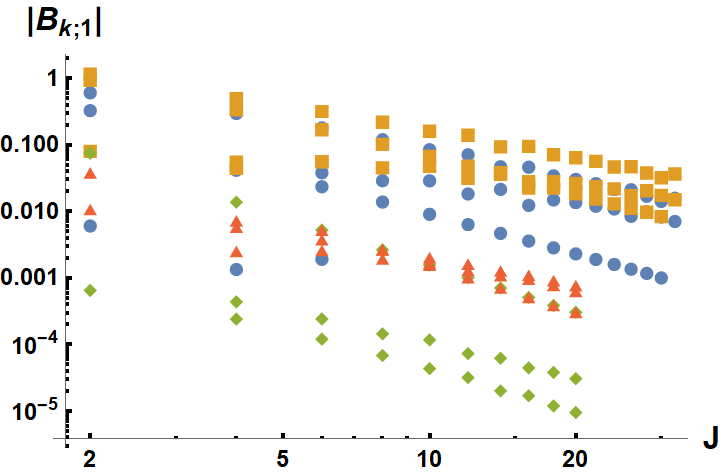}
\end{subfigure}
\hfill
\begin{subfigure}{0.49 \textwidth}
\includegraphics[width=\linewidth]{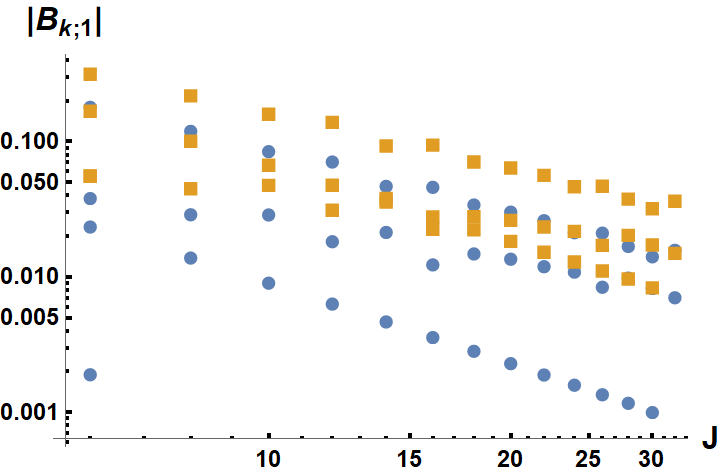}
\end{subfigure}
\caption{We plot the absolute value of the functional contribution for various operators as a function of spin. The green and red plots correspond to the $B_{(1,1,0,0);1}$ and $B_{(2,0,0,0);1}$ functionals applied to the $\langle \sigma \sigma \sigma \sigma \rangle$ correlator. The blue and orange plots correspond to the $B_{(1,1,0,0);1}$ and $B_{(2,0,0,0);1}$ functionals applied to the $\langle \epsilon \epsilon \epsilon \epsilon \rangle$ correlator respectively. The right panel is a close-up of the $B_{\kv;1}$ functionals applied to the $\langle \epsilon \epsilon \epsilon \epsilon \rangle$ correlator. In the right panel, one double-twist family converging more quickly for the $(1,1,0,0)$ subtraction scheme than others: this is the $[\sigma \sigma]_{0,J}$ family.}
\label{fig:eeee convergence}
\end{figure}

One could also apply these functionals on the $\langle \epsilon \epsilon \epsilon \epsilon \rangle$ correlator. However, the OPE sum converges more slowly than for the $\langle \sigma \sigma \sigma \sigma \rangle$ correlators as illustrated in fig.~\ref{fig:eeee convergence}.
Indeed, by comparing the red and green curves to the blue and orange ones, large spin operators are necessary to provide convergence and therefore, one cannot truncate the data as effectively as was done for the $\langle \sigma \sigma \sigma \sigma \rangle$ correlator.
This leads to slowly convergent sum rules as shown in table~\ref{tab:eeee operators}.

\begin{figure}[h]
\centering
\includegraphics[width= \linewidth]{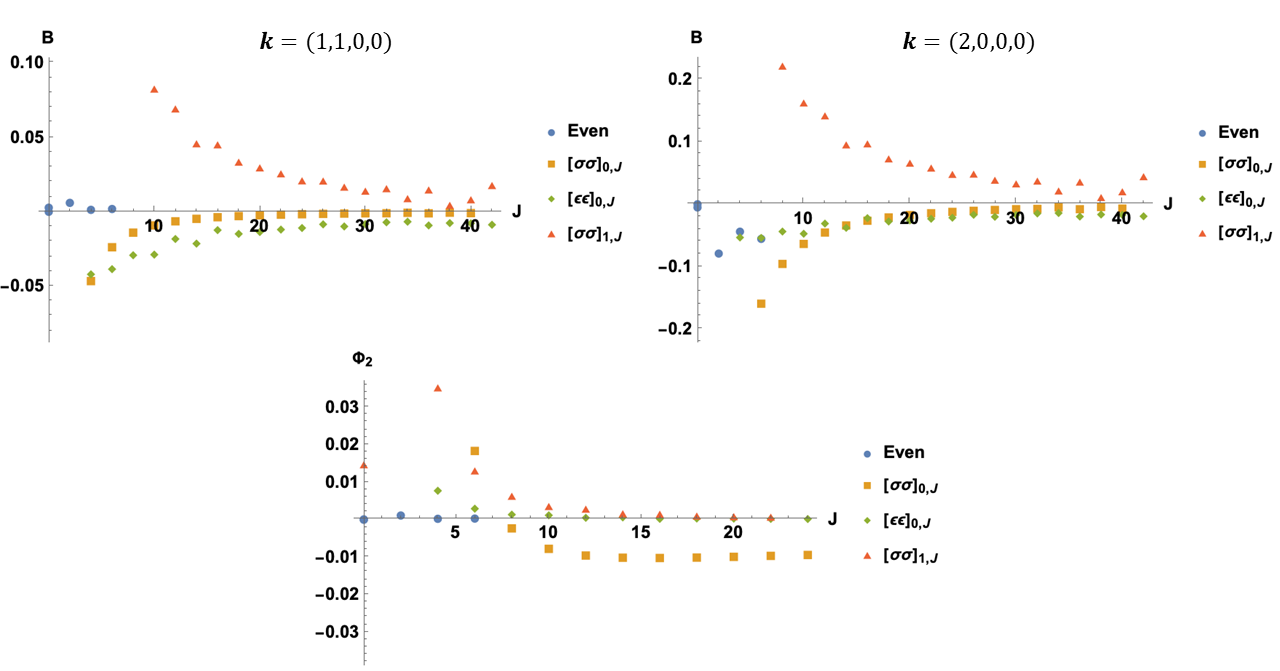}
\caption{Action of the $B_{\kv;1}$ and $\Phi_2$ functional on exchanged operators in the OPE of the $\langle \epsilon \epsilon \epsilon \epsilon \rangle$ correlator.  The ``even" family in the legend corresponds to other stable $\mathbb{Z}_2$ even operators. 
The functional acting on the $[\sigma\sigma]_{0,J}$ and low twist $[\epsilon\epsilon]_{0,J}$ families are defined by analytic continuation. For the $\Phi_2$ functional, the projection lifts the negativity of the $[\epsilon \epsilon]_{0,J}$ family at the cost of losing convergence of the $[\sigma \sigma]_{0,J}$ family for higher spin operators.}
\label{fig:eeee operators}
\end{figure}

This slow convergence can be better understood from fig.~\ref{fig:eeee operators}.
We see that both $[\sigma \sigma]_{0,J}$ and $[\epsilon\epsilon]_{0,J}$ families are negative due to their twists falling below the collinear thresholds tabulated in table~\ref{tab:s and t collinear} such that they are defined by analytic continuation (by shifting the Mellin-mandelstam $\mt$ contour).
Therefore, the slow convergence can be remedied by adding heavier operators in our sum.
This is only true of the $\kv=(1,1,0,0)$ subtraction scheme where heavier higher twist operators would contribute positively to the sum to support the contributions from the $[\sigma \sigma]_{1,J}$ against lower twist operator families $[\sigma \sigma]_{0,J}$ and $[\epsilon \epsilon]_{0,J}$.
On the other hand, the $(2,0,0,0)$ subtracted functional converges well since it mostly balances the $[\sigma \sigma]_{1,J}$ family against all other operators.
As for the $\Phi_2$ functional, the sum rule appears to be non-convergent as shown in fig.~\ref{fig:eeee convergence}.
It would be interesting to further test convergence properties of the $\Phi_2$ functional to better understand this observation.

\begin{figure}[h]
\centering
\includegraphics[width=\textwidth]{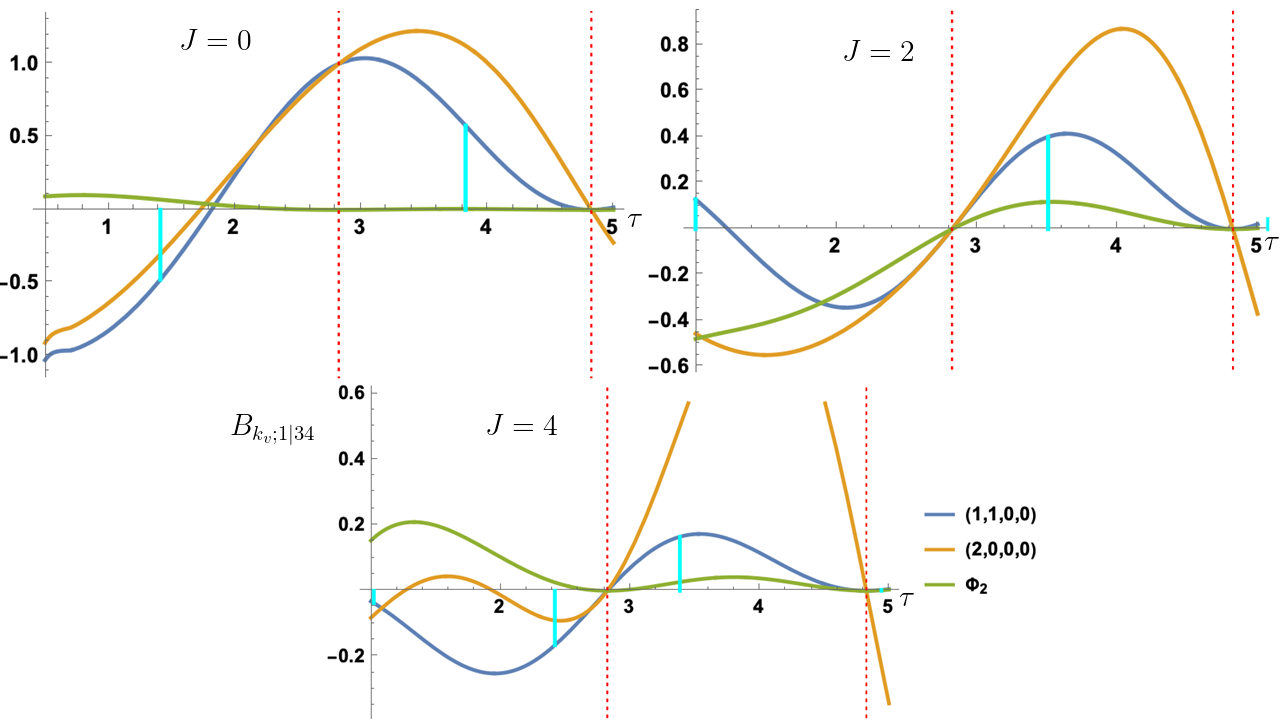}
\caption{Action of $B_{\kv;1|34}$ for the $\langle \epsilon \epsilon \epsilon \epsilon \rangle$ acting on conformal blocks $G_{\D,J}$ of different spin $J$, twist $\tau=\D-J$ and subtraction schemes $\kv=(1,1,0,0)$ and $(2,0,0,0)$ represented as blue and orange curves respectively. Red dotted vertical lines denote double-twist values while vertical cyan lines denote the Ising model operators appearing in the OPE of $\langle \epsilon \epsilon \epsilon \epsilon \rangle$.}
\label{spinGraph eeee}
\end{figure}

The slow convergence prevents any reasonable truncation of the OPE as shown by the results in table~\ref{tab:eeee operators}.
The derivative sum rule $B_{\kv;1|34}'$ on the other hand is dominated by the identity, the stress tensor, $\epsilon$ and $\epsilon'$.
The presence of the $\epsilon'$ operator prevents us from using these sum rules to derive partial sums satisfying crossing in the next section, however they are a good candidate to be used in future bootstrap computations.
The improved convergence for the derivative functional can be understood from eq.~\eqref{derivative functional}: higher powers in $\mt$ can be viewed as higher spin Mack polynomials which translates into an accelerated convergence in spin.
\begin{table}[h]
\centering
\begin{tabular}{|c| c |c| c || c|c|c|c|c |}
\hline
$\OO$ & $\mathbb{Z}_2$ & $J_{\OO}$ & $\tau_{\OO}$ & $\Phi_2$ &  $B_{(1,1,0,0);1|34}$ & $B_{(2,0,0,0);1|34}$ & $B_{(1,1,0,0);1|34}'$ &  $B_{(2,0,0,0);1|34}'$ \\ \hline
$-\mathbf{1}_{\mU}$ & + & 0 & 0 & 0& $-0.731$ & $-0.379$ & 0 & 0 \\
$\mathbf{1}$ & + & 0 & 0& 0 & $-0.731$  & $-0.379$ & $-0.692$ & $-0.755$ \\
$T_{\mu \nu}$ &+ & 2 & 1 & $-1$  & $0.238$ & $-0.433$ & $1$ & $0.269$  \\
$\epsilon$ &+ & 0 &  $1.41263$ & $0.137$ & $-0.820$ & $-0.271$ & $-0.697$ & $-0.527$\\ 
$\epsilon'$ &+ & 0 & $3.82968$ & & $1$ & $1$ & $0.447$ & $1$  \\
$[\sigma \sigma]_{0,J}$ &+ &4 & $1.02267$ & $0.206$ & & $-0.120$ & &  \\
$[\sigma \sigma]_{1,J}$ &+ & 2 & $3.50915$ & $0.150$ & $0.449$  & $0.337$ & & \\
$[\sigma \sigma]_{1,J}$ &+ & 4 & $3.38568$ &  & $0.212$ & $0.185$ & & \\
$[\sigma \sigma]_{1,J}$ &+ & 6 & $3.32032$ &  & $0.127$ & $0.120$ & & \\ \hline
\multicolumn{4}{|c||}{Sum of all operators} & $-0.475$ &$-0.139$ &$0.05$ & $0.0004$ & $-0.005$ \\ \hline
\end{tabular}
\caption{Contribution of each operator to the $\langle \epsilon \epsilon \epsilon \epsilon \rangle$ sum rule with $v=1$; these results include the contribution of the OPE coefficient square $f_{\epsilon \epsilon \OO}^2$. Empty cells and all other operators contribute at order $O(10^{-2})$ and smaller. Finally, the sum of all operators must add up to 0 in accordance with eq.~\eqref{sum rule ssss}.}
\label{tab:eeee operators}
\end{table}
Lastly, we record the action of $\Phi_2$ and $B_{\kv;1|34}$ on the $\langle \epsilon \epsilon \epsilon \epsilon \rangle$ correlator for spins $J=0,2,4$ in fig.~\ref{spinGraph eeee}.

\subsubsection{$\langle \epsilon \sigma \epsilon \sigma \rangle$}
Since the $\langle \epsilon \sigma \epsilon \sigma \rangle$ correlator is manifestly $\ms \leftrightarrow \mt$-channel crossing-symmetric, the sum rule reads:
\begin{equation} \label{sum rule eses}
- \mathbf{1}_{\mU} + \sum_{\OO \in \mathbb{Z}_2^-}(-1)^J f_{\epsilon \sigma \OO}^2 \ \omega [ \GG ] = 0.
\end{equation}
For this first mixed correlator, as indicated in table~\ref{Tab:convergence}, we require at least one $s$-channel subtraction to obtain non-log functionals while subtractions in the $t$-channel are unconstrained. Moreover, the two $t$-channel trajectories are indistinguishable. and therefore, we present results for the following $(k_{12},k_{23},k_{34},k_{14})$ subtraction schemes in table~\ref{tab:eses operators}: $(2,0,0,0)$ and $(1,1,0,0)$. 

\begin{table}[h]
\centering
\begin{tabular}{|c|c|c|c||c|c|c|c|c|}
\hline
$\OO$ & $\mathbb{Z}_2$ & $J$ & $\tau_\OO$ & $B_{(2,0,0,0);1|34}$ & $B_{(1,1,0,0);1|34}$ & $B_{(2,0,0,0);1|34}'$ & $B_{(1,1,0,0);1|34}'$ \\ \hline
$-\mathbf{1}_\mU$ & $+$ & 0 & 0 & $-1$  & $-1$ & 0 & 0 \\
$\sigma$ & $-$ & 0 & $0.518149$ & $0.865$  & $0.608$ & $1$  & $-0.892$ \\
$[\sigma \epsilon]_{0,J}$ & $-$ & 2 & $2.18031$&$0.162$ & $0.295$ & $-0.676$ & $1$ \\
$[\sigma \epsilon]_{0,J}$ & $-$ & 3 & $1.63804$ &$-0.038$ & $-0.077$ & $0.042$ & $-0.207$ \\
$[\sigma \epsilon]_{0,J}$ & $-$ & 4 & $2.11267$ &$0.067$ & $0.111$ & $-0.056$ & $0.130$ \\ \hline
\multicolumn{4}{|c||}{Truncated sum} & $0.057$ &$-0.062$ & $0.314$ & $0.031$  \\
\multicolumn{4}{|c||}{Sum of all operators} & $0.07$ &$-0.03$ & $-0.2$ & $-0.005$ \\ \hline
\end{tabular}
\caption{Contribution of each operator to the $\langle \epsilon \sigma \epsilon \sigma \rangle$ sum rule with $v=1$ normalized relative to the largest contributor of the sum rule; these results include the contribution of the OPE coefficient square $(-1)^J f_{\epsilon \sigma \OO}^2$. Sum rules from these subtracted functionals should add up to 0 in accordance with eq.~\eqref{sum rule eses}.  Other operators contribute at order $O(10^{-2})$ and smaller; we show the $[\sigma \epsilon]_{0,3}$ operator to illustrate the negativity for odd spin operators in these sum rules.}
\label{tab:eses operators}
\end{table}

\begin{figure}[h]
\centering
\includegraphics[width=\textwidth]{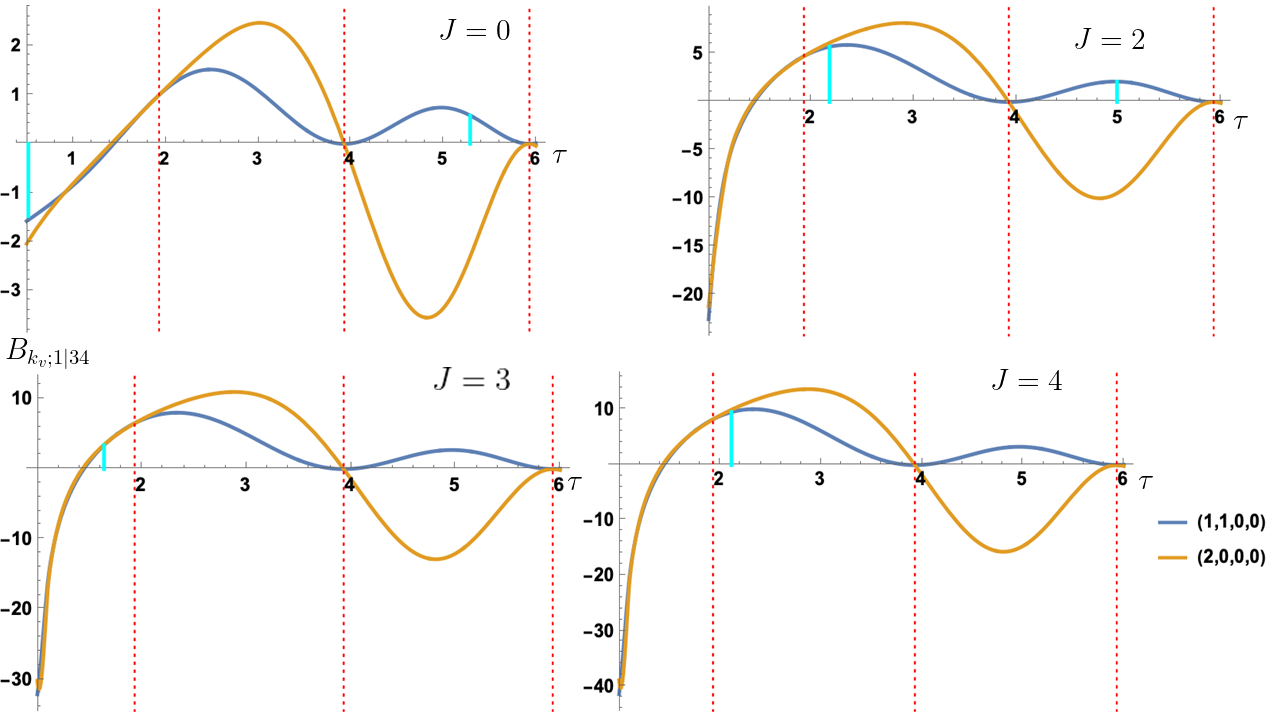}
\caption{Action of $B_{\kv;1|34}$ for the $\langle \epsilon \sigma \epsilon \sigma \rangle$ acting on conformal blocks $G_{\D,J}$ of different spin $J$, twist $\tau=\D-J$ and subtraction schemes $\kv=(1,1,0,0)$ and $(2,0,0,0)$ represented as blue and orange curves respectively. Red dotted vertical lines denote double-twist values while vertical cyan lines denote the Ising model operators appearing in the OPE of $\langle \epsilon \sigma \epsilon \sigma \rangle$.}
\label{spinGraph eses}
\end{figure}

From table~\ref{tab:eses operators}, we first note that odd spin operators have negative contributions while even spin operators contribute positively in accordance with eq.~\eqref{sum rule eses}.
That being said, the pair of operators with spin $J_{odd}$ and $J_{odd}+1$ always act positively in order to balance the contribution of the $\mU$-channel identity.
Moreover, the sum rule associated with the derivative $B_{(2,0,0,0);1|34}'$ functional does not satisfy eq.~\eqref{sum rule eses}.
\begin{figure}[h]
\centering
\begin{subfigure}[b]{0.42\linewidth}
\centering
\includegraphics[width=\textwidth]{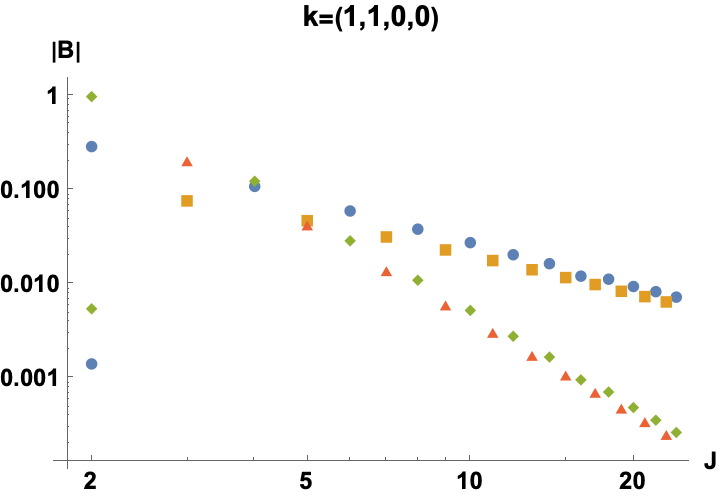}
\end{subfigure}
\begin{subfigure}[b]{0.55\linewidth}
\centering
\includegraphics[width=\textwidth]{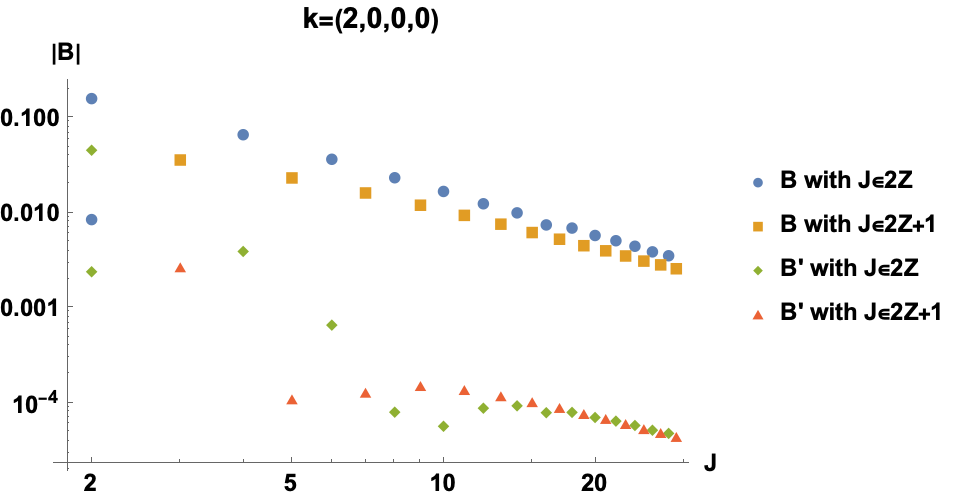}
\end{subfigure}
\caption{Log-log plot of the absolute value of the collinear $B_{\kv;1|34}$ and derivative $B_{\kv;1|34}'$ functionals acting on exchanged operators of the $\langle \epsilon \sigma \epsilon \sigma \rangle$ correlator. The left and right plots corresponds to the $\kv=(1,1,0,0)$ and $\kv=(2,0,0,0)$ subtraction schemes respectively.}
\label{fig:eses convergence}
\end{figure}
According to fig.~\ref{fig:eses convergence}, the large spin tail is convergent and therefore we conclude that this functional is sensitive to heavier operators not included in the OPE sum.
Fig.~\ref{spinGraph eses} shows the action of the $f_{\epsilon \sigma \OO}^{2}B_{\kv;1|34}$ functional for the $\langle \epsilon \sigma \epsilon \sigma \rangle$ correlator.

\subsubsection{$\langle \epsilon \sigma \sigma \epsilon \rangle$}
We now explore dispersive sum rules associated with the $\langle ABBA \rangle$ correlator:
\begin{equation} \label{sum rule esse}
\sum_{\OO \in \mathbb{Z}_2^+} f_{\sigma \sigma \OO} f_{\epsilon \epsilon \OO} B_{\OO}^t + \sum_{\PP \in \mathbb{Z}_2^-} f_{\epsilon \sigma \PP}^2 B_{\PP}^s = 0.
\end{equation}
Since double-twist operators are exchanged in the $s$-channel, we require at least one $s$-channel subtraction to obtain non-log functionals.
In the $t$-channel, depending on the operator ordering, we need at least one subtraction along one of the $t$-channel twist trajectory.
Therefore, WLOG, we study the $\langle \epsilon \sigma \sigma \epsilon \rangle$ correlator.
Given this operator ordering and the conditions in table~\ref{Tab:convergence}, we are left with a single allowable subtraction scheme: $\kv=(k_{12}, k_{23}, k_{34}, k_{14})=(1,1,0,0)$. Contributions to this sum rule is recorded in table~\ref{tab:esse operators}.

Due to the two double-twist families $[\sigma \sigma]_{n,J}$ and $[\epsilon \epsilon]_{n,J}$,  the $t$-channel collinear functional is not guaranteed to be sign-definite.
However, this only occurs for $2\D_\sigma+2 \geq \tau \geq 2\D_\epsilon$ where no operators are exchanged.
Moreover, the subtraction scheme guarantees that the functional possess the same sign for $t$-channel operators outside that rnage.
Therefore, the sum rule is satisfied by balancing sign-definite contributions from $s$-channel operators against sign-definite contribution from the $t$-channel separately.
This feature will not be preserved for the $\langle AABB \rangle$ correlator.
The action of the functional for varying twist at fixed spin is plotted in fig.~\ref{fig:ssee_odd} and \ref{fig:ssee_even} along with the next correlators.

\begin{table}[h]
\centering
\begin{tabular}{|c|c|c|c||c|c|}
\hline
$\OO$ & $\mathbb{Z}_2$ & $J$ & $\tau_\OO$ & $B_{(1,1,0,0);1|34}$ & $B_{(1,1,0,0);1|34}'$ \\ \hline
$\sigma$ & $-$ & 0 & $0.518149$ & $1$ & $1$  \\
$[\sigma \epsilon]_{0,J}$ & $-$ & 2 & $2.18031$ & $0.027$ & $-0.089$  \\
$\mathbf{1}$ & $+$ & 0 & 0 & $-0.465$ & $-0.628$  \\
$\epsilon$ & $+$ &0 & $1.41263$ & $-0.563$ & $-0.280$  \\
$T_{\mu \nu}$ & $+$ &2 & $3$ & $-0.011$ & $0.042$  \\ \hline
\multicolumn{4}{|c||}{$\sigma+\mathbf{1}+\epsilon$} & $-0.028$ & $0.092 $ \\
\multicolumn{4}{|c||}{Sum of all operators $B^s+B^t$} & $-0.002$ & $0.005$ \\ \hline
\end{tabular}
\caption{Contribution of each operator to the $\langle \epsilon \sigma \sigma \epsilon \rangle$ sum rule with $v=1$; these results include the contribution of the OPE coefficients $f_{O_i O_j \OO}f_{O_l O_k \OO}$. The values are normalized with respect to the largest contribution to the sum rule i.e. the $\sigma$ operator. All other operators contribute at order $O(10^{-3})$ and less. }
\label{tab:esse operators}
\end{table}
According to table~\ref{tab:esse operators}, 
this sum rule mostly balances the $\sigma$ operator against the identity and the $\epsilon$ operator.
The derivative sum rule enhances the identity contribution at the cost of lowering the contribution of the $\epsilon$ operator.


\subsubsection{$\langle \sigma \sigma \epsilon \epsilon \rangle$ and $\langle \epsilon \epsilon \sigma \sigma \rangle$}
Finally, we examine correlators of the form $\langle AABB \rangle$.
The sum rule for this correlator is the same as that of eq.~\eqref{sum rule esse}, but with $B^t \leftrightarrow B^s$ exchanged.
However, significant differences follow from the discussion about convergence in section \ref{sec:convergence}.

Based on table~\ref{Tab:convergence}, it is preferable to set $(\D_A,\D_B)=(\D_{\sigma},\D_\epsilon)$ i.e. to study the $\langle \sigma \sigma \epsilon \epsilon \rangle$ correlator;
this ordering minimizes the number of necessary $t$-channel subtractions by requiring that $k_{23},k_{14}>-0.447$ whereas the other operator ordering would require that $k_{23},k_{14}>0.447$.
Furthermore, the two $t$-channel subtractions are indistiguishable, while $s$-channel subtractions are unconstrained even for non-log functionals.
We therefore detail the action of the $B_{\kv;v|34}$ functional with subtraction schemes $(2,0,0,0)$,$(0,1,0,1)$, $(1,1,0,0)$ and $(0,2,0,0)$ for the $\langle \sigma \sigma \epsilon \epsilon \rangle$ correlator in the table~\ref{tab:ssee B operators}.
We present the action of the derivative functional $B_{\kv;v|34}'$ in table~\ref{tab:ssee B prime operators}.
\begin{table}[h]
\centering
\begin{tabular}{|c|c|c|c||c|c|c|c|}
\hline
$\OO$ & $\mathbb{Z}_2$ & $J$ & $\tau_\OO$ & $B_{(2,0,0,0);1|34}$ & $B_{(0,1,0,1);1|34}$ & $B_{(1,1,0,0);1|34}$ & $B_{(0,2,0,0);1|34}$ \\ \hline
$\sigma$ & $-$ & 0 & $0.518149$ & $1$ & $-0.311$ & $1$ & $-0.593$ \\
$[\sigma \epsilon]_{0,J}$ & $-$ & 2 & $2.18031$ & & $-0.207$ &  & $-0.224$ \\
$\mathbf{1}$ & $+$ & 0 & 0 & 0  & $1$ &0 & 0 \\
$\epsilon$ & $+$ &0 & $1.41263$ & $-0.621$ & $-0.496$ & $-0.586$ & $1$ \\
$T_{\mu \nu}$ & $+$ &2 & $3$ & $-0.226$ &  & $-0.418$ & \\
$ \epsilon' $ & $+$ & 0 & $3.82968$ & $-0.115$  &  &   & $-0.111$ \\ \hline
\multicolumn{4}{|c||}{Truncated sum }& $0.038$ & $-0.014$ & $0.006$ &$0.072$ \\
\multicolumn{4}{|c||}{Sum of all operators $B^s+B^t$} & $-0.05$ & $0.0003$ & $0.0003$ & $0.005$ \\ \hline
\end{tabular}
\caption{Contribution of each operator to the $B$ functional acting on the $\langle \sigma \sigma \epsilon \epsilon \rangle$ correlator with $v=1$; these results include the contribution of the OPE coefficients $f_{O_i O_j \OO}f_{O_l O_k \OO}$. The values are presented as ratios of the individual operator relative to the largest contribution to the sum rule. Empty cells contribute at order $O(10^{-2})$ and less.}
\label{tab:ssee B operators}
\end{table}

\begin{table}[h]
\centering
\begin{tabular}{|c|c|c|c||c|c|c|c|}
\hline
$\OO$ & $\mathbb{Z}_2$ & $J$ & $\tau_\OO$ & $B_{(2,0,0,0);1|34}'$ & $B_{(0,1,0,1);1|34)}'$ & $B_{(1,1,0,0);1|34}'$ & $B_{(0,2,0,0);1|34}'$ \\ \hline
$\sigma$ & $-$ & 0 & $0.518149$ & $1$ & $-0.116$ & $1$ & $-0.322$ \\
$[\sigma \epsilon]_{0,J}$ & $-$ & 2 & $2.18031$ & & $-0.412$ &  & $-0.586$ \\
$[\sigma \epsilon]_{0,J}$ & $-$ & 3 & $1.63804$ & & $-0.169$  &  & $0.191$ \\
$\mathbf{1}$ & $+$ & 0 & 0 & 0  & $1$ &0 & 0 \\
$\epsilon$ & $+$ &0 & $1.41263$ & $-0.268$ & $-0.358$ & $-0.296$ & $1$ \\
$T_{\mu \nu}$ & $+$ &2 & $3$ & $-0.695$ &  & $-0.719$ & $-0.231$ \\ \hline
\multicolumn{4}{|c||}{Truncated sum }& $0.037$ & $-0.055$ & $-0.015$ &$0.052$ \\
\multicolumn{4}{|c||}{Sum of all operators $B^s+B^t$} & $0.008$ & $0.003$ & $0.00007$ & $-0.005$ \\ \hline
\end{tabular}
\caption{Contribution of each operator to the derivative $B'$ functional acting on the $\langle \sigma \sigma \epsilon \epsilon \rangle$ correlator with $v=1$; these results include the contribution of the OPE coefficients $f_{O_i O_j \OO}f_{O_l O_k \OO}$. The values are presented as ratios of the individual operator relative to the largest contribution to the sum rule. Empty cells contribute at order $O(10^{-2})$ and less.}
\label{tab:ssee B prime operators}
\end{table}

One could also consider the $\langle \epsilon \epsilon \sigma \sigma \rangle$ correlator which would require a subtraction in both $t$-channel trajectories thereby restricting the space of twice-subtracted functionals to the $\kv=(0,1,0,1)$ subtraction scheme; we present numerical results for this sum rule in table~\ref{tab:eess operators}.
\begin{table}[h]
\centering
\begin{tabular}{|c|c|c|c||c|c|}
\hline
$\OO$ & $\mathbb{Z}_2$ & $J$ & $\tau_\OO$ & $B_{(0,1,0,1);v|34}$ & $B_{(0,1,0,1);v|34}'$ \\ \hline
$\sigma$ & $-$ & 0 & $0.518149$ & $1$ & $1$ \\
$\mathbf{1}$ & $+$ & 0 & 0 & $-0.473$ & $-0.628$ \\
$\epsilon$ & $+$ &0 & $1.41263$ & $-0.552$ & $-0.261$ \\ \hline
\multicolumn{4}{|c||}{Truncated sum }& $-0.025$ & $0.111$  \\
\multicolumn{4}{|c||}{Sum of all operators $B^s+B^t$} & $-0.0002$ & $-0.102$  \\ \hline
\end{tabular}
\caption{Contribution of each operator to the $\langle \epsilon \epsilon \sigma \sigma \rangle$ sum rule with $v=1$; these results include the contribution of the OPE coefficients $f_{O_i O_j \OO}f_{O_l O_k \OO}$. The values are presented as ratios of the individual operator relative to the identity contribution. Empty cells contribute at order $O(10^{-2})$ and less.}
\label{tab:eess operators}
\end{table}

\begin{figure}[h]
\centering
\includegraphics[width=0.95\textwidth]{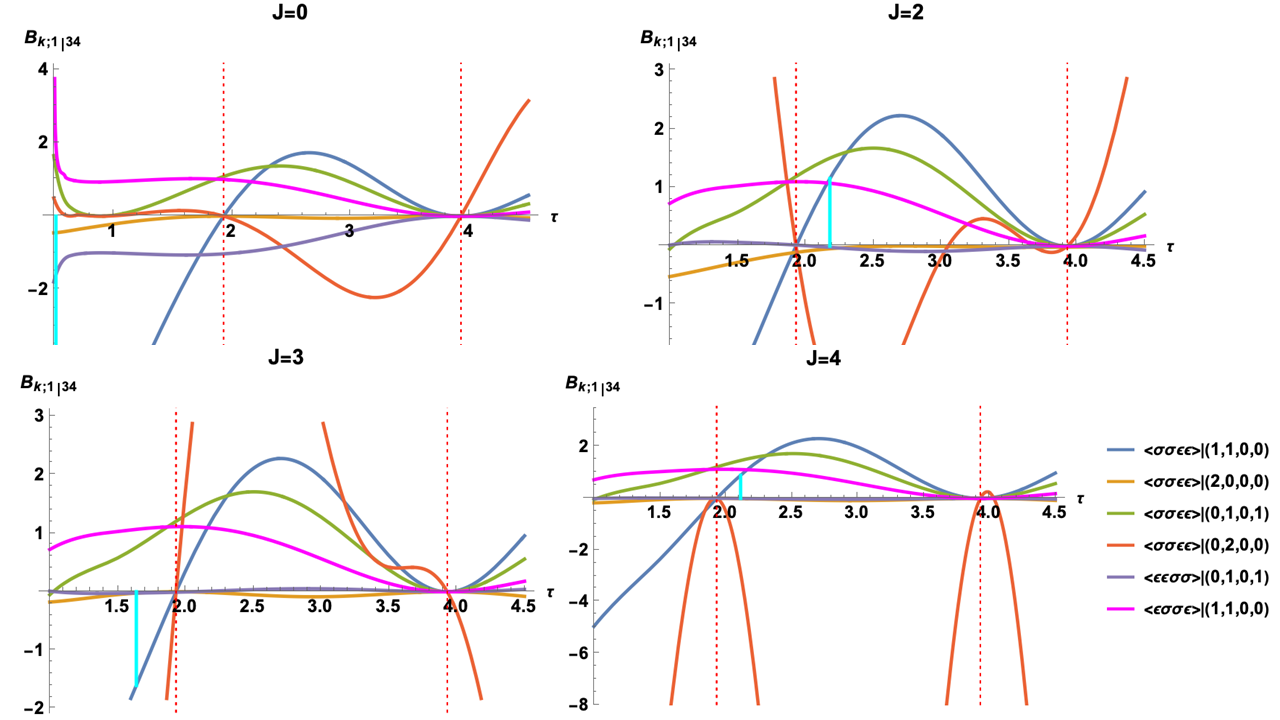}
\caption{$B_{\kv;1|34}$ for fixed spin $J$ as a function of twist $\tau$ for channels where odd $\mathbb{Z}_2$ operators are exchanged.  We have included all three correlators $\langle \sigma \sigma\epsilon\epsilon \rangle$, $\langle\epsilon\epsilon\sigma\sigma \rangle$ and $\langle \epsilon \sigma\sigma \epsilon \rangle$ to highlight differences between subtraction schemes and operator ordering.}
\label{fig:ssee_odd}
\end{figure}

\begin{figure}[h]
\centering
\includegraphics[width=0.95\textwidth]{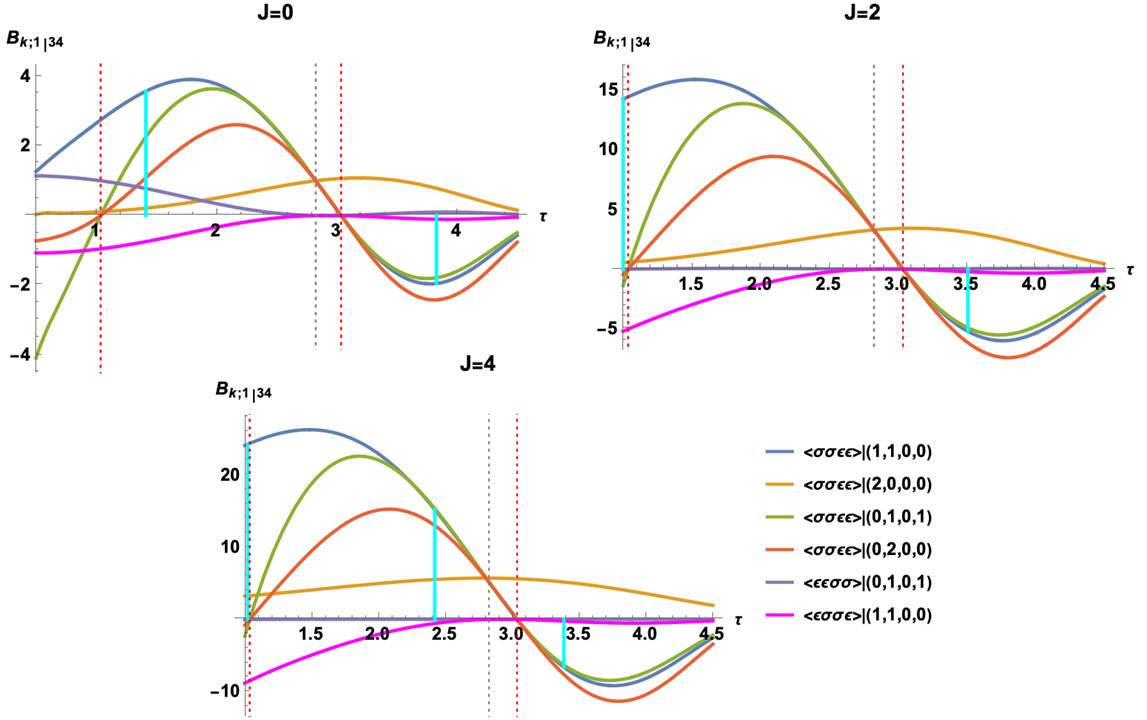}
\caption{$B_{\kv;1|34}$ for fixed spin $J$ as a function of twist $\tau$ for channels where even $\mathbb{Z}_2$ operators are exchanged.  We have included all three correlators $\langle \sigma \sigma\epsilon\epsilon \rangle$, $\langle\epsilon\epsilon\sigma\sigma \rangle$ and $\langle \epsilon \sigma\sigma \epsilon \rangle$ to highlight differences between subtraction schemes and operator ordering.}
\label{fig:ssee_even}
\end{figure}

To better understand the distinguishing features of these functionals, we plot their action as a function of twist for fixed spins in fig.~\ref{fig:ssee_odd} and \ref{fig:ssee_even}.
We highlight four features of these figures: their sign, the largest contributor to the $\langle \sigma \sigma \epsilon \epsilon \rangle$ correlator, the presence or absence of the identity exchange, and the difference between the $\langle \sigma \sigma \epsilon \epsilon \rangle$ and $\langle \epsilon \epsilon \sigma \sigma \rangle$ correlator with $(0,1,0,1)$ subtraction scheme.

\begin{figure}[h]
\centering
\includegraphics[width=0.7\textwidth]{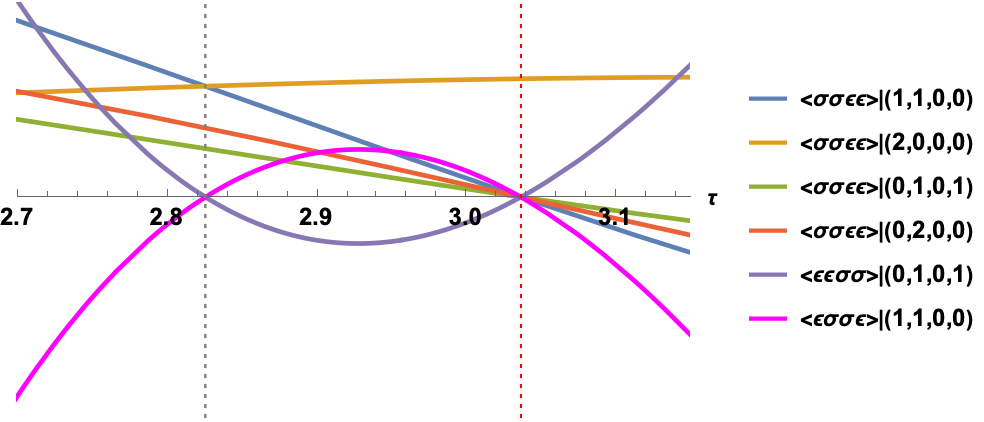}
\caption{Close-up of the action of $B_{\kv;1|34}$ functionals for even $\mathbb{Z}_2$ operators at $J=0$ near the leading $2\D_\epsilon$ (grey vertical dotted line) and subleading $2\D_\sigma+2$ (red vertical dotted line) double-twist values.}
\label{fig:ssee_j0_zoom}
\end{figure}

Let us first discuss the sign of these functionals.
The latter returns simple zeros on the two $s$-channel double-twist families.
By evaluating the functional on the $(mn)$ double-twist family, the funcitonal becomes non-zero on the corresponding double-twist trajectory as shown by fig.~\ref{fig:ssee_j0_zoom}.
Therefore, for $\tau< 2\D_\epsilon+2$, the sign of the functional is fixed by the number of $k_{12}$ subtractions.
In particular, only the $(2,0,0,0)$ subtraction scheme yields a sign definite functional acting on $\tau < 2\D_{\epsilon}+2$.
In the $t$-channel, sign definiteness is guaranteed for the $[\sigma\epsilon]_{0,J}$ family by including at least one $t$-channel subtraction.
Otherwise, there is a sign change due to a simple zero at $\tau=\D_\epsilon+\D_\sigma$ as shown by the orange curve in fig.~\ref{fig:ssee_odd}.
Therefore, we conclude that regardless of the subtraction scheme, at least one of the $s$- or $t$-channel functional will be sign-definite for most operators:
at large enough twist, the functionals become sign-definite except between the two double-twist families $[\epsilon \epsilon]_{n,J}$ and $[\sigma \sigma]_{n+1,J}$.

The zero structure sheds light into the contribution of various operators to the sum rule: by enhancing double-twist families with poles from $P_\kv$, we further enhance contributions from all operators exchanged in that channel.
For instances, as shown in table~\ref{tab:ssee B operators}, $s$-channel subtractions mostly balances a single odd $\mathbb{Z}_2$ operator against a small number of even $\mathbb{Z}_2$ operators.
While we cannot predict which operator will dominate the sum rule, we observe empirically
that the largest contributor to the sum rule is given by a low twist operator appearing in the channel where fewer subtractions are applied.

Let us briefly comment on the presence or absence of the identity in table~\ref{tab:ssee B operators}.
The identity saturates the lower bound in the Mellin-mandelstam space $\mt$ as defined by eq.~\eqref{mellin t contour}, and therefore, we expect the pole associated with the identity exchange to be located at the boundary $\mt=\D+\D_A-\D_B-2k_{34}$.
Such a singularity in the Mellin-Mandelstam $\mt$-plane only appears if we include $t$-channel subtractions which generates poles of the form $(\D+\D_A-\D_B-2k_{34}-\mt)$ to cancel double-zeros from the Mack polynomial.
Therefore, can cancel the double-zero by adding double poles with two $t$-channel subtractions $k_{23}$ and $k_{14}$ since the two trajectories are degenerate. 
By doing so, we can enhance the integrand with a double-poles which yields a residue of the form
\begin{equation}
\frac{\Gamma(\D)}{\Gamma(\D/2)^2} f(\D_A,\D_B) v^{-\D_B} \psi^{(0)}(\D/2).
\end{equation}
The above is non-zero as $\D \rightarrow 0$ due to the $\psi^{(0)}(\D/2)$ factor.
Thus, a non-vanishing $s$-channel identity only appears if we include symmetric subtractions along the $\tau_{23}$ and $\tau_{14}$ trajectories to enhance the Mellin integrand with a double-pole such that the identity contribution is
\begin{equation}
\widehat{B}_{(0101);v|34}^{s}[G_{0,0}^s]=-\D_B v^{-\D_B} \frac{\Gamma(\D_A-\D_B)\Gamma(\D_B)}{\Gamma(\D_A)}.
\end{equation}
This calculation also illustrates how the $\langle \sigma \sigma \epsilon \epsilon \rangle$ correlator differs from the $\langle \epsilon \epsilon \sigma \sigma \rangle$ one;
while not apparent from fig.~\ref{fig:ssee_odd} and \ref{fig:ssee_even} due to the scale, the asymmetric dependence on $\D_B$ suppresses the overall behaviour of $\widehat{B}_{\kv;v|34}$ for the $\langle \epsilon \epsilon \sigma \sigma \rangle$ correlator.
This non-linear suppression leads to better truncated sum rules as shown in table~\ref{tab:eess operators}.

Finally, we note that the values in table~\ref{tab:eess operators} are similar to those in table~\ref{tab:esse operators} and therefore, we do not consider them linearly independent sum rules.
This resemblance extends to apparent miror images between the $\langle \epsilon \epsilon \sigma \sigma \rangle$ and $\langle \epsilon \sigma \sigma \epsilon \rangle$ correlators for the spin $J=0$ curves in fig.~\ref{fig:ssee_odd}-\ref{fig:ssee_even}.
It is unclear why these two correlators and their repsective subtraction schemes yield such similar results despite possessing clearly distinct Mellin-space integrands.
Nevertheless, this emphasize that these mixed correlators can encode similar information through different operator ordering and subtraction schemes.

\subsection{Approximate solutions to crossing from truncated sum rules} \label{sec:approximate sol}
The previous section highlights the value of these dispersive functionals by suppressing the double-twist sectors thereby restricting the sum rule to a small number of low twist operators.
Using these partial sum rules, this section aims to derive approximate solutions to crossing.

Deriving approximate solutions to crossing is an important step towards bounding CFT data using numerical minimization methods.
Consider a linear combination of functionals denoted by $\mathcal{A}$ which approximately satisfy crossing for a finite number of of exchanged operators:
\begin{equation} \label{crossing functional}
\mathcal{A} = \sum_{i=1}^{n} a_i B_{\kv;v|34}^{s,t}[\OO_i] = 0,
\end{equation}
for coefficients $a_i$.
By adding functionals that act positively on heavy operators and negatively for a small finite number of light operators, one could use optimization methods to bound CFT data.
We take a first step towards tackling this optimization problem when using dispersive functionals by first evaluating $\mathcal{A}$ using insight from the previous subsection.

As shown in the first half of table~\ref{tab:sum rule summary}, there are 7 linearly independent sum rules that are restricted to three exchanged operators: $\sigma$, $\epsilon$ and $T_{\mu \nu}$.
These sum rules follow from three correlators:
\begin{subequations}
\begin{align}
\langle \sigma \sigma \sigma \sigma \rangle \quad :& \quad \mathbf{1}-\mathbf{1}_{\mU} + f_{\sigma \sigma \epsilon}^2 \ B[\epsilon] + f_{\sigma \sigma T}^2 \ B[T_{\mu \nu}] = 0 \\
& \quad f_{\sigma \sigma \epsilon}^2 \ \Phi_2[\epsilon] + f_{\sigma \sigma T}^2 \ \Phi_2[T_{\mu \nu}] = 0 \\
\langle \epsilon \sigma \sigma \epsilon \rangle \quad :& \quad \mathbf{1} + f_{\sigma \sigma \epsilon}f_{\epsilon\epsilon\epsilon} \ B[\epsilon] + f_{\sigma \sigma \epsilon}^2 \ B[\sigma] = 0 \\
\langle \sigma \sigma \epsilon \epsilon \rangle \quad :& \quad f_{\sigma \sigma \epsilon}^2 \ B[\sigma] + f_{\sigma \sigma \epsilon} f_{\epsilon \epsilon \epsilon} \ B[\epsilon]+ f_{\sigma \sigma T}f_{\epsilon \epsilon T} \ B[T_{\mu \nu}] = 0
\end{align}    
\end{subequations}
There are therefore 5 unknowns in this system of equations:
\begin{equation}
\D_\sigma, \quad \D_\epsilon, \quad, f_{\sigma \sigma \epsilon}, \quad f_{\epsilon \epsilon \epsilon}, \quad \sqrt{c} = \frac{\sqrt{\D_i \D_j }}{f_{ijT}}.
\end{equation}
Using our sum rules, we can find approximate solutions to crossing for 3D Ising model operators by minimizing the stress-tensor coupling according to a $c$-minimization procedure \cite{El-Showk:2014dwa}.

To implement this procedure,
we cancel the $f_{\epsilon \epsilon \epsilon}$ coupling by taking linear combination of the $B_{\kv;1|mn}$ and its derivative. Unlike the equal operator sum rule, the mixed correlator ones are not sign definite, and therefore we allow for the sum rule to be satisfied up to $\pm10^{-1}$ error given the errors tabulated in the previous section.
Using the notation $B_{\langle \phi_1 \phi_2 \phi_3 \phi_4 \rangle}[\OO]$ to denote the action of the $B_{(1,1,0,0);1|34}^{s,t}$ collinear functional with fixed operator ordering and exchange operator $\OO$, we then construct the following inequalities:
\begin{subequations} \label{ising inequalities}
\begin{align}
0 \leq & \quad B_{\sigma \sigma \sigma \sigma}\mathbf{1}]-\mathbf{1}_{\mU} + f_{\sigma \sigma \epsilon}^2 \ B_{\sigma \sigma \sigma \sigma}[\epsilon] + f_{\sigma \sigma T}^2 \ B_{\sigma \sigma \sigma \sigma}[T_{\mu \nu}]  \\
0 \leq  & \quad B_{\sigma \sigma \sigma \sigma}'[\mathbf{1}]-\mathbf{1}_{\mU} + f_{\sigma \sigma \epsilon}^2 \ B_{\sigma \sigma \sigma \sigma}'[\epsilon] + f_{\sigma \sigma T}^2 \ B_{\sigma \sigma \sigma \sigma}'[T_{\mu \nu}]  \\
0 \leq  & \quad f_{\sigma \sigma \epsilon}^2 \ \Phi_2[\epsilon] + f_{\sigma \sigma T}^2 \ \Phi_2[T_{\mu \nu}]  \\
-0.1 \leq & \quad  \left( B_{\epsilon \sigma \sigma \epsilon}[\mathbf{1}] - \frac{ B_{\epsilon \sigma \sigma \epsilon}[\epsilon] }{B_{\epsilon \sigma \sigma \epsilon}'[\epsilon]} B_{\epsilon \sigma \sigma \epsilon}'[\mathbf{1}] \right) \nonumber \\
& \qquad \quad + f_{\sigma \sigma \epsilon}^2 \left( B_{\epsilon \sigma \sigma \epsilon}[\sigma] - \frac{B_{\epsilon \sigma \sigma \epsilon}[\epsilon] }{ B_{\epsilon \sigma \sigma \epsilon}'[\epsilon]} B_{\epsilon \sigma \sigma \epsilon}'[\sigma]\right) \leq 0.1 \\
-0.1 \leq & \quad f_{\sigma \sigma \epsilon}^2 \left( B_{\sigma \sigma \epsilon \epsilon}[\sigma]  - \frac{B_{\sigma \sigma \epsilon \epsilon}[\epsilon] }{ B_{\sigma \sigma \epsilon \epsilon}'[\epsilon]} B_{\sigma \sigma \epsilon \epsilon}'[\sigma]\right) \nonumber \\
& \qquad \quad  + f_{\sigma \sigma T}f_{\epsilon \epsilon T} \left( B_{\sigma \sigma \epsilon \epsilon}[T_{\mu \nu}] - \frac{B_{\sigma \sigma \epsilon \epsilon}[\epsilon] }{ B_{\sigma \sigma \epsilon \epsilon}'[\epsilon]} B_{\sigma \sigma \epsilon \epsilon}'[T_{\mu \nu}]\right)  \leq 0.1
\end{align}
\end{subequations}
Since the functional $B_{\kv;v|34}^{s,t}$ is a non-linear function of $\D_\sigma, \D_\epsilon$, we sample points within a grid $(\D_\sigma,\D_\epsilon) \in ( [0.5, 0.66], [1.34,1.48])$ and interpolate its action within that domain. 
By minimizing the stress-tensor coupling $c$ using the \texttt{NMinimize} function in Mathematica, the best approximation to the Ising model data is recorded in tab~\ref{tab:approximate solution} for the four remaining unknowns in our system of equations.
\begin{table}[h] 
\centering
\begin{tabular}{c || c c c c}
  & $\D_\sigma$ & $\D_\epsilon$ & $f_{\sigma \sigma \epsilon}$ & $c$ \\  \hline
Results from \cite{ElShowk:2012ht} & $0.518149$ & $1.41263$ & $1.05185$ & $2.5241$ \\
Results from truncated dispersive sum rules & $0.512390$ & $1.42934$ & $1.01279$ & $2.6370$ 
\end{tabular}
\caption{3D Ising model data derived from 7 truncated dispersive sum rules that combine to provide 5 inequalities.  These results were obtained by sampling $N_{\D_\sigma} \times N_{\D_\epsilon} = 42 \times 40$ points distributed quadratically around their expected values. }
\label{tab:approximate solution}
\end{table}
We underscore the necessity to use all five inequalities to obtain reasonable crossing solutions despite the fact that the system of equations only includes 4 unknowns.

Despite the small grid size, our results are comparable to those in \cite{ElShowk:2012ht}.
We observed empirically that better accuracy can be obtained by increasing the number of data points within the grid.
One could also perform a two-dimensional quadratic (or higher order) fit of the $B_{\kv;v|34}$ functionals within a domain near the expect Ising model values, however
the mixed correlator functionals diverge near the unitarity bound thereby forcing one to increase the size of the dataset to obtain high precision fits; we found the use of interpolation functions to be better suited (although less rigorous) for smaller datasets.
This exercise simply served to illustrate the effectiveness of these dispersive functionals.
We expect higher precision is achievable by increasing the number of sampling data points, by increasing the number of functionals and inequalities, or by using combinations of functionals with only single-sided inequalities.

\section{Discussion}

In this paper, we extend the study of dispersive CFT functionals to mixed correlators.
We first conduct our analysis in position-space (section~\ref{sec:mixed pos}) where our main result is the derivation of the mixed correlator dispersive kernel given by eq.~\eqref{B34 sub ab}.
Convergence, positivity and analyticity properties are examined in section~\ref{sec:convergence} where we found that the first two depend on the subtraction scheme $\kv=(k_{12},k_{23},k_{34},k_{14})$ and the operator ordering; we summarized the convergence conditions in table~\ref{Tab:convergence}.
We conduct an analogous derivation of the functionals in
Mellin-space (section~\ref{sec:mixed mellin}) where the Mellin representation of the functional is given by eq.~\eqref{Bkt to Bkv}. 
The benefits of working in Mellin-space are threefold: the zero and pole structure is manifest, projection functionals similar to $\Phi_\ell$ are easier to construct, and they are easier to evaluate numerically.
This comes at the cost of obscuring convergence and positivity properties that were clearer in position-space.
Equipped with both the position-space and Mellin-space representation of the dispersive collinear functional $B_{\kv;v|mn}$, we extended the work of \cite{Caron-Huot:2021enk} by constructing mixed correlator holographic functionals $C_{\kv;\nu|mn}$ in sections~\ref{sec:pos Regge} and \ref{sec:mellin regge} in position- and Mellin-space respectively.

Our sum rules follow from eq.~\eqref{B sum rule} such that for large enough twist, each $s$- and $t$-channel functional become sign definite.
Therefore, cancellations occur between low twist operators which allows one to construct truncated sum rules.
We verify this statement in section~\ref{sec:Ising sum rule} by studying spin-2 convergent non-log sum rules applied to five 3D Ising correlators.
As mentioned previously, our dispersive sum rules are not invariant under operator ordering due to expanding in the collinear limit, and therefore we conduct an extensive analysis for sum rules with different subtraction schemes and operator ordering to find those dominated by the fewest lowest twist operators.
Our results
 indicate that crossing-symmetric subtraction schemes are the most effective in that regard since they enhance the fewest operators in each channel.
One may build other functionals similar to those in \cite{Penedones:2019tng,Carmi:2020ekr} to isolate other operators in the sum rule rather than the lowest twist ones.
Asymmetric subtraction schemes also appear to be useful for correlators of heavy external operators as indicative by results for the $\langle \epsilon \epsilon \epsilon \epsilon \rangle$ correlator in section~\ref{sec:Ising eeee}.

In section~\ref{sec:approximate sol}, we aimed to derive approximate solutions to crossing by using a set of truncated dispersive sum rules.
Using only 5 inequalities, two of which follow from mixed correlators, we were able to obtain reasonable approximations to the 3D Ising model as shown in table~\ref{tab:approximate solution} by evaluating the functionals within a grid of $(\D_\sigma, \D_\epsilon)$, and interpolating its action before minimizing the stress-tensor coupling.

While our method appears promising,
errors are difficult to constrain due to the lack of sign definiteness for mixed correlators sum rules, and our methodology requires sampling a large number of data points to find high precision minima. 
A much more effective method is to leverage positivity properties of these functionals by constructing a semidefinite matrix amenable to the use of semidefinite programming similar to the work of \cite{Kos:2014bka}. The authors constructed the following function as a basis element for their mixed correlator bootstrap implementation:
\begin{equation}
F^{ij,kl}_{\pm,\D,\ell}(u,v) \equiv v^{\frac{\D_k+\D_j}{2}} G_{\D,\ell}^{ij,kl}(u,v) \pm u^{\frac{\D_k+\D_j}{2}} G_{\D,\ell}^{ij,kl}(v,u).
\end{equation}
It would be interesting to find an analogous linear combination that would maximize the strengths of these dispersive functionals.
Such a construction would also serve to build better functionals described at the beginning of the previous subsection.
Moreover, it would be interesting to test whether navigator functions \cite{Reehorst:2021ykw,Reehorst:2021hmp} built from dispersive functionals may be more effective.

Beyond their numerical implementations, there are other avenues to explore.
Consider the $\Phi_2$ functional for example.
Convergence properties of this functional should be further studied in light of our results for the $\langle \epsilon \epsilon \epsilon \epsilon \rangle$ correlator in section~\ref{sec:Ising eeee}.
Furthermore, the numerical evaluation becomes expensive for large spin operators given the oscillatory nature of the Mack polynomials.
A position-space version of these functionals could prove more scalable since conformal blocks are better behaved at large spin.
One could further explore other types of projection operators for mixed correlators building upon the early analysis conducted in appendix~\ref{app:projection}.

Finally, the derivation of mixed correlator holographic functionals in sections~\ref{sec:mixed pos} and \ref{sec:mellin regge} provides a new pathway to probe certain features of AdS quantum gravity.
In particular, these functionals could serve helpful to study heavy-heavy-light-light CFT correlators which are interpreted as a probe particle travelling in a black hole geometry \cite{Karlsson:2019qfi,Li:2019zba,Li:2020dqm}.
Furthermore, the simplicity in eq.~\eqref{Ckv position} and \eqref{Ckv mellin} encourages the construction of spinning functionals that would allow one to bootstrap gravitons in AdS which may further shed light on the relations between flat-space and AdS scattering.

\section*{Acknowledgements}
We thank Simon Caron-Huot, Frank Coronado, Cl\'ement Virally and Zahra Zahraee for discussions throughout this project.
The work of AT is supported by the National Science and Engineering Council of Canada and the Simons Collaboration on the Nonperturbative Bootstrap.

\begin{appendix}
\section{Detailed derivation of the $\BB_{\kv|mn}^{\mathfrak{a},\mathfrak{b}}$ kernel}
\label{app:B kernel}
To be explicit, we derive the $\BB_{\kv|34}^{\mathfrak{a},\mathfrak{b}}$ kernel by starting with eq.~\eqref{Kernel general definition} which is a triple integral in the Mellin mandelstam variables $\ms', \mt'$ and $\ms$.
We first deform the $\ms$ contour to evaluate the residue at $s=\D_3+\D_4 + 2k_{34}$, followed by explicitly resuming the $\mt'$ residues at $\D_2+\D_3+2k_{23}+2m$ and $\D_1+\D_4+2k_{14}+2m$ by deforming the contour to the right..
Finally, the $s'$ contour is deformed to the left to pick up the poles at $\D_1+\D_2+k_{12}-2m$.
Let us consider explicitly the resumed poles at $\mt'=\D_2+\D_3+2k_{23}+2m$:
\begin{equation}
\begin{split}
&\BB_{\kv|34}^{\mathfrak{a},\mathfrak{b}} \bigg|_{\mt'{=}\D_2{+}\D_3{+}2k_{23}{+}2m} = \sum \displaylimits_{m_s=0}^{\infty} \frac{ \left(u'\right)^{m_s-k_{12}} \left(v'\right)^{p+q+k_{12}-k_{23}-k_{34}-m_s-2} v^{k_{23}} (-1)^{m_s-k_{12}} u^{-p-q+k_{34}} }{2 \pi  \Gamma \left(p+q+k_{12}-k_{34}-m_s-1\right)\Gamma \left(m_s+1\right)} \\
& \quad \times \frac{\Gamma \left(p-q+k_{14}-k_{23}\right) \Gamma \left(p+q+k_{12}-k_{34}\right) \csc \left(\pi  \left(p+q+k_{12}-m_s-1\right)\right)}{\left(p+q+k_{12}-k_{34}-m_s-1\right) \Gamma \left(-p-q-k_{12}+k_{34}+m_s+1\right)} \\
& \quad \times \frac{{}_2F_1\big({-}p{-}q{-}k_{12}{+}k_{34}{+}m_s{+}2,{-}2 p{-}k_{12}{-}k_{14}{+}k_{23}{+}k_{34}{+}m_s{+}2;{-}p{+}q{-}k_{14}{+}k_{23}{+}1;\frac{v}{v'}\big)}{\Gamma \left(2 p+k_{12}+k_{14}-k_{23}-k_{34}-m_s-1\right)},
\end{split}
\end{equation}
where $m_s$ labels the poles originating from the $\ms'$ contour deformation, and
\begin{equation}
p=\frac{\D_{13}}{2}, \qquad q=\frac{\D_{24}}{2}.
\end{equation}
Unfortunately, this sum cannot be performed straightforwardly.
To progress, we write an ansatz of a hypergeometric as a function of $\chi$ as defined by eq.~\eqref{chit var} and we match its arguments term-by-term by expanding around $u',v \rightarrow 0$.
By doing so, we find
\begin{equation}
\begin{split}
& \BB_{\kv|34}^{\mathfrak{a},\mathfrak{b}} \bigg|_{\mt'{=}\D_2{+}\D_3{+}2k_{23}{+}2m} = \frac{(-1)^{1-k_{12}} 2^{2 p+k_{12}+k_{14}-k_{23}-k_{34}-4} }{\pi ^2 \Gamma \left(2 p+k_{12}+k_{14}-k_{23}-k_{34}-1\right)}  \left(-u'+v'+v\right)  \\
&\quad \times v^{\frac{1}{2} \left(2 p+k_{12}+k_{14}+k_{23}-k_{34}-3\right)} \chi ^{\frac{1}{2} \left(-2 p-k_{12}-k_{14}+k_{23}+k_{34}+3\right)} u^{-p-q+k_{34}} \left(u'\right)^{-k_{12}}\left(v'\right)^{\frac{1}{2} \left(2 q+k_{12}-k_{14}-k_{23}-k_{34}-3\right)} \\
& \quad \times \csc \left(\pi  \left(p+q+k_{12}\right)\right) \Gamma \left(p-q+k_{14}-k_{23}\right) \Gamma \left(p+q+k_{12}-k_{34}\right) \sin \left(\pi  \left(p+q+k_{12}-k_{34}\right)\right) \\
& \quad \times \, _2F_1\left(\tfrac{1}{2} \left(2 q{+}k_{12}{-}k_{14}{+}k_{23}{-}k_{34}\right),\tfrac{1}{2} \left({-}2 p{-}k_{12}{-}k_{14}{+}k_{23}{+}k_{34}{+}3\right);{-}p{+}q{-}k_{14}{+}k_{23}{+}1;{-}\chi \right).
\end{split}
\end{equation}
We can now combine the two $\mt'$ towers of poles by using the following identity:
\begin{equation}
\begin{split}
{}_2F_1(a,b,c,z) &= \frac{(-z)^{-a} (\Gamma (c) \Gamma (b-a))}{\Gamma (b) \Gamma (c-a)}  \, _2F_1\left(a,a-c+1;a-b+1;\frac{1}{z}\right) \\
& \qquad +\frac{(-z)^{-b} (\Gamma (c) \Gamma (a-b))}{\Gamma (a) \Gamma (c-b)} \, _2F_1\left(b-c+1,b;-a+b+1;\frac{1}{z}\right),
\end{split}
\end{equation}
which allows us to obtain our final quoted result in eq.~\eqref{B34 sub ab}.

To derive the full dispersion kernel for mixed correlators, one would have to resum poles originating from the $\ms$-contour.
A well-posed ansatz to resum these poles would be a hypergeometric function with the ratio defined on the LHS of eq.~\eqref{chit var} as an argument.
One could fix the $p,q$ dependence by matching the series expansion.

\section{$\BB_{\kv|mn}^{\mathfrak{a},\mathfrak{b}}$ kernel simplification} \label{app:kernel simp}
We explain simplicity of the equal operator kernel found in \cite{Caron-Huot:2020adz}.
For concreteness, let us focus on the $\BB_{\kv|34}^{\mathfrak{a},\mathfrak{b}}(u',v')$ kernel which contains the following hypergeometric function:
\begin{equation}
\BB_{\kv|34}^{\mathfrak{a},\mathfrak{b}} \supset \, _2F_1\left(\tfrac{k_{12}+k_{14}-k_{23}+\D_{13}}{2},\tfrac{k_{12}-k_{14}+k_{23}+\D_{24}}{2};k_{12}+\tfrac{\D_{13}+\D_{24}}{2}-\tfrac{1}{2};-\tfrac{1}{\chi} \right).
\end{equation}
Without loss of generality, we set $k_{34}=0$ in the above.
Since hypergeometric functions ${}_2F_1$ are symmetric in its first two arguments, we see that subtractions in $k_{23}$ and $k_{14}$ are antisymmetric relative to a fixed point determined by the $\mU$-channel double-twist values $\tfrac{\D_{13}}{2}$ and $\tfrac{\D_{24}}{2}$.

For pairwise external operators, this hypergeometric function simplifies for the three cases where double-twist operators can be exchanged in the $s$-channel; this corresponds to the cases where $\D_{13}=-\D_{24}$ or $\D_{13}=\D_{24}=0$.
In such cases, the third argument of the hypergeomtric function reduces to $k_{12}-\tfrac{1}{2}$.
When $k_{12}=1$, the hypergeomtric function further reduces to
\begin{equation}
\begin{split}
\BB_{(1,k_{23},0,k_{14})|34}^{\mathfrak{a},\mathfrak{b}} \bigg|_{\D_{13}=-\D_{24}} &\supset {}_2F_1(\tfrac{1}{2}+y,\tfrac{1}{2}-y,\tfrac{1}{2},-\tfrac{1}{\chi}) = \frac{\left(\frac{\sqrt{\chi +1}-1}{\sqrt{\chi }}\right)^{y}+\left(\frac{\sqrt{\chi +1}-1}{\sqrt{\chi }}\right)^{-y}}{2 \sqrt{\frac{1}{\chi }+1}}.
\end{split}
\end{equation}
where $y=k_{14}-k_{23}+ \D_{13}$.
This explains why the equal operator kernel in \cite{Caron-Huot:2020adz} was simpler: a single subtraction along the $\tau_{12}$ trajectory allows for this reduction.

For the $\langle AABB \rangle$ where $\D_{13}=\D_{24}$, there is no
possible (generic) simplification since $\D_{13} \in \mathbb{R}$ and the $\D_{13}$ dependence in the third argument of the hypergeometric function prevents any further simplifications.

\section{Mack Polynomials} \label{app:Mack}
We write explicit formulae for Mack polynomials with unequal external operators which play a key role in describing Mellin amplitudes $M(\ms,\mt)$.
Mack Polynomials are well-documented \cite{Mack:2009mi,Costa:2012cb,Gopakumar:2016cpb,Gopakumar:2018xqi}.
We hope to provide a more compact notation that may hopefully help readers\footnote{We thank Simon Caron-Huot for experimentations that led to this formulae, and for sharing the Mathematica notebook.}.

Mellin amplitudes have poles when the Mellin mandelstam variable $\ms$ approaches the twist of exchanged operators $\OO$:
\begin{equation}
M(\ms,\mt) \propto M^s(\ms,\mt) \sim \frac{\QQ_{\D_\OO,J_\OO}^{m,\D_i}(\mt)}{\ms-\tau_\OO - 2m},
\end{equation}
where the kinematical polynomial $\mathcal{Q}_{\D,J}^{m,\D_i}$ takes the form
\begin{equation}
\mathcal{Q}_{\D,J}^{m,\D_i}(\mt) = k_{\D,J}^{m,\D_i} \ Q_{\D,J}^{m,a,b}(\mt), 
\end{equation}
where $a,b$ is given by eq.~\eqref{a,b for blocks}, and
\begin{align}
k_{\D,J}^{m,\D_i} &=  \frac{1}{\Gamma \left(\frac{1}{2} \left(J+\Delta +\Delta _1-\Delta _2\right)\right) \Gamma \left(\frac{1}{2} \left(J+\Delta -\Delta _1+\Delta _2\right)\right) \Gamma \left(\frac{1}{2} \left(J-\Delta +\Delta _1+\Delta _2\right)-m\right) } \nonumber \\
&\times \frac{1}{ \Gamma \left(\frac{1}{2} \left(J+\Delta -\Delta _3+\Delta _4\right)\right) \Gamma \left(\frac{1}{2} \left(J+\Delta +\Delta _3-\Delta _4\right)\right) \Gamma \left(\frac{1}{2} \left(J-\Delta +\Delta _3+\Delta _4\right)-m\right)} \nonumber \\
&\times  \frac{2 \Gamma (J+\Delta -1) \Gamma (J+\Delta )}{\Gamma (\Delta -1) \Gamma (m+1) \left(-\frac{d}{2}+\Delta +1\right)_m }. \label{kfactor Q}
\end{align}
$Q_{\D,J}^{m,a,b}(\mt)$ is the so-called Mack polynomial.

Below, we reorganize results from \cite{Costa:2012cb,Gopakumar:2016cpb} by first generalizing the Mack polynomials from \cite{Gopakumar:2016cpb}, and then verifying against the appendices in \cite{Costa:2012cb}.
The sums in \cite{Gopakumar:2016cpb} were then re-organized in order to isolate the $J$ dependence at low $m$ for $Q_{\D,J}^{0,a,b}$.
The recursion relation reads
\begin{equation}
\begin{split}
Q_{\D,J}^{m,a,b}(t) &= \sum \limits_{q=0}^{\text{min}(m,J)} \sum \limits_{p=0}^q \mathfrak{q}_{\D,J}^{a,b}(p,q) \  (-m)_q  \ Q_{\D + p,J-q}^{0,a - \tfrac{q - p}{2},b+\tfrac{q-p}{2}}(\mt),
\end{split}
\end{equation}
where
\begin{equation}
\begin{split}
\mathfrak{q}_{\D,J;d}^{a,b}(p,q)&= \frac{J! (-1)^{p+q}}{p! (J-q)! (q-p)!} \frac{\left(\frac{d-2}{2}+J\right)_{-q} (J+\Delta -1)_{p-q}}{\left(a+\frac{J+\Delta }{2}\right)_{p-q} \left(\frac{J+\Delta }{2}-b\right)_{p-q}} \\
& \qquad \times \frac{\left(\frac{1}{2} (-d-J+\Delta +2)+a\right)_p \left(\frac{1}{2} (-d-J+\Delta +2)-b\right)_p}{(-d-J+\Delta +2)_p},
\end{split}
\end{equation}
and the $m=0$ Mack polynomial is given by the following closed-form expression
\begin{equation}
\begin{split}
Q_{\D,J}^{0,a,b}(\mt) &= \frac{\left(a+\frac{\Delta -J}{2}\right)_J \left(b+\frac{\Delta -J}{2}\right)_J}{(\Delta -1)_J} \, _3F_2\left(-J,-\frac{\mt}{2},\Delta -1;a+\frac{\Delta }{2}-\frac{J}{2},b+\frac{\Delta }{2}-\frac{J}{2};1\right).
\end{split}
\end{equation}

\section{Mixed correlator analytic double-twist functionals} \label{app:projection}
In \cite{Caron-Huot:2020adz}, a projection functional $\Phi_\ell$ was introduced with the following properties:
\begin{enumerate}
\item It is positivie definite above a threshold $\tau^*$.
\item It returns double-zeros on all double-twists with spin $J \in 2\mathbb{Z} \neq \ell$.
\item It returns a simple zero on the leading double-twist $n=0$ for $J=\ell \in 2\mathbb{Z}$.
\end{enumerate}
These properties are well suited for the numerical bootstrap.
To study these projection operators, it is convenient to expand the $\widehat{B}_{\kv;\mt|mn}$ Mellin functionals using the basis of analytic double-twist functionals introduced earlier in eq.~\eqref{B analytic funct exp} which we rewrite here schematically\footnote{For mixed correlators, there are additional labels to identify the expansion trajectory and the double-twist family.} in Mellin-space for convenience:
\begin{equation}
M_{\kv}^{s,t}(\ms,\mt) = \sum_{n,\ell} \alpha_{n,\ell}^{s,t}[M_{\kv}] \widehat{a}_{n,\ell}^{\D_i,s,t}(\ms,\mt) + \beta_{n,\ell}^{s,t} [M_{\kv}] \widehat{b}_{n,\ell}^{\D_i,s,t}(\ms,\mt).
\end{equation}
By evaluating the Mellin amplitude (correlator) in the $s$-channel collinear limit, we restrict ourselves to the leading double-twist $n=0$ sector.
In general, for the collinear $\widehat{B}_{\kv;\mt|mn}$ functionals, we have for arbitrary subtraction schemes
\begin{equation}
\widehat{B}_{\kv;\mt|mn}^{s,t} = \sum_{\ell} \alpha_{0,\ell}^{s,t} \; \widehat{a}_{\ell|mn}^{\D_i,s,t}(\mt) + \beta_{0,\ell}^{s,t} \; \widehat{b}_{\ell|mn}^{\D_i,s,t}(\mt),
\end{equation}
where the expansion coefficients were defined in eq.~\eqref{B analytic funct exp}we will drop the descendant label $n$ since $n=0$ for $\widehat{B}_{\kv;\mt|mn}$ functionals.
Some coefficients $\widehat{a}$ or $\widehat{b}$ may be absent depending on the correlator configuration and the subtraction scheme.
For example,
the $\langle AABB \rangle$ correlator with $\kv=(2,0,0,0)$ can be expanded in the following basis:
\begin{align}
\widehat{B}^t_{(2,0,0,0);v|34} &= O((\D-\D_A-\D_B-J)^2), \\
\widehat{B}^s_{(2,0,0,0);v|34} &= \sum_{\ell} \alpha_{\ell,2\D_A}^{s} \widehat{a}_{\ell|34,2\D_A}^{\D_i,s}(\mt) + \beta_{\ell,2\D_A}^s \widehat{b}_{\ell|34,2\D_A}^{\D_i,s}(\mt) \nonumber \\
& + \alpha_{\ell,2\D_B}^{s} \widehat{a}_{\ell|34,2\D_B}^{\D_i,s}(\mt) + \beta_{\ell,2\D_B}^s \widehat{b}_{\ell|34,2\D_B}^{\D_i,s}(\mt),
\end{align}
where we have added an additional subscript to $\widehat{a},\widehat{b}$ to label the double-twist family.
The presence of $\widehat{a},\widehat{b}$ can be understood from the zero structure of Mack polynomials and subtraction schemes as discussed at the end of section~\ref{sec:mellin derivation}.
This foreshadows a prominent challenge in constructing a projection functional for mixed correlators: it is difficult to find a basis that diagonalizes the sum rule in both $s$- and $t$-channels.
This subsection is devoted to elaborating on this remark.

Before tackling the mixed correlator case, let us briefly sketch and review the derivation of the projection functional $\Phi_\ell$ for equal operators first introduced in \cite{Caron-Huot:2020adz}.
Such a functional is obtained by integrating $\widehat{B}_{2,\mt}$ against a kernel $f_\ell(\mt)$
\begin{equation}
\Phi_{\ell} = \int\displaylimits_{\D_\phi-i \infty}^{\D_\phi+i \infty} \frac{d\mt}{2\pi i} f_{\ell}(\mt) \widehat{B}_{2,\mt},
\end{equation}
where $f_{\ell}(\mt)$ can be shown to be
\begin{equation}
f_{\ell}(\mt) = \frac{(2\D_\phi+2\ell-1)\Gamma(\D_\phi+\ell)^4 \Gamma(2\D_\phi+\ell-1)}{\Gamma(\ell+1)\Gamma(2\D_\phi+2\ell)^2} \int \displaylimits_{\D_\phi}^{\mt} dx \ \Gamma_{\D_\phi}^4(x) a_{\ell}(x).
\end{equation}
$\Phi_{\ell}$ is evaluated by using integration by parts.
The second property at the top of this subsection can be satisfied by leveraging the orthogonality relations between Mack polynomials.
The third property is satisfied if $f_\ell(\mt)$ is an odd function with respect to $\mt$:
\begin{equation} \label{odd sym}
f_{\ell}(\mt) = - f_{\ell}(2\D_\phi - \mt).
\end{equation}
Together, properties 2 and 3 suggest that $f_{\ell}(\mt) \propto \widehat{b}_\ell(\mt)$.
To obtain a concise closed-form expression for $f_{\ell}(\mt)$, one can relate the linear double-twist coefficient $\widehat{b}_{\ell}(\mt)$ to the zeroth order coefficient $\widehat{a}_\ell(\mt)$ as follows:
\begin{equation} \label{bhat to ahat}
\widehat{b}_\ell (\mt) - \widehat{b}_\ell(2\D_\phi-\mt) = - \frac{d \widehat{a}_\ell(\mt)}{d\mt},
\end{equation}
thereby allowing us to use orthogonality properties of the Mack polynomial.
To better understand the importance of this symmetry, let us consider the equal operator case such that we lose the double-twist family subscript $(mn)$.
One can show that for the spin-2 subtracted $\widehat{B}_{2;\mt}$ equal operator functional, the individual $s$- and $t$-channel Mack polynomial allow for the following expansion around double-twist values:
\begin{align}
\widehat{B}^{s}_{2;\mt} \big|_{\D=2\D_\phi+\ell} &= \widehat{a}^s_{\ell}(\mt) + \widehat{b}^s_{\ell}(\mt) ( \D - 2\D_\phi - \ell)  + O((\D - 2\D_\phi - \ell)^2),\\
\widehat{B}^{t}_{2;\mt} \big|_{\D=2\D_\phi+\ell} &= \widehat{b}^t_{\ell}(\mt) ( \D - 2\D_\phi - \ell)  + O((\D - 2\D_\phi - \ell)^2).
\end{align}
Crucially, it is the combination $\widehat{b}_{\ell}^{s} + \widehat{b}_{\ell}^{t} = \widehat{b}_{\ell}$ that benefits from the symmetry of eq.~\eqref{bhat to ahat}, and not the individual $s$- or $t$-channel terms.
Moreover, the symmetry that kills even spin, eq.~\eqref{odd sym}, was already present at the level of the $\widehat{a}_{\ell}^s=\widehat{a}_{\ell}$ coefficient which is entirely captured by the $s$-channel Mack polynomials.

For mixed correlators, the prominent barrier to constructing a kernel $f_{\ell|mn}(\mt)$ is the lack of symmetry that could relate odd (even) spins $\ell$.
To see this, let us explicitely evaluate the $\widehat{B}_{\kv;v|mn}$ functional for the $\langle ABAB \rangle$ correlator with $(mn)=(34)$ and $\kv=(1,1,0,0)$ for which the analytic double-twist functional expansion is the same as for $\widehat{B}_{2;\mt}$ for equal operators, but evaluated at $\D_A+\D_B+\ell$ instead.
Doing so, we find
\begin{equation}
\begin{split}
\widehat{a}_{\ell|34}^{\D_A,\D_B,\D_A,\D_B} &= {}_3F_2(\{-\ell,-1+\ell+\D_A+\D_B,\tfrac{\D_A+\D_B-t}{2}\}; \{\D_A,\D_B\}; 1) \\
& (-1)^{\ell/2} \frac{\Gamma(\ell+\D_A)\Gamma(\ell+\D_B)}{\Gamma(\D_A)\Gamma(\ell+\D_B)\Gamma(\ell+1)}.
\end{split}
\end{equation}
In contrast to the equal operator case, this hypergeometric function lacks any symmetry with respect to the Mandelstam variable $\mt$ which would allow us to write a variation of eq.~\eqref{odd sym}.
One could rewrite $\widehat{b}_{\ell}^{s,t}$ in terms of $\widehat{a}_{\ell}^{s,t}$ similar to eq.~(4.56) of \cite{Caron-Huot:2020adz}, but the lack of symmetry in Mandelstam $\mt$ reflects the inability for these Mack polynomials to diagonalize the OPE in both channels for all spin.
It may be possible to derive projection operators for even and odd spins separately for the $\langle ABAB \rangle$ correlator, but
we leave such investigations for future work.


\section{Varying $v$ for the $\langle \sigma \sigma \sigma \sigma \rangle$ sum rule} \label{app:varying v}

\begin{figure}[h]
\centering
\includegraphics[width=0.8 \textwidth]{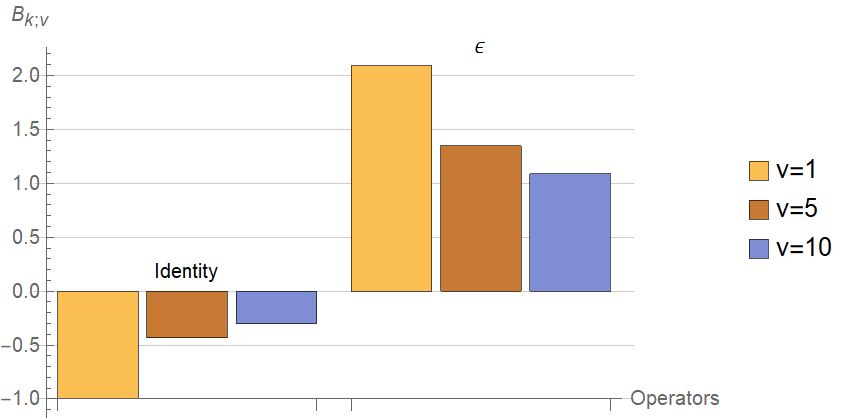}
\caption{Contribution of certain operators to the sum rule of $\langle \sigma \sigma \sigma \sigma \rangle$ with the subtraction scheme $(k_{12},k_{23},k_{34},k_{14})=(1,1,0,0)$. Contribution of other operators are of order $O(10^{-2})$ and less. The sum of all operators lead to $0.998$, $0.999$ and $0.999$ for the three values of $v$ in the plot respectively.}
\label{fig:ssss Bsum barchart}
\end{figure}

\begin{figure}[h]
\centering
\begin{subfigure}{0.48 \textwidth}
\includegraphics[width=\linewidth]{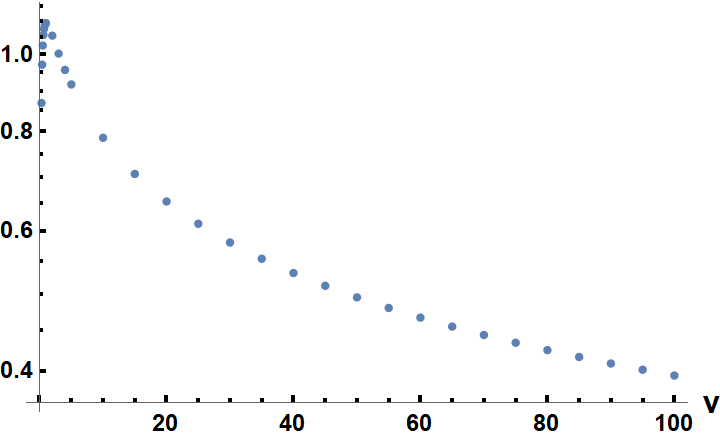}
\end{subfigure}
\hfill
\begin{subfigure}{0.48 \textwidth}
\includegraphics[width=\linewidth]{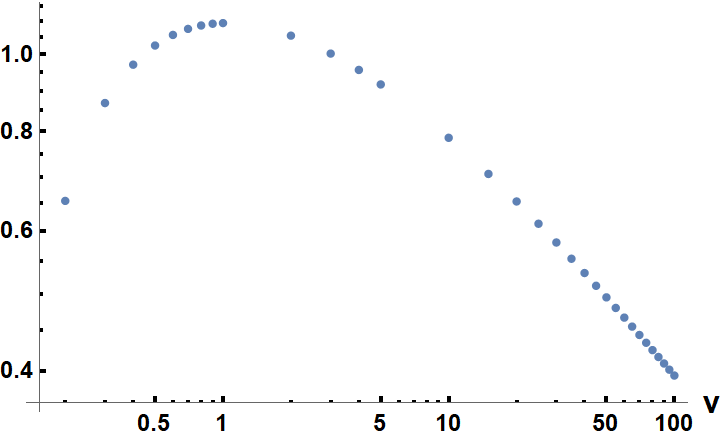}
\end{subfigure}
\caption{Ratio of the truncated sum over operators belonging to $N_3$ over the contribution of all operators $N_\infty$. The right panel is a log-log plot of the same data to highlight the small $v$ limit.}
\label{fig:ssss varyingV}
\end{figure}

We consider the $\langle \sigma \sigma \sigma \sigma \rangle$ correlator discussed in section \ref{sec:Ising sum rule}.
Figure~\ref{fig:ssss Bsum barchart} shows that the contribution of the most dominant operators diminish as we increase $v$. 
This is supported by table~\ref{tab:ssss N3 operators} which shows the contribution of the sum rule for the subset of operators $N_3=\{ \mathbf{1},\epsilon,T_{\mu \nu} \}$, while figure~\ref{fig:ssss varyingV} shows how much these operators saturate the sum rule as we vary $v$.
Increasing $v$ translates into integrating a smaller subset of the lightcone. As we shrink the integration domain, the contribution of all operators vanishes as a power-law.

\begin{table}[h]
\centering
\begin{tabular}{c|| c c c c}
& $v=1$ & $v=5$ & $v=10$ & $v=100$ \\ \hline
$\frac{ \sum_{\{\OO_i \} \in N_3 } f_{\sigma \sigma \OO_i}^2 B_{\kv;v|34} }{\sum_{\{\OO_i \} \in N_\infty } f_{\sigma \sigma \OO_i}^2 B_{\kv;v|34}}$ &$1.093$ &$0.915$ & $0.784$ & $0.405$
\end{tabular}
\caption{$N_3 = \{ \mathbf{1}, \epsilon, T_{\mu \nu} \}$ and $N_\infty$ denote the set of all possible exchanged operators.}
\label{tab:ssss N3 operators}
\end{table}

\end{appendix}

\clearpage

\bibliographystyle{JHEP}
\bibliography{references}

\providecommand{\href}[2]{#2}\begingroup\raggedright\begin{thebibliography}{10}

\bibitem{Rattazzi:2008pe}
R.~Rattazzi, V.~S. Rychkov, E.~Tonni, and A.~Vichi, {\it {Bounding scalar
  operator dimensions in 4D CFT}},  {\em JHEP} {\bf 12} (2008) 031,
  [\href{http://arxiv.org/abs/0807.0004}{{\tt arXiv:0807.0004}}].

\bibitem{Poland:2018epd}
D.~Poland, S.~Rychkov, and A.~Vichi, {\it {The Conformal Bootstrap: Theory,
  Numerical Techniques, and Applications}},  {\em Rev. Mod. Phys.} {\bf 91}
  (2019) 015002, [\href{http://arxiv.org/abs/1805.04405}{{\tt
  arXiv:1805.04405}}].

\bibitem{Kravchuk:2020scc}
P.~Kravchuk, J.~Qiao, and S.~Rychkov, {\it {Distributions in CFT. Part I.
  Cross-ratio space}},  {\em JHEP} {\bf 05} (2020) 137,
  [\href{http://arxiv.org/abs/2001.08778}{{\tt arXiv:2001.08778}}].

\bibitem{Kravchuk:2021kwe}
P.~Kravchuk, J.~Qiao, and S.~Rychkov, {\it {Distributions in CFT. Part II.
  Minkowski space}},  {\em JHEP} {\bf 08} (2021) 094,
  [\href{http://arxiv.org/abs/2104.02090}{{\tt arXiv:2104.02090}}].

\bibitem{ElShowk:2012ht}
S.~El-Showk, M.~F. Paulos, D.~Poland, S.~Rychkov, D.~Simmons-Duffin, and
  A.~Vichi, {\it {Solving the 3D Ising Model with the Conformal Bootstrap}},
  {\em Phys. Rev. D} {\bf 86} (2012) 025022,
  [\href{http://arxiv.org/abs/1203.6064}{{\tt arXiv:1203.6064}}].

\bibitem{El-Showk:2014dwa}
S.~El-Showk, M.~F. Paulos, D.~Poland, S.~Rychkov, D.~Simmons-Duffin, and
  A.~Vichi, {\it {Solving the 3d Ising Model with the Conformal Bootstrap II.
  c-Minimization and Precise Critical Exponents}},  {\em J. Stat. Phys.} {\bf
  157} (2014) 869, [\href{http://arxiv.org/abs/1403.4545}{{\tt
  arXiv:1403.4545}}].

\bibitem{Kos:2014bka}
F.~Kos, D.~Poland, and D.~Simmons-Duffin, {\it {Bootstrapping Mixed Correlators
  in the 3D Ising Model}},  {\em JHEP} {\bf 11} (2014) 109,
  [\href{http://arxiv.org/abs/1406.4858}{{\tt arXiv:1406.4858}}].

\bibitem{Caron-Huot:2020ouj}
S.~Caron-Huot, Y.~Gobeil, and Z.~Zahraee, {\it {The leading trajectory in the
  2+1D Ising CFT}},  \href{http://arxiv.org/abs/2007.11647}{{\tt
  arXiv:2007.11647}}.

\bibitem{Kos:2013tga}
F.~Kos, D.~Poland, and D.~Simmons-Duffin, {\it {Bootstrapping the $O(N)$ vector
  models}},  {\em JHEP} {\bf 06} (2014) 091,
  [\href{http://arxiv.org/abs/1307.6856}{{\tt arXiv:1307.6856}}].

\bibitem{Kos:2016ysd}
F.~Kos, D.~Poland, D.~Simmons-Duffin, and A.~Vichi, {\it {Precision Islands in
  the Ising and $O(N)$ Models}},  {\em JHEP} {\bf 08} (2016) 036,
  [\href{http://arxiv.org/abs/1603.04436}{{\tt arXiv:1603.04436}}].

\bibitem{Chester:2019ifh}
S.~M. Chester, W.~Landry, J.~Liu, D.~Poland, D.~Simmons-Duffin, N.~Su, and
  A.~Vichi, {\it {Carving out OPE space and precise $O(2)$ model critical
  exponents}},  {\em JHEP} {\bf 06} (2020) 142,
  [\href{http://arxiv.org/abs/1912.03324}{{\tt arXiv:1912.03324}}].

\bibitem{Heemskerk:2009pn}
I.~Heemskerk, J.~Penedones, J.~Polchinski, and J.~Sully, {\it {Holography from
  Conformal Field Theory}},  {\em JHEP} {\bf 10} (2009) 079,
  [\href{http://arxiv.org/abs/0907.0151}{{\tt arXiv:0907.0151}}].

\bibitem{Penedones:2010ue}
J.~Penedones, {\it {Writing CFT correlation functions as AdS scattering
  amplitudes}},  {\em JHEP} {\bf 03} (2011) 025,
  [\href{http://arxiv.org/abs/1011.1485}{{\tt arXiv:1011.1485}}].

\bibitem{Rastelli:2016nze}
L.~Rastelli and X.~Zhou, {\it {Mellin amplitudes for $AdS_5\times S^5$}},  {\em
  Phys. Rev. Lett.} {\bf 118} (2017), no.~9 091602,
  [\href{http://arxiv.org/abs/1608.06624}{{\tt arXiv:1608.06624}}].

\bibitem{Alday:2017vkk}
L.~F. Alday and S.~Caron-Huot, {\it {Gravitational S-matrix from CFT dispersion
  relations}},  {\em JHEP} {\bf 12} (2018) 017,
  [\href{http://arxiv.org/abs/1711.02031}{{\tt arXiv:1711.02031}}].

\bibitem{Caron-Huot:2018kta}
S.~Caron-Huot and A.-K. Trinh, {\it {All tree-level correlators in
  AdS$_{5}$\texttimes{}S$_{5}$ supergravity: hidden ten-dimensional conformal
  symmetry}},  {\em JHEP} {\bf 01} (2019) 196,
  [\href{http://arxiv.org/abs/1809.09173}{{\tt arXiv:1809.09173}}].

\bibitem{Goncalves:2019znr}
V.~Gon\c{c}alves, R.~Pereira, and X.~Zhou, {\it {$20'$ Five-Point Function from
  $AdS_5\times S^5$ Supergravity}},  {\em JHEP} {\bf 10} (2019) 247,
  [\href{http://arxiv.org/abs/1906.05305}{{\tt arXiv:1906.05305}}].

\bibitem{Kaviraj:2015cxa}
A.~Kaviraj, K.~Sen, and A.~Sinha, {\it {Analytic bootstrap at large spin}},
  {\em JHEP} {\bf 11} (2015) 083, [\href{http://arxiv.org/abs/1502.01437}{{\tt
  arXiv:1502.01437}}].

\bibitem{Simmons-Duffin:2016wlq}
D.~Simmons-Duffin, {\it {The Lightcone Bootstrap and the Spectrum of the 3d
  Ising CFT}},  {\em JHEP} {\bf 03} (2017) 086,
  [\href{http://arxiv.org/abs/1612.08471}{{\tt arXiv:1612.08471}}].

\bibitem{Mazac:2016qev}
D.~Mazac, {\it {Analytic bounds and emergence of AdS$_{2}$ physics from the
  conformal bootstrap}},  {\em JHEP} {\bf 04} (2017) 146,
  [\href{http://arxiv.org/abs/1611.10060}{{\tt arXiv:1611.10060}}].

\bibitem{Mazac:2018mdx}
D.~Mazac and M.~F. Paulos, {\it {The analytic functional bootstrap. Part I: 1D
  CFTs and 2D S-matrices}},  {\em JHEP} {\bf 02} (2019) 162,
  [\href{http://arxiv.org/abs/1803.10233}{{\tt arXiv:1803.10233}}].

\bibitem{Mazac:2018ycv}
D.~Mazac and M.~F. Paulos, {\it {The analytic functional bootstrap. Part II.
  Natural bases for the crossing equation}},  {\em JHEP} {\bf 02} (2019) 163,
  [\href{http://arxiv.org/abs/1811.10646}{{\tt arXiv:1811.10646}}].

\bibitem{Mazac:2018qmi}
D.~Maz\'a\v{c}, {\it {A Crossing-Symmetric OPE Inversion Formula}},  {\em JHEP}
  {\bf 06} (2019) 082, [\href{http://arxiv.org/abs/1812.02254}{{\tt
  arXiv:1812.02254}}].

\bibitem{Paulos:2019gtx}
M.~F. Paulos, {\it {Analytic functional bootstrap for CFTs in $d > 1$}},  {\em
  JHEP} {\bf 04} (2020) 093, [\href{http://arxiv.org/abs/1910.08563}{{\tt
  arXiv:1910.08563}}].

\bibitem{Mazac:2019shk}
D.~Maz\'a\v{c}, L.~Rastelli, and X.~Zhou, {\it {A Basis of Analytic Functionals
  for CFTs in General Dimension}},  \href{http://arxiv.org/abs/1910.12855}{{\tt
  arXiv:1910.12855}}.

\bibitem{Paulos:2020zxx}
M.~F. Paulos, {\it {Dispersion relations and exact bounds on CFT correlators}},
   \href{http://arxiv.org/abs/2012.10454}{{\tt arXiv:2012.10454}}.

\bibitem{Kaviraj:2021cvq}
A.~Kaviraj, {\it {Crossing antisymmetric Polyakov blocks + Dispersion
  relation}},  \href{http://arxiv.org/abs/2109.02658}{{\tt arXiv:2109.02658}}.

\bibitem{Carmi:2019cub}
D.~Carmi and S.~Caron-Huot, {\it {A Conformal Dispersion Relation: Correlations
  from Absorption}},  {\em JHEP} {\bf 09} (2020) 009,
  [\href{http://arxiv.org/abs/1910.12123}{{\tt arXiv:1910.12123}}].

\bibitem{Caron-Huot:2020adz}
S.~Caron-Huot, D.~Mazac, L.~Rastelli, and D.~Simmons-Duffin, {\it {Dispersive
  CFT Sum Rules}},  \href{http://arxiv.org/abs/2008.04931}{{\tt
  arXiv:2008.04931}}.

\bibitem{Caron-Huot:2021enk}
S.~Caron-Huot, D.~Mazac, L.~Rastelli, and D.~Simmons-Duffin, {\it {AdS Bulk
  Locality from Sharp CFT Bounds}},
  \href{http://arxiv.org/abs/2106.10274}{{\tt arXiv:2106.10274}}.

\bibitem{Penedones:2019tng}
J.~Penedones, J.~A. Silva, and A.~Zhiboedov, {\it {Nonperturbative Mellin
  Amplitudes: Existence, Properties, Applications}},  {\em JHEP} {\bf 08}
  (2020) 031, [\href{http://arxiv.org/abs/1912.11100}{{\tt arXiv:1912.11100}}].

\bibitem{Carmi:2020ekr}
D.~Carmi, J.~Penedones, J.~A. Silva, and A.~Zhiboedov, {\it {Applications of
  dispersive sum rules: $\epsilon$-expansion and holography}},
  \href{http://arxiv.org/abs/2009.13506}{{\tt arXiv:2009.13506}}.

\bibitem{Gopakumar:2021dvg}
R.~Gopakumar, A.~Sinha, and A.~Zahed, {\it {Crossing Symmetric Dispersion
  Relations for Mellin Amplitudes}},  {\em Phys. Rev. Lett.} {\bf 126} (2021),
  no.~21 211602, [\href{http://arxiv.org/abs/2101.09017}{{\tt
  arXiv:2101.09017}}].

\bibitem{Meltzer:2021bmb}
D.~Meltzer, {\it {Dispersion Formulas in QFTs, CFTs, and Holography}},  {\em
  JHEP} {\bf 05} (2021) 098, [\href{http://arxiv.org/abs/2103.15839}{{\tt
  arXiv:2103.15839}}].

\bibitem{Caron-Huot:2020cmc}
S.~Caron-Huot and V.~Van~Duong, {\it {Extremal Effective Field Theories}},
  {\em JHEP} {\bf 05} (2021) 280, [\href{http://arxiv.org/abs/2011.02957}{{\tt
  arXiv:2011.02957}}].

\bibitem{Caron-Huot:2021rmr}
S.~Caron-Huot, D.~Mazac, L.~Rastelli, and D.~Simmons-Duffin, {\it {Sharp
  Boundaries for the Swampland}},  {\em JHEP} {\bf 07} (2021) 110,
  [\href{http://arxiv.org/abs/2102.08951}{{\tt arXiv:2102.08951}}].

\bibitem{Mack:2009mi}
G.~Mack, {\it {D-independent representation of Conformal Field Theories in D
  dimensions via transformation to auxiliary Dual Resonance Models. Scalar
  amplitudes}},  \href{http://arxiv.org/abs/0907.2407}{{\tt arXiv:0907.2407}}.

\bibitem{Penedones:2016voo}
J.~Penedones, {\it {TASI lectures on AdS/CFT}},  in {\em {Theoretical Advanced
  Study Institute in Elementary Particle Physics}: {New Frontiers in Fields and
  Strings}}, pp.~75--136, 2017.
\newblock \href{http://arxiv.org/abs/1608.04948}{{\tt arXiv:1608.04948}}.

\bibitem{Gopakumar:2016wkt}
R.~Gopakumar, A.~Kaviraj, K.~Sen, and A.~Sinha, {\it {Conformal Bootstrap in
  Mellin Space}},  {\em Phys. Rev. Lett.} {\bf 118} (2017), no.~8 081601,
  [\href{http://arxiv.org/abs/1609.00572}{{\tt arXiv:1609.00572}}].

\bibitem{Gopakumar:2016cpb}
R.~Gopakumar, A.~Kaviraj, K.~Sen, and A.~Sinha, {\it {A Mellin space approach
  to the conformal bootstrap}},  {\em JHEP} {\bf 05} (2017) 027,
  [\href{http://arxiv.org/abs/1611.08407}{{\tt arXiv:1611.08407}}].

\bibitem{Gopakumar:2018xqi}
R.~Gopakumar and A.~Sinha, {\it {On the Polyakov-Mellin bootstrap}},  {\em
  JHEP} {\bf 12} (2018) 040, [\href{http://arxiv.org/abs/1809.10975}{{\tt
  arXiv:1809.10975}}].

\bibitem{Haldar:2019prg}
P.~Haldar and A.~Sinha, {\it {Froissart bound for/from CFT Mellin amplitudes}},
   {\em SciPost Phys.} {\bf 8} (2020) 095,
  [\href{http://arxiv.org/abs/1911.05974}{{\tt arXiv:1911.05974}}].

\bibitem{Caron-Huot:2017vep}
S.~Caron-Huot, {\it {Analyticity in Spin in Conformal Theories}},  {\em JHEP}
  {\bf 09} (2017) 078, [\href{http://arxiv.org/abs/1703.00278}{{\tt
  arXiv:1703.00278}}].

\bibitem{Kologlu:2019bco}
M.~Kologlu, P.~Kravchuk, D.~Simmons-Duffin, and A.~Zhiboedov, {\it {Shocks,
  Superconvergence, and a Stringy Equivalence Principle}},  {\em JHEP} {\bf 11}
  (2020) 096, [\href{http://arxiv.org/abs/1904.05905}{{\tt arXiv:1904.05905}}].

\bibitem{Hogervorst:2013sma}
M.~Hogervorst and S.~Rychkov, {\it {Radial Coordinates for Conformal Blocks}},
  {\em Phys. Rev. D} {\bf 87} (2013) 106004,
  [\href{http://arxiv.org/abs/1303.1111}{{\tt arXiv:1303.1111}}].

\bibitem{Reehorst:2021ykw}
M.~Reehorst, S.~Rychkov, D.~Simmons-Duffin, B.~Sirois, N.~Su, and B.~van Rees,
  {\it {Navigator Function for the Conformal Bootstrap}},  {\em SciPost Phys.}
  {\bf 11} (2021) 072, [\href{http://arxiv.org/abs/2104.09518}{{\tt
  arXiv:2104.09518}}].

\bibitem{Reehorst:2021hmp}
M.~Reehorst, {\it {Rigorous bounds on irrelevant operators in the 3d Ising
  model CFT}},  \href{http://arxiv.org/abs/2111.12093}{{\tt arXiv:2111.12093}}.

\bibitem{Karlsson:2019qfi}
R.~Karlsson, M.~Kulaxizi, A.~Parnachev, and P.~Tadi\'c, {\it {Black Holes and
  Conformal Regge Bootstrap}},  {\em JHEP} {\bf 10} (2019) 046,
  [\href{http://arxiv.org/abs/1904.00060}{{\tt arXiv:1904.00060}}].

\bibitem{Li:2019zba}
Y.-Z. Li, {\it {Heavy-light Bootstrap from Lorentzian Inversion Formula}},
  {\em JHEP} {\bf 07} (2020) 046, [\href{http://arxiv.org/abs/1910.06357}{{\tt
  arXiv:1910.06357}}].

\bibitem{Li:2020dqm}
Y.-Z. Li and H.-Y. Zhang, {\it {More on heavy-light bootstrap up to
  double-stress-tensor}},  {\em JHEP} {\bf 10} (2020) 055,
  [\href{http://arxiv.org/abs/2004.04758}{{\tt arXiv:2004.04758}}].

\bibitem{Costa:2012cb}
M.~S. Costa, V.~Goncalves, and J.~Penedones, {\it {Conformal Regge theory}},
  {\em JHEP} {\bf 12} (2012) 091, [\href{http://arxiv.org/abs/1209.4355}{{\tt
  arXiv:1209.4355}}].

\end{thebibliography}\endgroup

\end{document}